\documentclass{aa}  

\usepackage[colorlinks=true,
            linkcolor=blue,
            citecolor=blue,
            urlcolor=blue]{hyperref}
            
\usepackage{xcolor} 
\usepackage{longtable,array,multirow}
\usepackage{graphicx}
\usepackage{txfonts}
\usepackage{lipsum}
\usepackage{multirow}
\usepackage{subcaption}         
\usepackage{lscape}             
\usepackage{placeins}           

\makeatletter
\renewcommand\tablefoottext[2]{%
  \par\addvspace{2pt}%
  \noindent
  \textsuperscript{#1}\,#2\par
}
\makeatother 
                                
\begin{document}

   \title{Jets and winds from the most powerful compact object in NGC\,5408}


   \author{ R. Soria
            \inst{1,2}\fnmsep\thanks{Corresponding author}
            \and
            M. W. Pakull
          \inst{3}
          \and
          C. Motch\inst{3}
          \and
          T. D. Russell\inst{4}
          \and
          F. Gris\'e\inst{5}
          }

   \institute{INAF-Osservatorio Astrofisico di Torino, Strada Osservatorio 20, I-10025 Pino Torinese, Italy\\
                 \email{roberto.soria@inaf.it}
                 \and
              Sydney Institute for Astronomy, School of Physics A28, The University of Sydney, Sydney, NSW 2006, Australia
              \and
   Observatoire Astronomique, CNRS, Universit\'e de Strasbourg, UMR 7550, F-67000 Strasbourg, France
              \and
              INAF, Istituto di Astrofisica Spaziale e Fisica Cosmica, Via U. La Malfa 153, I-90146 Palermo, Italy
         \and
         Department of Astronomy and Astrophysics, The Pennsylvania State University, 525 Davey Lab, University Park, PA 16802, USA
             }

   \date{Received 2026 June 18; accepted 2026 October 24}



%

 
  \abstract
   {Super-critical accretion onto stellar-mass compact objects is sometimes revealed by ultraluminous X-ray sources (ULXs).  In other cases, X-rays are heavily obscured along our line of sight, but a super-Eddington accretion power can be indirectly revealed by multiband studies.
}
   {We have identified a candidate super-critical source in NGC\,5408 (labelled X-2). We argue that: it is powered by accretion; it is among the most powerful in the local universe, despite its apparent X-ray faintness; it has exceptional optical properties (both in the continuum and in the lines).
} 
  %
   {We carried out a multiband imaging and spectral study with the Australia Telescope Compact Array, the Very Large Array, the {\it{Hubble Space Telescope}}, the Very Large Telescope and the {\it{Chandra X-ray Observatory}}.
  }
  %
   {We found an unresolved source with: variable radio emission (5-GHz luminosity of a few $\times 10^{35}$ erg s$^{-1}$); red optical continuum with absolute magnitudes  $M_{V,0} \approx -9.1$, $M_{I,0} \approx -10.6$; exceptionally broad low-ionization emission lines (full width at half maximum $\approx$1700 km s$^{-1}$) with P-Cygni profiles; narrow, photoionized emission lines including strong He\,{\footnotesize{II}} $\lambda$4686; point-like hard X-ray emission; marginally resolved soft X-ray emission. We also found an $\approx$20-pc He$^{++}$ nebula near X-2, associated with a candidate Wolf-Rayet star. The optical continuum declined by 2 mag between 2010 and 2024, as did the intensity of broad and narrow lines. We argue that the system is powered by the radiative and mechanical output of a super-Eddington stellar mass black hole, and estimate an accretion power $\gtrsim$10$^{40}$ erg s$^{-1}$.
   }
%
   {We propose that X-2 is a Compton-thick ULX seen at high inclination, with a flaring jet and X-ray emission emerging from the polar funnel. We argue that the optical continuum and the broad emission lines are signatures of an optically thick outflow, launched either from the accretion disk, or from an LBV-like donor star, or, more speculatively, caused by the long sought-for, elusive phase of common envelope ejection. We suggest that we are looking at a short-lived phase of hyper-Eddington accretion.
}


   \maketitle
\nolinenumbers

\section{Introduction}

Multiband studies of galaxies in the local universe over the last two decades  have revealed a population of stellar-mass compact objects accreting well above their Eddington limit \citep{king23,pinto23}. Such sources provide a test of accretion physics in the super-critical regime, and of the connection between inflows, radiation and outflows \citep{poutanen07,ohsuga11,narayan17,asahina24}. They also help us model the fast growth of supermassive black holes (BHs) in the early universe, and their shock-ionization and photo-ionization effects on the surrounding gas \citep{volonteri15,regan19,maiolino24}.

In particular, the large majority of empirically-defined ultraluminous X-ray sources (ULXs), that is off-nuclear compact objects with X-ray luminosity $L_{\rm{X}} > 10^{39}$ erg s$^{-1}$, are stellar mass BHs or neutron stars in the super-critical accretion regime, receiving gas from a massive companion star \citep{bachetti16,king23,pinto23}. A small fraction of ULXs may instead be intermediate-mass BHs \citep{mezcua15,mezcua17}, young supernova remnants (SNRs), beamed sources or mis-identified background quasars. Conversely, many but not all stellar mass BHs and neutron stars in the super-critical accretion regime appear as ULXs. The main reason why some super-critical sources are not very X-ray bright is that they are seen at high inclination, so that most of their X-ray photons along our line of sight are absorbed or scattered by an optically thick outflow or a geometrically thick disk \citep{poutanen07,narayan17}; many of them would look like ULXs if they were seen face-on, along the polar funnel. It is also plausible that some stellar-mass compact objects have a long-term-average accretion rate above the Eddington rate, but they undergo limit cycles instabilities and are currently in a low accretion, low luminosity state.

Theoretical models and numerical simulations suggest that compact objects in super-critical accretion regime release a comparable amount of power in the form of kinetic energy of winds and jets, as via photon emission \citep{poutanen07,jiang14,hashizume15,shen16}. In fact, the launching of massive, fast outflows is one of the hallmarks of super-critical accretion. As a consequence, many super-critical compact objects are associated with a large bubble of shock-ionized gas, expanding into the local interstellar medium (ISM) and powered by the kinetic energy of the outflow. Such shock-ionized bubbles are usually referred to as ULX bubbles \citep{pakull02,pakull03,pakull06,pakull10,soria10,cseh12,soria21,gurpide22}. This terminology does not imply that we always see the accreting compact object as a ULX; the compact object inside a ULX bubble may be temporarily in an off/low state, or it may be X-ray luminous but occulted along our line of sight \citep{pakull10}.

   \begin{figure}[t!]
   \centering
   \includegraphics[width=\hsize]{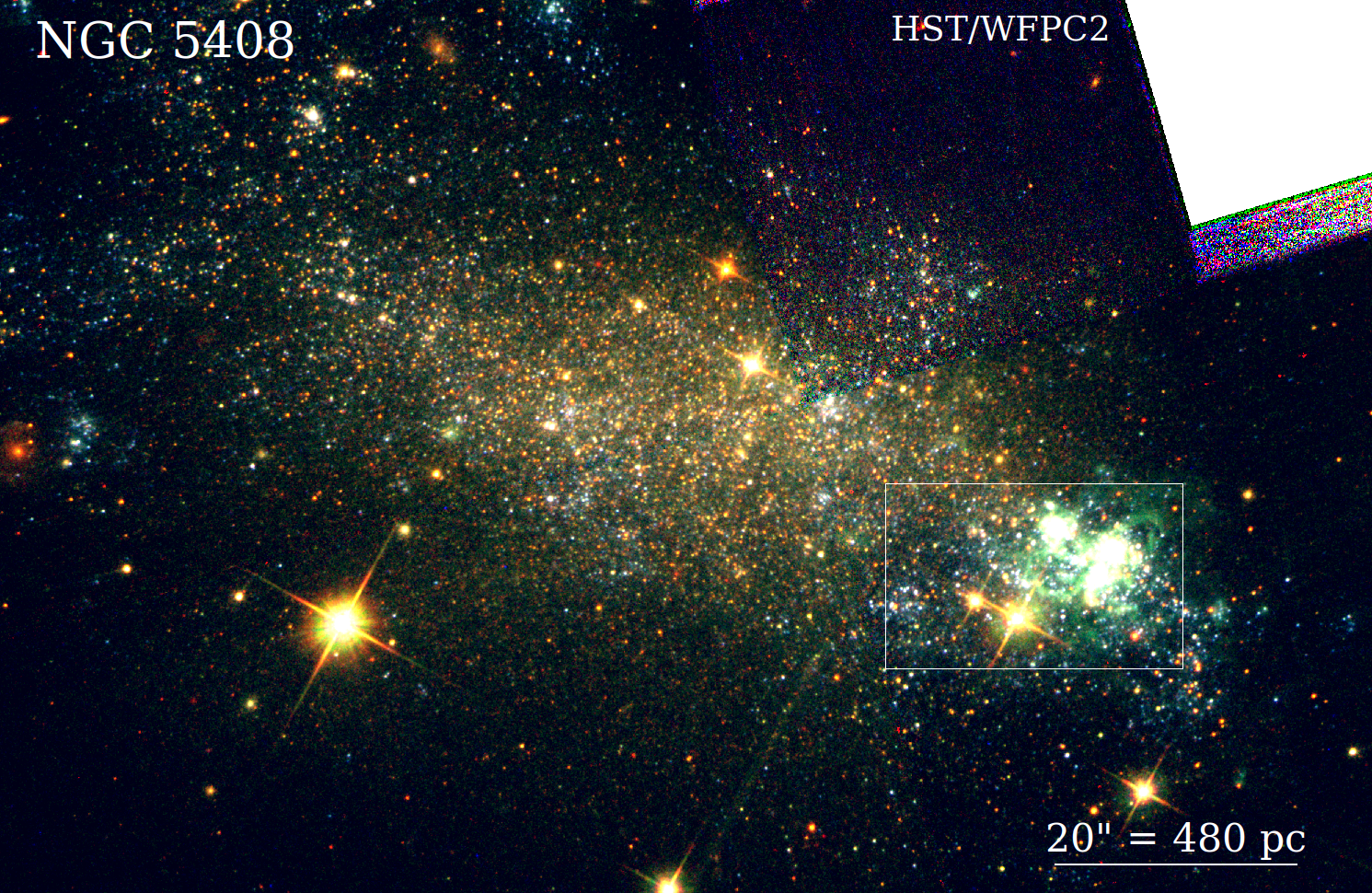}\\
   \includegraphics[width=\hsize]{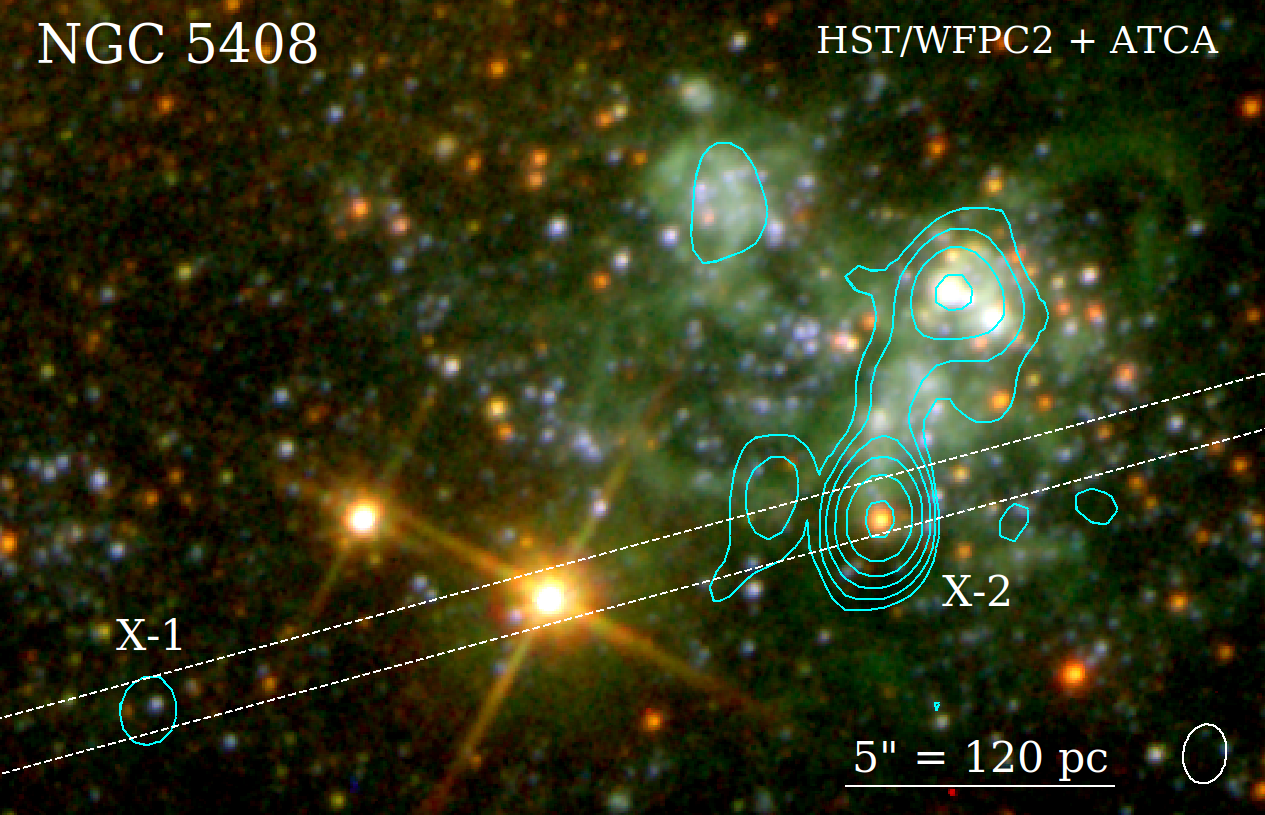}
      \caption{
      Top panel: true-colour optical image of NGC\,5408, from {\it{HST}}/WFPC2; red is the F814W band ($\approx$I), green is F555W ($\approx$V), blue is F336W ($\approx$U). All three bands were observed on 2009 April 4. Here and in all other images in this paper, north is up and east to the left. The starburst region at the south-west end of NGC\,5408 (inside the white box) is displayed in greater details in the bottom panel. 
      Bottom panel: zoomed-in view of the most actively star-forming region of NGC\,5408, with radio contours (cyan) superposed.
      The ATCA radio contours combine the data from the 5.5 GHz and 9.0 GHz bands from 2009 August 22; a Briggs robust parameter of $-$2 was used. Flux density contours correspond to $2^{n/2} \times$ rms noise level, where $n = 3,5,7,9,11,13$ and the rms noise level is $\approx$20 $\mu$Jy. The size of the synthesized beam is shown at the bottom right, and corresponds to the 9-GHz beam in Table 1. X-1 is a well-studied ULX, the brightest X-ray source in NGC\,5408. X-2, the subject of our study, is the red, bright star-like object exactly coincident with the brightest radio source in the galaxy. The dotted line indicates the size and position of the 1\farcs0 FORS2 slit during the 2010 VLT observation: designed to study X-1, it fortuitously covered also X-2.}
         \label{fig1}
   \end{figure}

ULX bubbles look somewhat similar to SNRs, but are often larger (diameters of $\sim$100--300 pc) and more energetic (energy content of $\sim$10$^{52}$ erg). While in SNRs all the energy is injected at the initial moment of the stellar collapse, in ULX bubbles the kinetic power is supplied gradually by winds and/or jets (with $P_{\rm{kin}} \gtrsim 10^{39}$ erg s$^{-1}$) over the duration of the super-critical accretion phase, which can last for up to several times $10^5$ yr. Analogous to SNRs, ULX bubbles often have also optically thin radio synchrotron emission, with peak radio emission at the interface of the jet/wind with the ISM. The nearest ULX bubble is SS\,433/W50 in the Milky Way, with a size of $\approx$100 $\times$ 200 pc at a distance of $\approx$5.5 kpc \citep{dubner98,blundell04,lockman07,broderick18}. 
The X-ray luminosity observed from the accreting compact object (SS\,433) is only $\sim$10$^{36}$ erg s$^{-1}$ \citep{medvedev10}, but that is because we are looking at the system at high inclination. If seen face-on, SS\,433 would probably appear to us as a ULX \citep{middleton21,fabrika21}. The detection of semi-relativistic jets ($v \approx 0.26 c$) with $P_{\rm{kin}} \gtrsim 10^{39}$ erg s$^{-1}$ \citep{fabrika04,blundell04,jeffrey16}, and of a large bubble where the jets impact the ISM, are typical signatures of super-critical accretion. Other notable ULX bubbles in nearby galaxies are NGC\,7793-S26 \citep{pakull10,soria10}, Holmberg IX X-1 \citep{berghea20}, NGC 1313 X-1 \citep{gurpide22} and X-2 \citep{pakull02,zhou22}, NGC\,5585 X-1 \citep{soria21}; all of them require kinetic energies $\approx$10$^{39}$--10$^{40}$ erg s$^{-1}$.

ULX bubbles can be used as calorimeters, to determine the mechanical power of the compact object, in two ways \citep{pakull10}: i) from the luminosity of characteristic diagnostic lines, such as H$_\beta$ or [Fe\,{\footnotesize{II}}] $\lambda$1.64 $\mu$m; ii) from their size and expansion speed ($\approx$ shock velocity $v_{\rm{s}}$, measurable from the line width). Thus, the identification and study of ULX bubbles from radio surveys or optical spectroscopic surveys enables us not only to compare radiative and kinetic power in ULXs, and estimate the current age of their super-critical accretion phase, but also to discover compact objects that are (or were until very recently) in the super-critical accretion regime but appear X-ray faint for the reasons mentioned above.

In this paper, we present the discovery and multiband study of a candidate super-critical accreting source with exceptional optical and radio properties, located in the nearby starburst galaxy NGC\,5408 (Figure \ref{fig1}). We adopt a distance $d \approx 4.9$ Mpc (distance modulus $\approx 28.45$ mag), median of three tip-of-the-red-giant-branch measurements from \cite{karachentsev02}, \cite{mccall12}, \cite{tully13}. This galaxy is already known to host one of the best studied ULXs, NGC\,5408 X-1 
(RA(J2000) $= 14^{\text{h}} 03^{\text{m}} 19^{\text{s}}.62$, Dec(J2000) $= -41^{\circ} 22^\prime 58\farcs9$), with an X-ray luminosity $L_{\rm{X}} \approx 10^{40}$ erg s$^{-1}$ \citep{sutton13,pintore17}, an $\approx$0.2\,$c$ super-Eddington outflow \citep{pinto16}, broad (full width at half maximum (FWHM) $\approx 750$ km s$^{-1}$) He\,{\footnotesize{II}} and H$\beta$ emission, possibly from the accretion disk \citep{cseh11,cseh13}, narrow He\,{\footnotesize{II}}, H$\beta$ and [O\,{\footnotesize{III}}] nebular emission from X-ray photo-ionized gas \citep{kaaret09,cseh11}, and a radio synchrotron nebula \citep{kaaret03,soria06,lang07}. During follow-up studies of X-1, we serendipitously discovered what we argue is an even more powerful (but apparently X-ray fainter) source in the same galaxy (henceforth, NGC\,5408 X-2, following the terminology of \citealt{pakull17}). Here, we present the observational properties of X-2 and propose possible interpretations.

We adopt cgs units for numerical results and we assume a
flat cosmology with $H_0 = 69.6$ km s$^{-1}$ Mpc$^{-1}$, $\Omega_{\mathrm{M}} = 0.286$ and $\Omega_\Lambda = 0.714$ \citep{bennett14}. At 4.9 Mpc, 1\arcsec $\approx 24$ pc.

\section{Brief summary of the observations}

\subsection{Radio}

We analysed archival Australia Telescope Compact Array (ATCA) observations of NGC\,5408 obtained in 2000, 2003, 2004, and 2009, covering frequencies between 4.8 and 9~GHz, in the 6-km configuration. We also reanalysed archival Very Large Array (VLA) observations from 2003 and 2005. The ATCA observations provide the main time series of the radio emission, while the higher-resolution datasets allow the radio counterpart of X-2 to be separated from neighbouring sources. The data were reduced and imaged using standard procedures, and radio positions, flux densities, and spectral indices were measured from the resulting images. The log of the ATCA and VLA observations is given in Table \ref{log_atca}. Full details of the observations and data reduction are given in Appendix~\ref{app1}.

\subsection{Optical}

We assembled optical imaging and spectroscopic data spanning more than three decades. Archival long-slit spectroscopy was obtained: with the ESO New Technology Telescope (NTT) in 1993 and 2006 (Appendix~\ref{app2}); with the Very Large Telescope (VLT) FORS1 spectrograph in 2008, and with FORS2 in 2010, 2011, and 2012 (Appendix~\ref{app3}); and with the Southern African Large Telescope (SALT) Robert Stobie Spectrograph (RSS) in 2017 (Appendix~\ref{app4}). These data provide broad wavelength coverage and a range of spectral resolutions for measuring the broad and narrow emission-line components of X-2. In all the VLT FORS1 and FORS2 spectra, the intended target was X-1, and X-2 happened to be on the slit by chance, because of its alignment with a reference star (bottom panel of Figure \ref{fig1} and Appendix~\ref{app3}).

We also analysed archival VLT Multi Unit Spectroscopic Explorer  (MUSE) observations from 2024, retaining the May 1 and June 27 datasets. These observations provide spatially resolved optical continuum and emission-line maps, including He\,{\footnotesize{II}} $\lambda$4686. Full details of the optical observations, data reduction, and photometry and spectroscopy are given in Appendix~\ref{app5}.

Finally, we analysed {\it{Hubble Space Telescope}} ({\it{HST)}} images obtained with the Wide Field Planetary Camera 2 (WFPC2) in 2000 and 2009 and with the Wide Field Camera 3 (WFC3) Infrared (IR) imager in 2010 (Appendix~\ref{app6}). The observations cover the near-UV through near-IR, including the narrow-band F656N filter centred on H$\alpha$. We list in Table \ref{tab:hst_obslog} all the {\it HST} observations that cover X-2. We used these data to measure the unresolved optical counterpart, its spectral energy distribution, and its long-term photometric evolution.

   \begin{figure}[t!]
\hspace{-0.4cm}
   \includegraphics[width=0.71\hsize,angle=270]{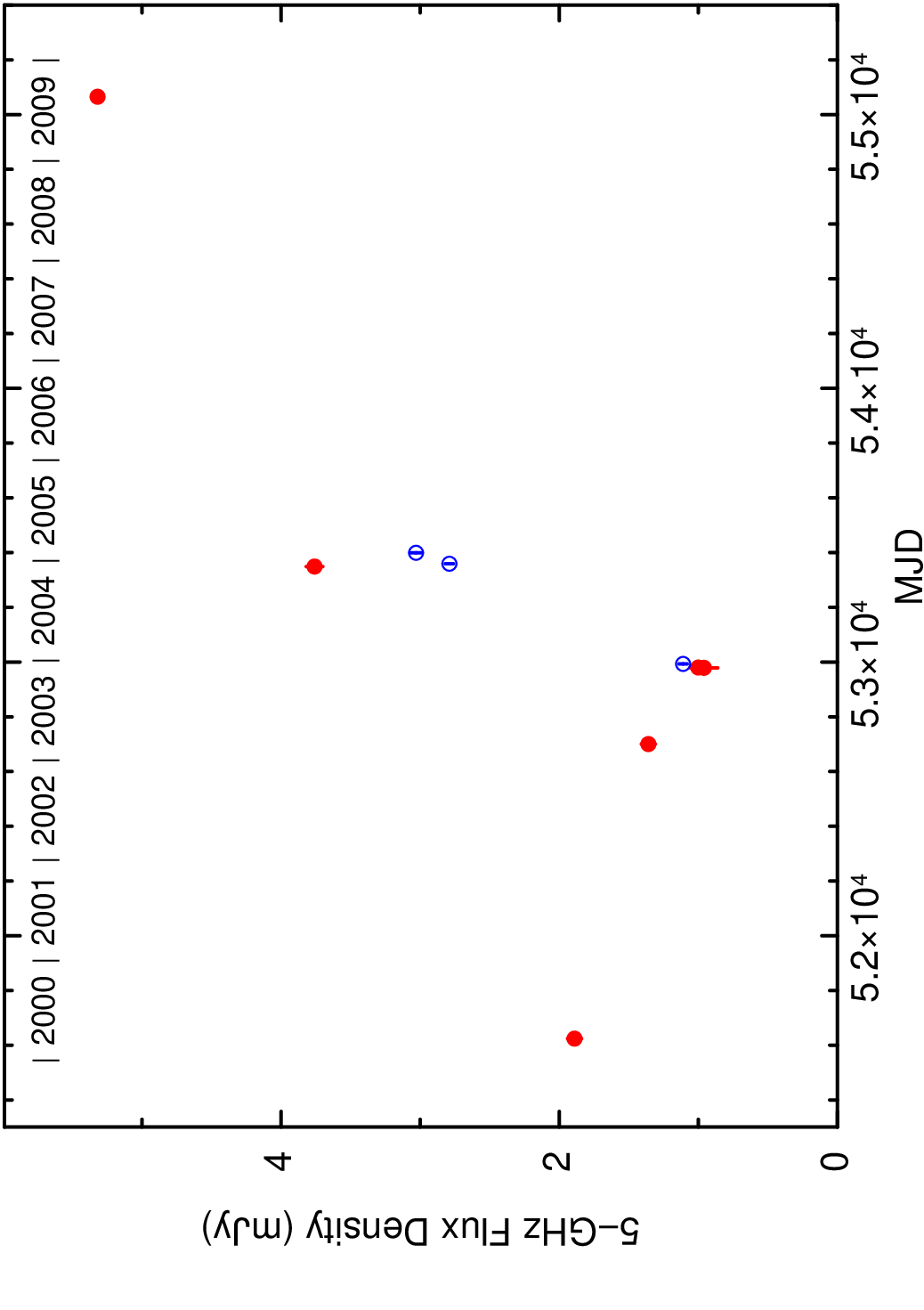}
   \vspace{0.8cm}
      \caption{5-GHz flux density evolution over the ATCA (filled red circles) and VLA (empty blue circles) datasets. Flux density values originally measured at 4.8, 4.9 or 5.5 GHz were rescaled to an equivalent 5-GHz value, with an assumed spectral index $\alpha = 1.1$.}
         \label{radio_lcurve}
   \end{figure}

   \begin{figure*}[t!]
   \centering
   \includegraphics[width=0.49\hsize]{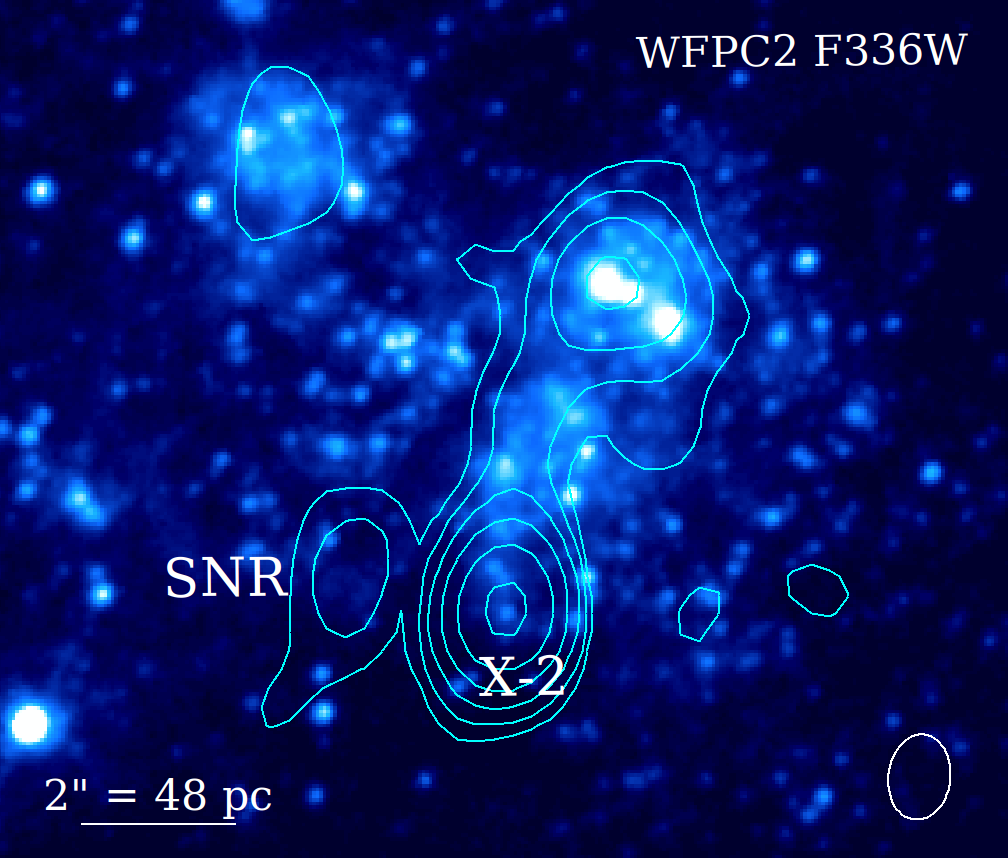}
   \includegraphics[width=0.49\hsize]{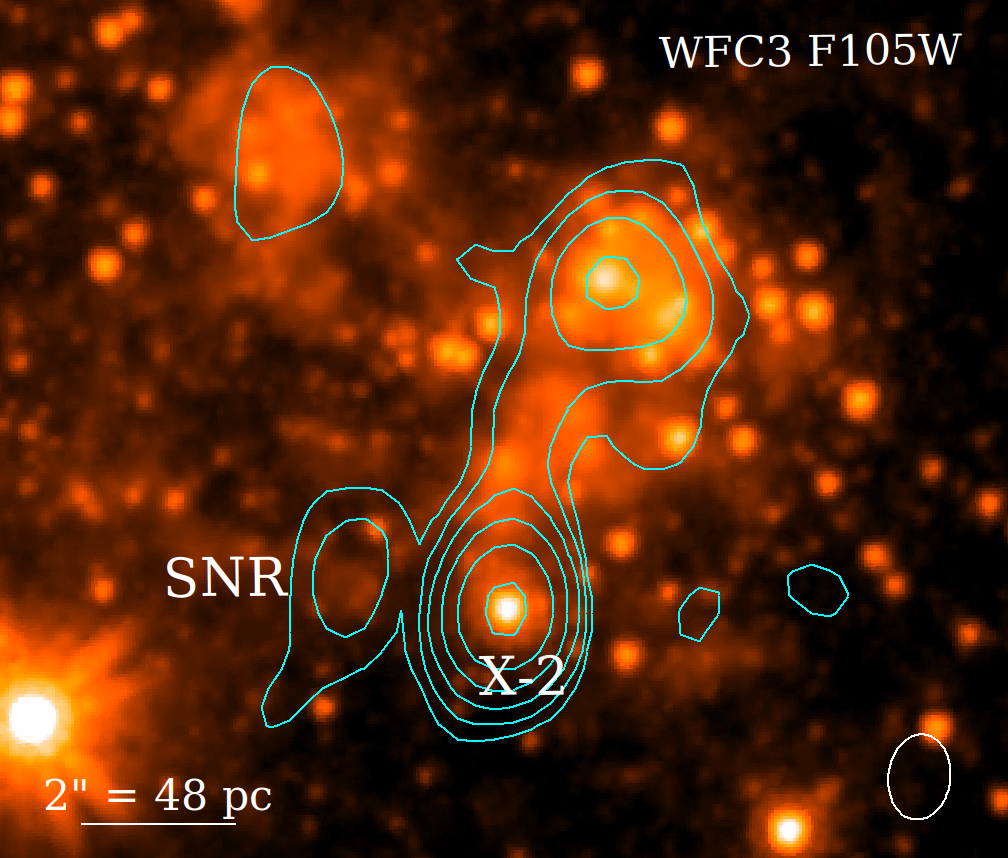}\\
   \includegraphics[width=0.49\hsize]{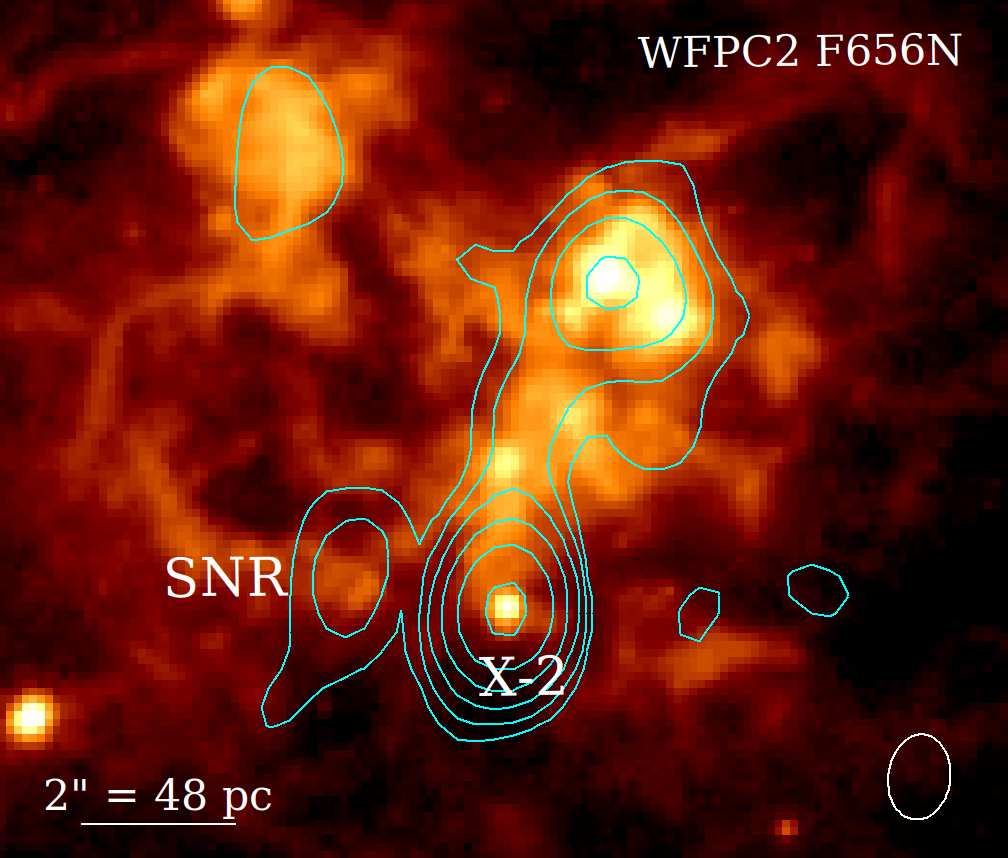}
   \includegraphics[width=0.49\hsize]{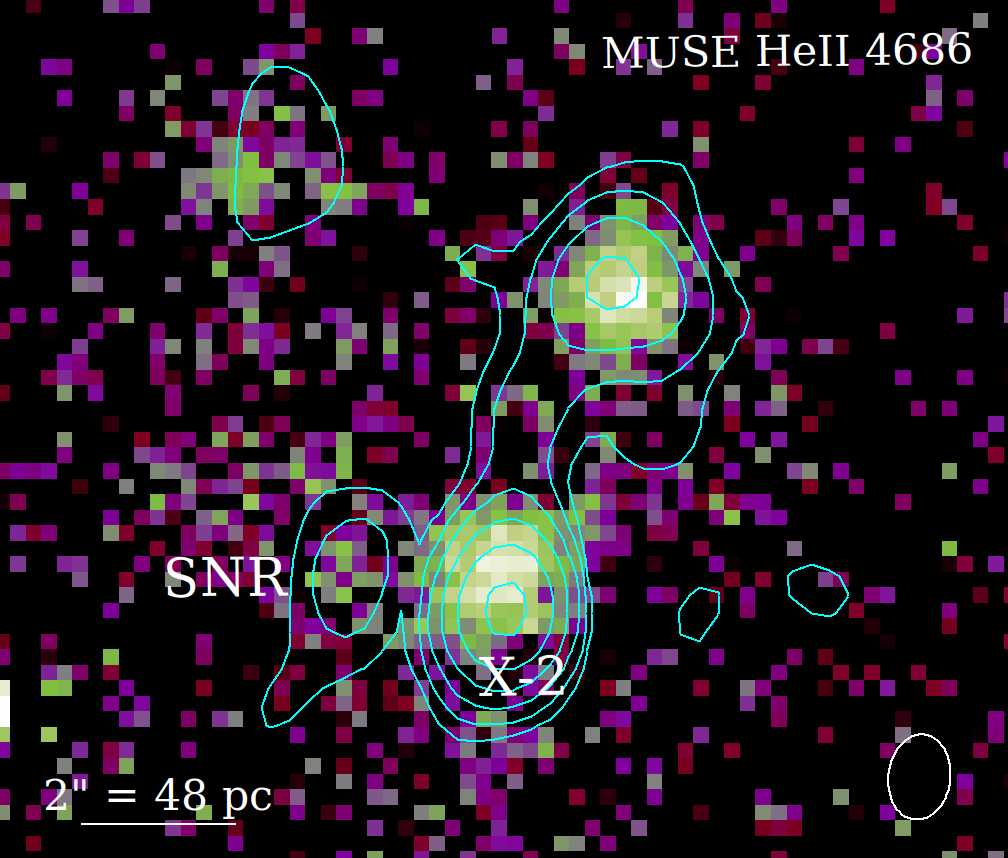}
      \caption{Near-UV, optical and near-IR images of the field around X-2, in different bands. Top left: {\it{HST}}/WFPC2 F336W. Top right: {\it{HST}}/WFC3 F105W. Bottom left: {\it{HST}}/WFPC2 F656N (narrow band filter centred on H$\alpha$, not including [N\,{\footnotesize{II}}] $\lambda\lambda$6548,6583). Bottom right: continuum-subtracted VLT/MUSE image in He\,{\footnotesize{II}} $\lambda$4686 (combined from 2024 May 1 and June 27). In each panel, we overplotted the stacked 5.5$+$9.0 GHz contours from the 2009 April 4 ATCA observations. The radio contours and the synthesized beam are defined as in Figure \ref{fig1}. For X-2, the optical continuum and H$\alpha$ emission are centred on the radio position. Instead, He\,{\footnotesize{II}} $\lambda$4686 emission is centred 0\farcs6 $\pm$ 0\farcs1 north of the radio position.}
         \label{fig2}
   \end{figure*}

   \begin{figure}[t!]
\hspace{-0.3cm}
\includegraphics[width=0.72\hsize,angle=270]{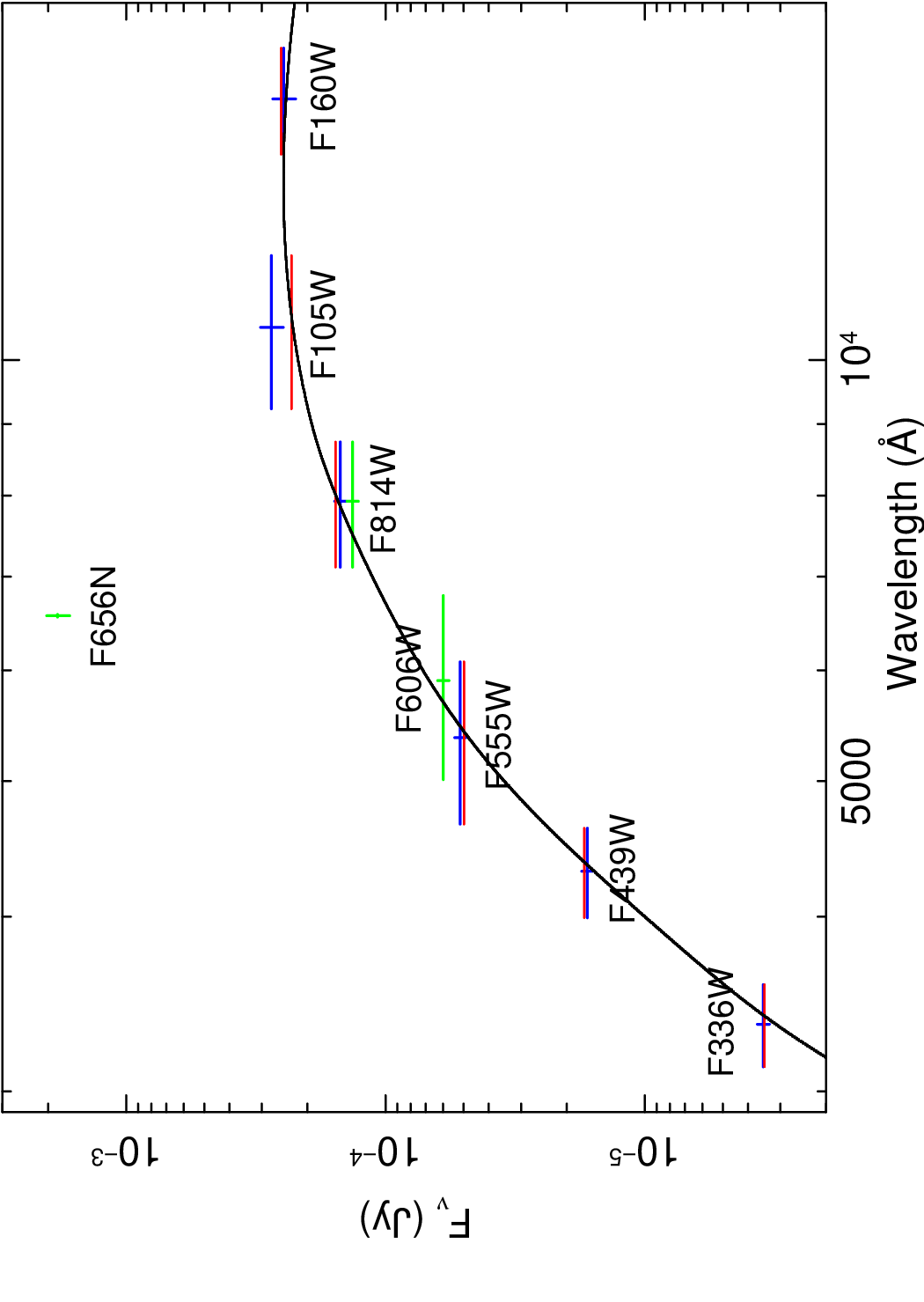}
   \vspace{0.9cm}
      \caption{Unfolded spectral energy distribution from the {\it{HST}} observations, fitted with a reddened blackbody in {\sc{xspec}}. Blue datapoints (observations taken between 2009--2010) were used for the fit, with the best-fitting flux density values plotted as red segments. Green datapoints were not included in the fit, because they were taken in 2000 (F606W and one of the two F814W observations) or because dominated by line emission (F656N).  The black line shows the best-fitting reddened blackbody model (Table \ref{hst_sed}).
      }
         \label{hst_sed_plot}
   \end{figure}

\subsection{X-rays}

We analysed eight archival {\it Chandra}/ACIS-S observations obtained between 2002 and 2010, with a total exposure of approximately 67~ks (Table \ref{log_chandra}). X-2 was located close to the ACIS-S aim point in all observations. The data were reprocessed and combined to improve the spatial localisation of the source, and spectra were extracted for the point-like and diffuse components. The {\it Chandra} and {\it HST} astrometry was also registered using common point sources to determine the relative positions of the X-ray, optical, and radio emission. Details of the Chandra observations and data analysis are given in Appendix~\ref{app7}.

   \begin{figure}[t!]
\includegraphics[width=\hsize]{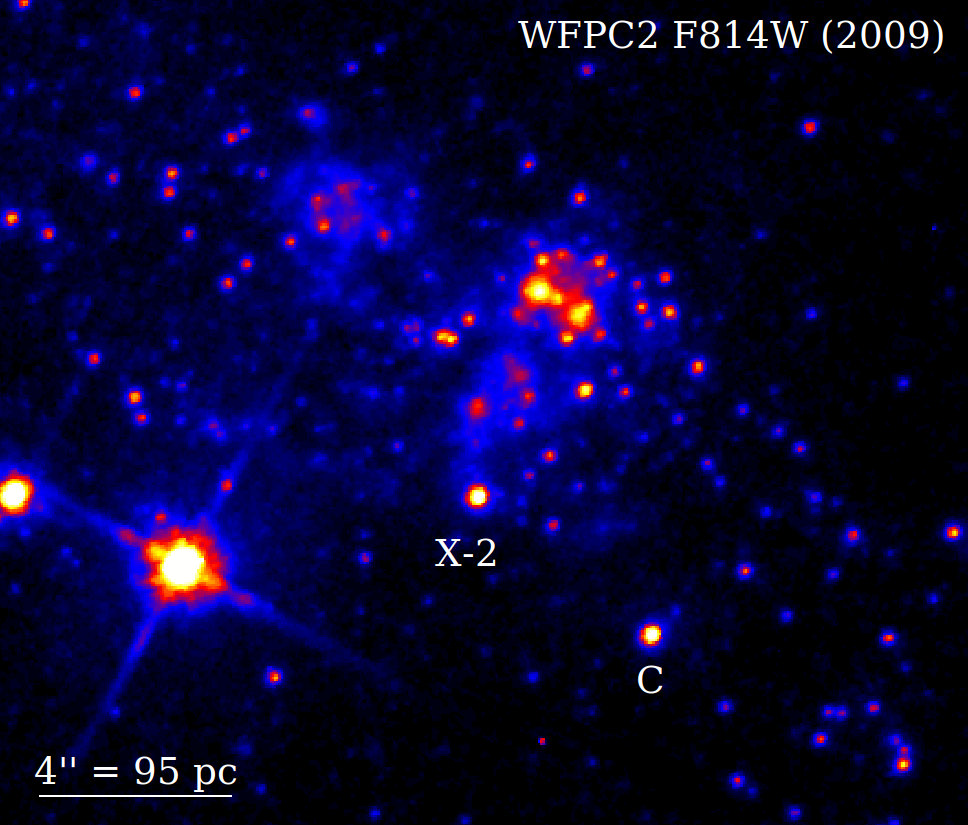}\\
\includegraphics[width=\hsize]{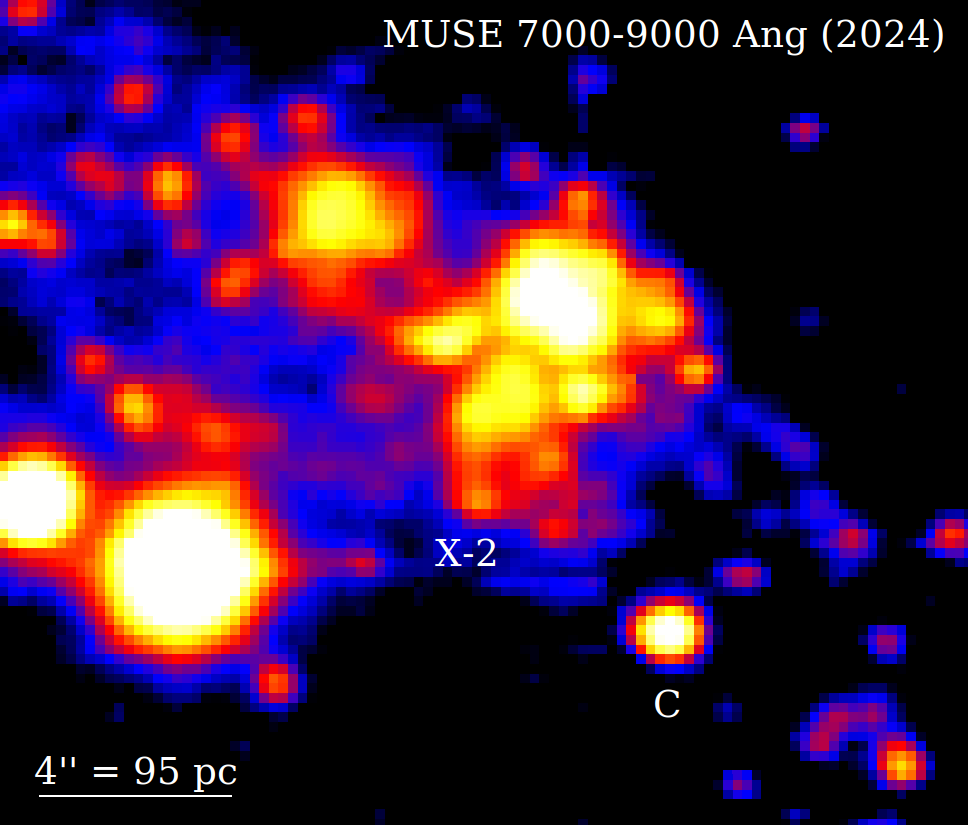}
      \caption{Top panel: {\it{HST}}/WFPC2 false-colour image from the 2009 April 4 observation, in the F814W filter. The apparent brightness of X-2 is $I \approx 18.0$ (Vega) mag. The comparison star C has $I \approx 18.1$ mag.
      Bottom panel: VLT/MUSE image in the 7000--9000 \AA\ band (roughly equivalent to F814W). X-2 has dimmed dramatically, relative to any other nearby star; star C has not changed but is now $\approx$2 mag brighter than X-2.
      }
         \label{muse_dimming}
   \end{figure}

\section{Results}

\subsection{A compact, variable radio source}
Previous radio studies of NGC\,5408 resolved the emission from the young star-forming region into at least four knots \citep{stevens02}. Three of them (including the source we now associate to X-2, labelled source B in \citealt{stevens02}) were interpreted as synchrotron emission from one or multiple SNRs (steeper spectral index), and one (source A in \citealt{stevens02}) as free-free emission from an H\,{\footnotesize{II}} region (flatter spectral index). We reprocessed and re-examined all archival ATCA and VLA datasets with sufficient spatial resolution to resolve X-2 from the three neighbouring sources.

The radio counterpart of X-2 is always unresolved (physical size $\lesssim$20 pc) at all epochs. The best fit position from the stacked 5.5$+$9.0 GHz, $r=-2$ ATCA image from 2009 is: RA(J2000) $= 14^{\text{h}} 03^{\text{m}} 18^{\text{s}}.428 \pm0\fs08$, Dec(J2000) $= -41^{\circ} 22^\prime 55\farcs21 \pm 0\farcs11$, where the error was taken as 10\% of the synthesized beam. This radio position is consistent with the {\it{Gaia}} position of the bright optical counterpart (within errors, being less than 0\farcs1 away).

We found that the radio flux of X-2 changed significantly from epoch to epoch (Table \ref{log_atca}, Figure \ref{radio_lcurve}). In particular, it was brighter by a factor of 5 in 2009 (reaching a 5.5-GHz flux density $\approx$4.8 mJy, corresponding to a luminosity of $\approx$7.6 $\times 10^{35}$ erg s$^{-1}$) compared with 2003 (4.8-GHz flux density $\approx$1.0 mJy, corresponding to a luminosity of $\approx$1.4 $\times 10^{35}$ erg s$^{-1}$).  We verified that other sources in the field (in particular, source A in \citealt{stevens02}) did not vary; this confirms that the flux variation of X-2 is not an instrumental artifact. Given the strong flux variation over few years, the emission must be dominated by a single, compact source, with a size $\lesssim$1 pc. 
The spectral index $\alpha$ (defined such that the flux density $S_\nu \propto \nu^{-\alpha}$) is consistently steep; the typical value measured from the four ATCA observations that were taken simultaneously in two bands is $\alpha \approx 1.1 \pm 0.1$ (Table \ref{alpha_atca}).

Considering the location in a young stellar environment, the high radio luminosity, and the steep spectral index, a core-collapse SN or young SNR are a plausible scenario to consider. The peak luminosity density among our observations is certainly extreme but not unphysical for an SNR. For a spectral index $\alpha = 1.1$, a 5.5-GHz flux density of 4.79 mJy corresponds to a 1.45-GHz luminosity density of $L_{1.45} \approx 6.0 \times 10^{26}$ erg s$^{-1}$ Hz$^{-1}$. Comparing with the SNR population study of \cite{chomiuk09}, we do find radio SNRs with similar luminosity in the much more powerful starburst galaxies NGC\,253 and M\,82, at the upper end of the SNR luminosity function in the local universe. The most luminous Milky Way SNR, Cas A, is an order of magnitude weaker than X-2: its observed 5-GHz flux density of $\approx$800 Jy at 3.4 kpc \citep{trotter17} corresponds to $\approx$0.39 mJy at 4.9 Mpc. The steep spectral index of X-2 is consistent with young SNRs \citep{urosevic14,luken20}. However, the rapid and non-monotonic variability (Table \ref{log_atca}, Figure \ref{radio_lcurve}) makes the SNR scenario very unlikely.

A decades-old SN would be perfectly consistent with the observed luminosity density of X-2 \citep[{\it{e.g.}\rm,}][{\rm{their Fig.\ 6}}]{gulati26}. Radio re-brightenings of core-collapse SNe $\sim$10--20 years after the event are sometimes observed and attributed to the shock wave encountering denser shells of circumstellar medium \citep{rose24,soria25}. In this scenario, the re-brightening started in early 2004 and was still ongoing in 2009 (Figure \ref{radio_lcurve}). There is no observational evidence of a SN exploding at the location of X-2 in the last decades of the 20th century; however, several nearby SNe were still missed at that time. A stronger argument against a SN interpretation comes from the appearance and evolution of the optical line spectra from 1993 to 2024 (Section 3.6), which are quite different from those of either aging SNe or young SNRs.

An alternative scenario is that the radio emission is caused by a powerful jet, marking X-2 as the second microquasar in super-critical accretion in NGC\,5408, in addition to X-1. In fact, the (unresolved) radio counterpart of X-2 is between 5 and 20 times more luminous than the (marginally resolved, diameter of $\approx$35--50 pc) radio nebula around X-1 \citep{kaaret09}; the latter source is already considered one of the most powerful microquasars in the local universe. Synchrotron-emitting ULX bubbles \citep{pakull02,pakull10,soria10,soria21}, caused by the interaction of a jet or of fast outflows with the ISM, are consistent with the observed luminosity and steep spectral index, and may appear unresolved ($\lesssim$20 pc at the distance of NGC\,5408) if they are very young and/or expanding in a dense medium; however, the shocked bubble scenario is difficult to reconcile with the rapid variability. Instead, most of the emission may be due to flaring (blob ejection) from the core. This would make the radio emission of X-2 qualitatively similar to the radio flaring seen in the ULX Holmberg II X-1 \citep{cseh14,cseh15} and in the intermediate mass BH ESO\,243-49 HLX-1 \citep{webb12,cseh15b}. The peak 5-GHz luminosity measured from NGC\,5408 X-2 is $\approx$30 times higher than that of Holmberg II X-1 ($L_{\rm{R}} \approx 2.3 \times 10^{34}$ erg s$^{-1}$; \citealt{cseh14,cseh15}) and only 5 times lower than the brightest flare of HLX-1 ($L_{\rm{R}} \approx 3.6 \times 10^{36}$ erg s$^{-1}$; \citealt{webb12}).

To make progress on the interpretation of the radio emission, we shall now  examine and integrate the main properties in the other electromagnetic bands.

   \begin{figure*}[t!]
   \centering
   \includegraphics[width=0.44\hsize]{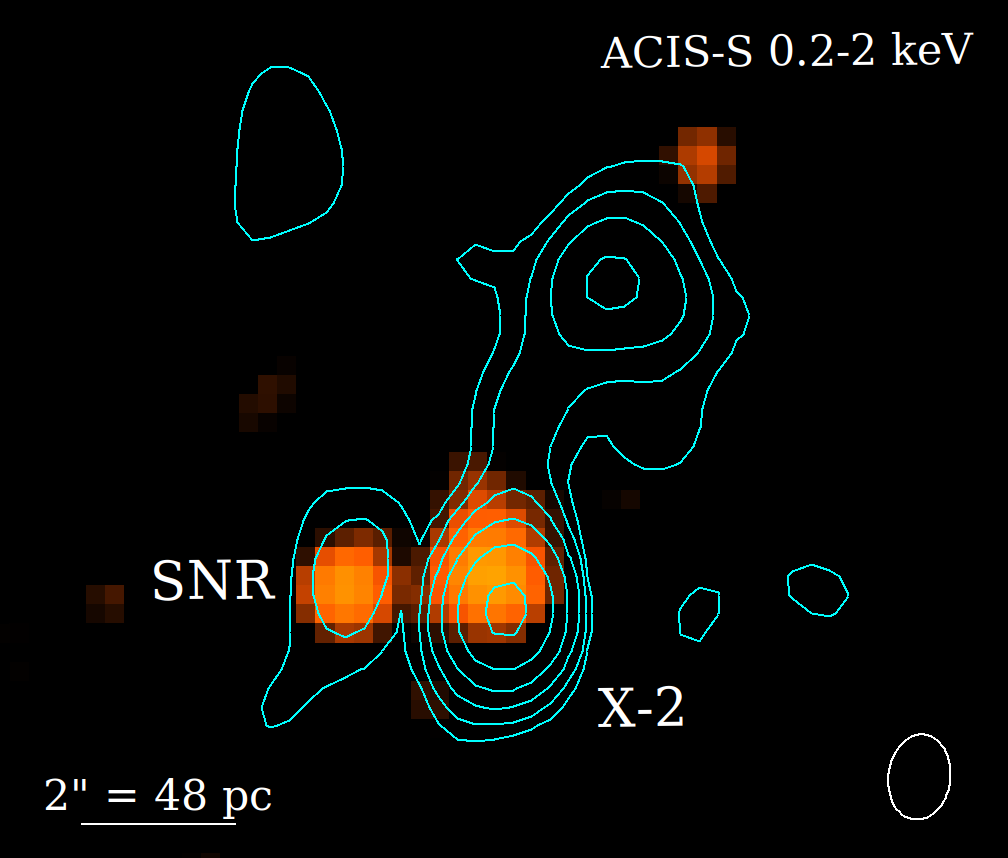}
   \includegraphics[width=0.44\hsize]{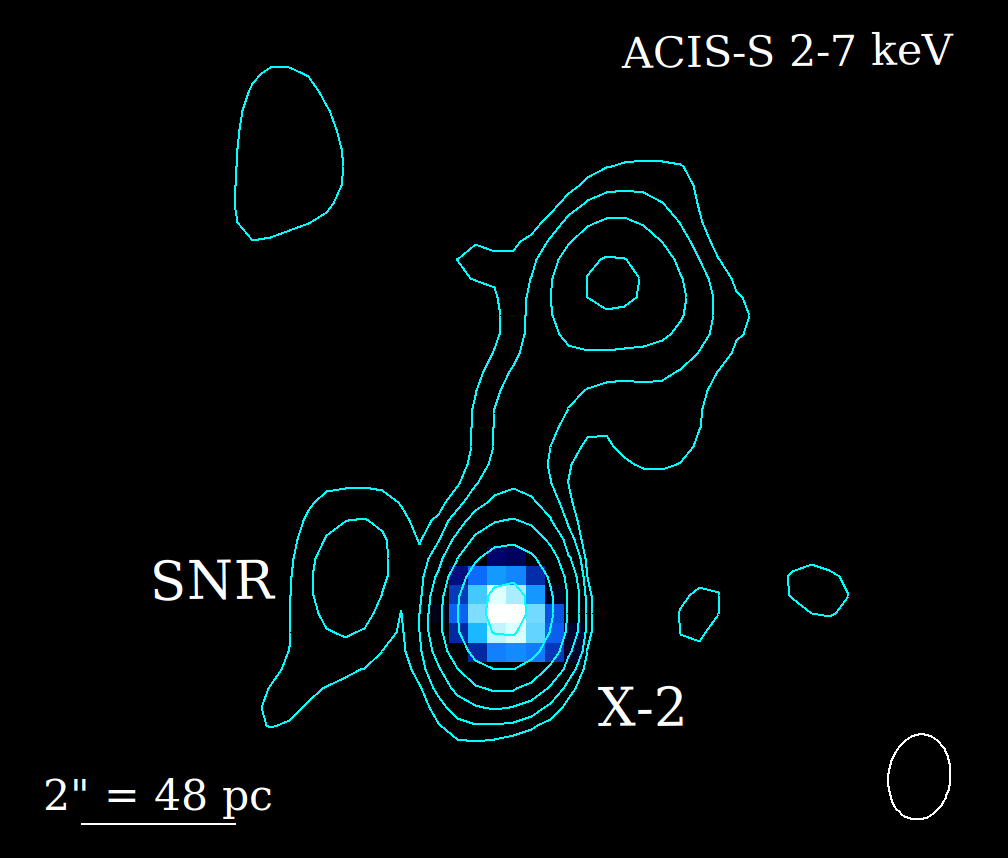}\\
      \includegraphics[width=0.44\hsize]{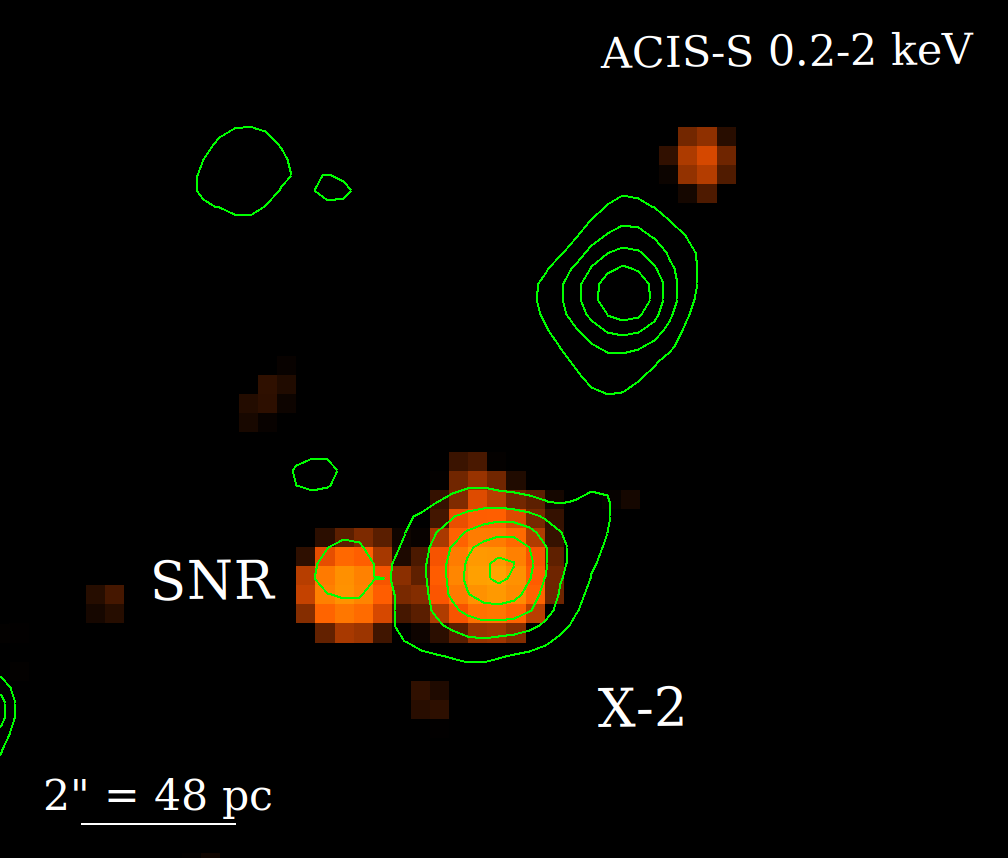}
   \includegraphics[width=0.44\hsize]{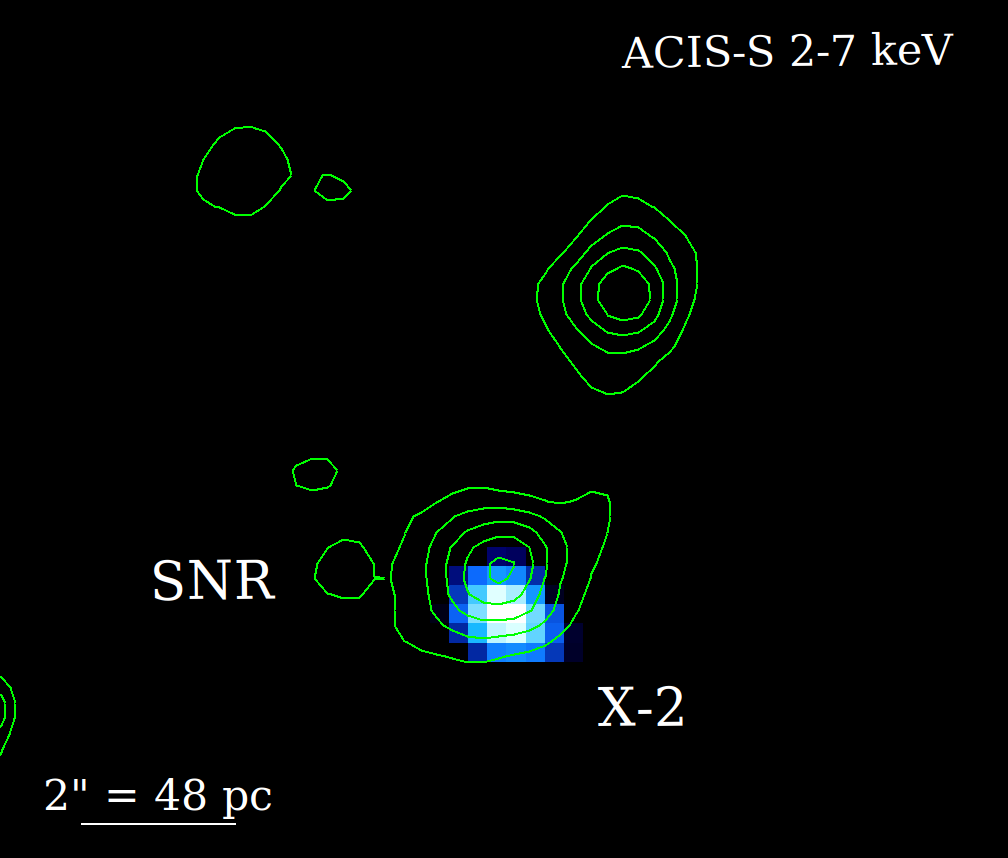}
      \caption{Top panels: comparison between X-ray and radio emission.  
      Top left: Gaussian-smoothed {\it{Chandra}}/ACIS-S image in the 0.2--2 keV band, showing that the soft X-ray emission from X-2 is extended (compared with the nearby, young SNR and with the theoretical PSF) and slightly offset to the north of the radio centroid. Top right: Gaussian-smoothed {\it{Chandra}}/ACIS-S image in the 2--7 keV band, showing that the hard X-ray emission from X-2 is point-like and coincident with the radio centroid and therefore also with the optical continuum. In each panel, we overplotted the stacked 5.5$+$9.0 GHz contours from the 2009 April 4 ATCA observations. The radio contours and the synthesized beam are defined as in Figures 1 and 2.
      Bottom panels: comparison between X-ray and He\,{\footnotesize{II}} $\lambda$4686 emission. Bottom left: Gaussian-smoothed {\it{Chandra}}/ACIS-S image in the 0.2--2 keV band, with superimposed, smoothed contours (green) from the VLT/MUSE He\,{\footnotesize{II}} $\lambda$4686 emission map. The contours are in linear scale with arbitrary zeropoint.
      Bottom right: Gaussian-smoothed {\it{Chandra}}/ACIS-S image in the 2--7 keV band, with VLT/MUSE He\,{\footnotesize{II}} $\lambda$4686 contours. The panels show that the soft X-ray emission is spatially associated with the peak of the He\,{\footnotesize{II}} $\lambda$4686 emission, and does not show a peak at the location of the hard X-ray peak.}
         \label{chandra_radio_he2}
   \end{figure*}

\subsection{Luminosity and colours of the point-like optical counterpart in {\it{HST}}}

Our photometric analysis of the {\it HST} data show an extremely bright, unresolved, red ``star'', with apparent brightness $m_{\rm{555W}} \approx V \approx 19.6$ mag (Vega system), $m_{\rm{814W}} \approx I \approx 18.0$ mag (Table \ref{tab:hst_obslog}). For F814W, there is marginal evidence of variability between the 2000 and 2009 data (the latter being $\approx$10 per cent brighter). At the distance of NGC\,5408, and correcting only for line-of-sight reddening ($E(B-V) \approx 0.060$ mag; \citealt{schlafly11}), the source has absolute magnitudes $M_{V,0} \approx -9.1$ mag, $M_{I,0} \approx -10.6$ mag. At first sight, it looks like a K-type supergiant, several times brighter than the other red supergiants in NGC\,5408, and located away from the {\it{centre of the}} main star-forming region (Figure 5). 

We used {\sc{ftools}}'s task {\it{flx2xsp}} to convert the measured flux densities into pha and rsp files suitable for {\sc{xspec}} fitting. We find that the spectral energy distribution from the near-UV to the near-IR is well fitted by a simple blackbody model with temperature $T_{\mathrm{bb}} \approx 4600$ K, radius $R_{\mathrm{bb}} \approx 1700 R_\odot$, and bolometric luminosity $\log (L_{\mathrm{bol}}/L_\odot) \approx 6.1$ (Table \ref{hst_sed} and Figure \ref{hst_sed_plot}). For this modelling, we used only the broad-band data from 2009--2010, to avoid additional uncertainty caused by possible decadal variability. If we also include the two broad-band datapoints from 2000, the modelling results do not change significantly (Figure \ref{hst_sed_plot}). Alternatively, we tried to fit the data with an irradiated disk model ({\it{diskir}} in {\sc{xspec}}; \citealt{gierlinski09}). However, we could not get an acceptable fit, because multi-temperature, irradiated disk models predict a rising flux density towards the UV, contrary to the observed UV faintness and red colour.

There is a well-known upper limit to the luminosity of red and yellow hypergiants: the Humphreys–Davidson limit \citep{sandage74,humphreys79}. It is the luminosity of the most massive star that passes through a red supergiant phase (initial mass $M \approx 40 M_\odot$, according to \citealt{gordon16}). Studies of nearby galaxies place this limit at $\log (L_{\mathrm{bol}}/L_\odot) \approx 5.5$--5.7, corresponding to $M_{I,0} \approx -10$--$-9.5$ mag \citep{humphreys83,gordon16,davies18,massey21,mcdonald22,jones25}. The red counterpart of X-2 is above the Humphreys–Davidson limit, for the assumed distance of NGC\,5408. There are several other point-like, red stars in the starburst region a few arcsec north of X-2 (Figure \ref{fig1}); we verified with a WFPC2 photometric measurement that they are all below the limit. Hypergiants at the Humphreys–Davidson limit, such as Var A in M\,33 \citep{humphreys06}, WOH G64 \citep{vanloon26} in the Large Magellanic Cloud, and V766 Cen in the Milky Way \citep{chesneau14,wittkowski17a,wittkowski17b} are rare objects, characterized by strong, episodic mass ejections, and possibly undergoing a common envelope phase with a binary companion.

In addition to the broad-band filters, we examined the WFPC2 observation in the narrow-band F656N filter (narrow H$\alpha$, effective width $\approx$28\AA). The H$\alpha$ emission has a strong peak at the exact location of the unresolved optical continuum and radio core. The observed narrow-band flux is $\approx$20 times higher than the underlying broad-band continuum (Table \ref{hst_sed} and Figure \ref{hst_sed_plot}). 
Considering the width of the filter, we already infer directly from the WFPC2 images that the equivalent width (EW) of the H$\alpha$ emission from X-2 must be $\gtrsim$500 \AA; this is a lower limit because we will see from the VLT/FORS2 spectra (Section 3.6) that the F656N filter band does not include all the broad line wings. In fact, we will show (Section 3.6) that the H$\alpha$ EW is $\approx$2000 \AA.

\subsection{Recent optical decline}

There is no substantial change between the brightness in the 2000 and 2009 WFPC2 observations, and even the 2010 WFC3-IR measurements are consistent with a constant flux (Table \ref{hst_sed}). By contrast, X-2 appears remarkably fainter in the 2024 MUSE images. For example, it is obvious by eye how much X-2 has dimmed compared with a reference red star\footnote{Although at first sight star C may look like a red supergiant in NGC\,5408, it is actually an M dwarf in the Milky Way, as proved by its high proper motion of $\approx$27 mas yr$^{-1}$ \citep{gaia23}. The M-type nature of star C is also clearly visible from its MUSE spectrum.} 5\arcsec\ to the south-west (labelled ``C'' in Figure \ref{muse_dimming}). Figure \ref{muse_dimming} illustrates the change in the 7000--9000 \AA\ band but analogous dimming is seen in other bands, especially the F656N one, which suggests a decrease not only of the red continuum but also of the H$\alpha$ emission. We estimate $I \approx 20.0$ mag in 2024. The dimming is reminiscent of the dramatic optical decline of another enigmatic ULX, NGC\,300 X-1 \citep{chene25,beasor26}, for which various speculative interpretations were proposed, from common envelope evolution to the silent collapse of the red supergiant donor into a BH.

We do not know when the optical decline of X-2 started, between 2012 and 2024. We can constrain the dimming by comparing the optical brightness in the broad $G$ band from subsequent {\it{Gaia}} data releases (DRs) \citep{gaia16,gaia23}. In DR3, which includes 189 observations between 2014--2017, X-2 is listed at $G = (19.62 \pm 0.02)$ mag, with $BP = (17.27 \pm 0.11)$ mag, $RP = (17.19 \pm 0.05)$ mag; this corresponds\footnote{\url{https://gea.esac.esa.int/archive/documentation/GDR3/}.} to $V = (19.65 \pm 0.05)$ mag, $I = (19.54 \pm 0.05)$ mag. In DR2, X-2 is at $G = (19.08 \pm 0.02)$ mag, from 146 observations between 2014--2016. In DR1, $G \approx 18.93$ mag, from 86 observations between 2014--2015. Thus, it appears that about half of the dimming happened between 2014 and 2017, and that the source has become bluer. More detailed studies of the brightness evolution in {\it{Gaia}} is left to follow-up work.

\subsection{Spatially resolved He\,{\footnotesize{III}} nebula}

We compared the {\it{HST}} images with the map of the nebular He\,{\footnotesize{II}} $\lambda$4686 emission from VLT/MUSE. For this, we selected a 5.8-\AA-wide band centred at $\lambda = 4693.8$. We find (Figures \ref{fig2}, \ref{hst_col_4686}) that the position of the optical and radio counterpart of X-2 is near but not exactly at the centre of one of the two main regions of He\,{\footnotesize{II}} $\lambda$4686 emission in this galaxy. The northern patch of He\,{\footnotesize{II}} $\lambda$4686 emission coincides with, and is easily explained by the most active starburst region in NGC\,5408 (the same region that produces the radio source A of \citealt{stevens02}). Instead, the southern patch (including X-2 at its southern edge) is associated with a region with weaker current star formation and without massive, young star clusters. Using only the MUSE data from 2024 May 1 because of their superior seeing, we estimated that the He\,{\footnotesize{II}} emission region near X-2 has a FWHM of $\approx$0\farcs85 ($\approx$20 pc). X-2 is located $\approx$0\farcs6 ($\approx$15 pc) south of the peak He\,{\footnotesize{II}} $\lambda$4686 emission.

Assuming that the 2024 May 1 observation is in photometric conditions, and correcting for the systematic 28 per cent loss of counts in the June 27 observations, we measured an observed flux $f_{\rm{4686,X-2}} 
= (6.1 \pm 0.6) \times 10^{-16}$ erg cm$^{-2}$ s$^{-1}$
from the emission patch next to X-2 within a circular 0\farcs8 aperture ($\approx$20 pc). This includes the location of X-2 itself, which does not stand out as a point-like emitter. By comparison, we also measure a flux $f_{\rm{4686,X-1}} = (3.1 \pm 0.2) \times 10^{-16}$ erg cm$^{-2}$ s$^{-1}$ from the nebula around the X-1 ULX, consistent with the total flux inferred by \cite{kaaret09}. Correcting for Galactic line-of-sight reddening ($E(B-V) \approx 0.060$ mag, corresponding to an extinction $A_{4686} \approx 0.23$ mag at $\lambda = 4686$ \AA), we obtain an unabsorbed flux 
$f_{\rm{4686,neb}} = (7.6 \pm 0.7) \times 10^{-16}$ erg cm$^{-2}$ s$^{-1}$, corresponding to a luminosity $L_{\rm{4686,neb}} = (2.2 \pm 0.2) \times 10^{36}$ erg s$^{-1}$.
This is a factor of 2 higher than the He\,{\footnotesize{II}} $\lambda$4686 luminosity of X-1 \citep{kaaret09} and similar to the line luminosity of the nebula around Holmberg II X-1 ($L_{\rm{4686}} \approx 2.5$--$2.7 \times 10^{36}$ erg s$^{-1}$: \citealt{pakull02,kaaret04}).

The obvious question to address is what creates the extended highly-ionized nebula near X-2. Given its huge radio power, the extreme brightness of its optical counterpart, and its exceptional set of optical lines (which we will discuss later, in Section 3.6), it is tempting to invoke X-2 as the main source of power also for the extended He\,{\footnotesize{II}} $\lambda$4686 emission. There are a few difficulties for this scenario, none of them strong enough to rule it out completely. First, the red optical continuum of X-2 is too cold to doubly-ionize helium and produce $\lambda$4686 photons; however, X-2 has associated X-ray emission (Section 3.5), perhaps mostly obscured along our line of sight, that may supply enough high-energy photons. Second, X-2 had faded by an order of magnitude in the 2024 MUSE image, to the point that it was mainly detected as a point source in the $I$ band; however, the He$^{++}$ nebula could be a fossil signature of its higher activity in the recent past. Third, X-2 is offset from the centre of the extended He\,{\footnotesize{II}} $\lambda$4686 emission; however, this is common in ULX nebulae (including the one near NGC\,5408 X-1), and can be explained by anisotropic illumination from the accreting source and anisotropic density distribution of the surrounding ISM. 

Although X-2 remains in the picture, there is probably a better alternative scenario for the extended He\,{\footnotesize{II}} $\lambda$4686 emission. Our {\it{HST}}/WFC2 imaging study reveals two blue stars at the centre of the nebula (stars A and B in Figure \ref{hst_col_4686}), separated by less than 0\farcs2. Star A's apparent brightness is 
$m_{\rm{336W}} = 21.3 \pm 0.1$ mag,  
$m_{\rm{439W}} = 22.3 \pm 0.1$ mag,
$m_{\rm{555W}} = 22.7 \pm 0.1$ mag,
$m_{\rm{814W}} = 23.2 \pm 0.2$ mag.
After a correction for line-of-sight reddening, star A has absolute magnitudes 
$M_{\rm{336W,0}} \approx -7.5$ mag, 
$M_{\rm{439W,0}} \approx -6.4$ mag, 
$M_{\rm{555W,0}} \approx -5.9$ mag, 
$M_{\rm{814W,0}} \approx -5.4$ mag. 
Star B has comparable brightness, with 
$m_{\rm{336W}} = 21.6 \pm 0.1$ mag,  
$m_{\rm{439W}} = 22.6 \pm 0.1$ mag, 
$m_{\rm{555W}} = 23.1 \pm 0.1$ mag,
$m_{\rm{814W}} = 23.2 \pm 0.2$ mag.
This corresponds to 
$M_{\rm{336W,0}} \approx -7.2$ mag, 
$M_{\rm{439W,0}} \approx -6.1$ mag, 
$M_{\rm{555W,0}} \approx -5.5$ mag, 
$M_{\rm{814W,0}} \approx -5.4$ mag. 
Colors and luminosities of both stars are consistent with an early O type, for example O3 (V--III) \citep{martins05}, or with Wolf-Rayet stars, in particular WN types with effective temperatures $\approx$1.0 $\times 10^5$ K \citep{hainich15}.

Assuming plausible values of $T_e \approx 10^4$ K and $n_e \approx 100$ cm$^{-3}$ (Sections 3.6, 3.7), a He\,{\footnotesize{II}} $\lambda$4686 luminosity $L_{\rm{4686,neb}} = (2.2 \pm 0.2) \times 10^{36}$ erg s$^{-1}$ requires an ionizing flux $Q(He^+) \approx 2.2 \times 10^{48}$ photons s$^{-1}$. An O star is not hot enough to generate such flux. Instead, the required flux is plausible for Wolf-Rayet stars; for example, WN3 stars, especially metal-poor ones with weaker winds \citep{gonzalez25}.

To make further progress on the possible sources of ionization for the gas at the location of X-2 and in the nearby nebula, and distinguish between broad and narrow components, we will examine (Sections 3.6, 3.7) the main features of their optical line spectra from FORS2 (for X-2) and from MUSE (for the extended nebula).

   \begin{figure}[t!]
\includegraphics[width=\hsize]{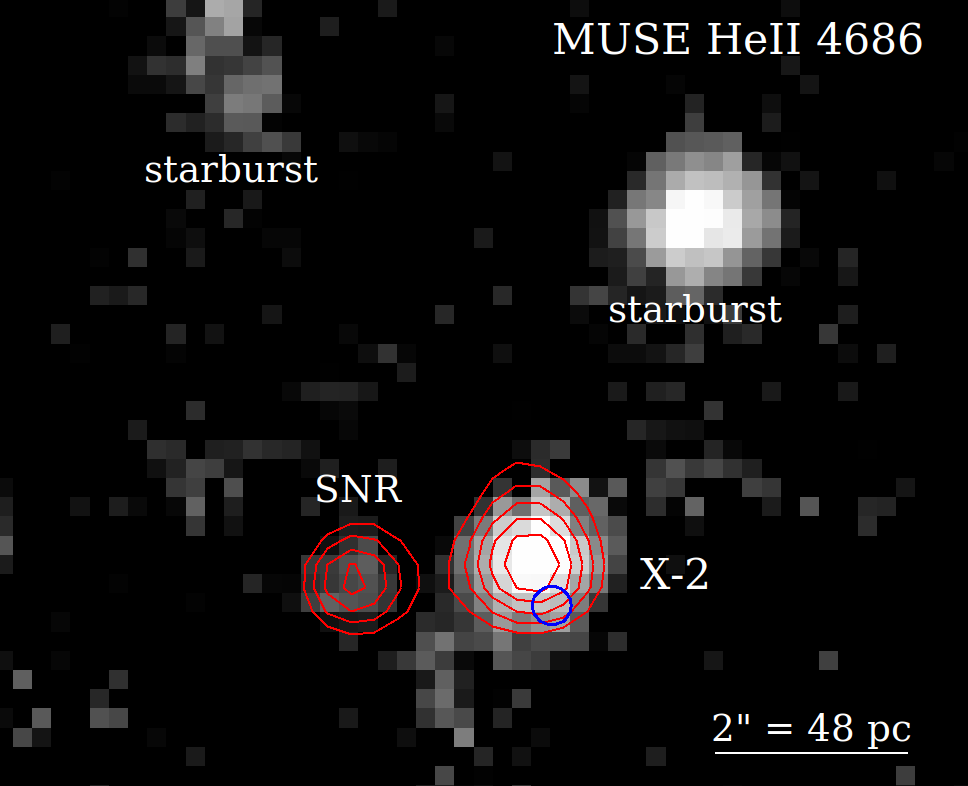}\\
\includegraphics[width=\hsize]{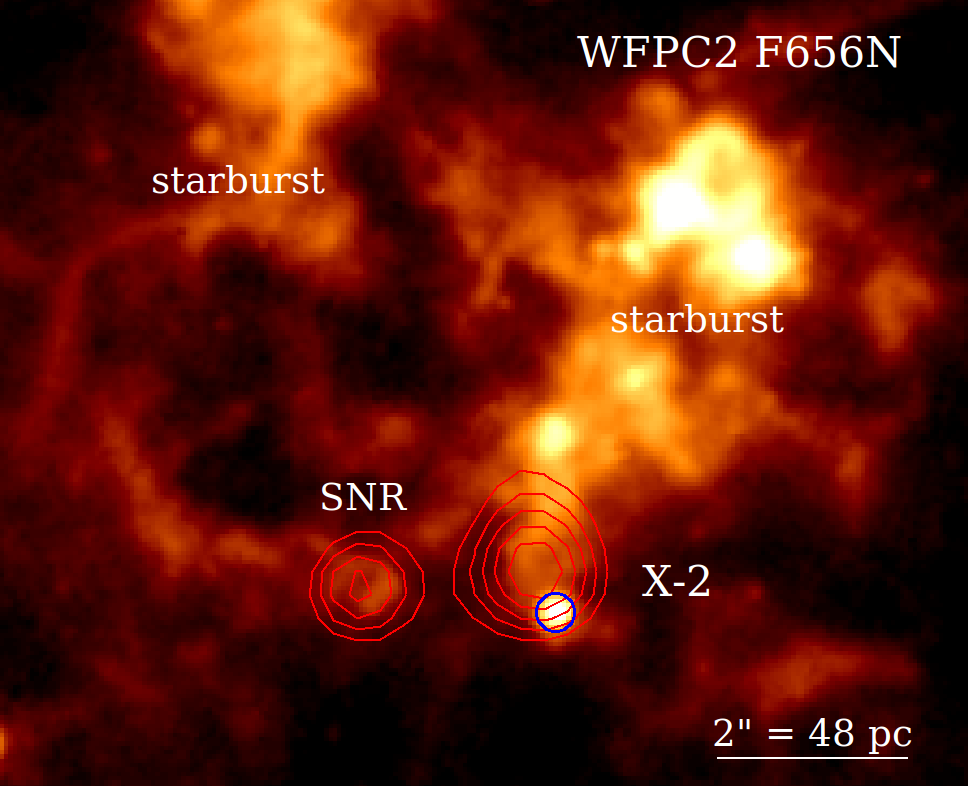}
      \caption{Top panel: VLT/MUSE He\,{\footnotesize{II}} $\lambda$4686 greyscale map, made from the 2024 May 1 data only (which were taken with the best seeing, 0\farcs65), compared with the 0.2--2 keV {\it{Chandra}}/ACIS map (red contours). The blue circle (0\farcs2) marks the position of the optical counterpart (from {\it{Gaia}}), also coincident with the point-like radio continuum (from ATCA) and with the centroid of the hard X-ray emission (2--7 keV from {\it{Chandra}}). Both the He\,{\footnotesize{II}} $\lambda$4686 emission and the soft X-ray emission regions are broader than their respective PSFs, consistent with an intrinsic size $\approx$20 pc.
      Bottom panel: {\it{HST}}/WFPC2 F656N (narrow-band H$\alpha$) image, compared with emission at other bands. Red contours and blue circle have the same meaning as in the top panel. 
      }
         \label{muse_0.2-2keV}
   \end{figure}

\subsection{Spatially resolved X-ray emission}

{\it{Chandra}} is the only X-ray instrument that can resolve X-2 from the X-ray/radio SNR located $\approx$2$^{{\prime}{\prime}}$ to the east (Figures \ref{chandra_radio_he2}, \ref{chandra1}). In the Chandra Source Catalog (CSC) 2.1 \citep{evans24}, X-2 corresponds to 2CXO J140318.4$-$412254 and the SNR to 2CXO J140318.6$-$412254. Also from the flux estimates in CSC 2.1, after correcting for the Galactic line-of-sight absorption ($N_{\rm{H}} \approx 5.9 \times 10^{20}$ cm$^{-2}$), we infer a 0.3--10 keV luminosity $L_{\rm{X}} \approx (5 \pm 1) \times 10^{37}$ erg s$^{-1}$ for X-2. 

However, this simple analysis does not tell the full story. When we split soft (0.2--2 keV) and hard (2--7 keV) bands (Figures \ref{chandra_radio_he2}, \ref{chandra1}), we find that:\\ 
i) the centroids of the soft and hard emission are offset by $\approx$0\farcs7, with the soft emission located to the north/north-east of the hard emission. The centroid of the hard component coincides, within 0\farcs1, with the ATCA (Figure \ref{chandra_radio_he2}) and {\it{Gaia}} positions for the radio and optical counterparts, respectively, and with the peak H$\alpha$ emission in the {\it{HST}} map (Figure \ref{muse_0.2-2keV}). The centroid of the soft component coincides with the peak intensity of the He\,{\footnotesize{II}} $\lambda$4686 emission seen by VLT/MUSE.
(Figure \ref{muse_0.2-2keV});\\
ii) the hard component of X-2 is point-like (consistent with the ACIS-S PSF); instead, the soft X-ray emission is extended (Figures \ref{chandra_radio_he2}, \ref{chandra1}). Using the {\sc{ciao}} task {\it{srcextent}}, we estimate a characteristic radius of 0\farcs96 $\pm$ 0\farcs28, well comparable to the characteristic size of the He\,{\footnotesize{II}} $\lambda$4686 emission nebula (Figure \ref{muse_0.2-2keV}).

Soft and hard sources do not have enough counts and are not sufficiently resolved to permit separate spectral analysis. Instead, we modelled the combined emission with a thermal-plasma (\texttt{apec}) plus power-law model, where the two components are associated to the soft and hard sources, respectively. For the thermal component, we assumed line-of-sight Galactic absorption only; for the power-law component, we allowed for additional intrinsic absorption. The likely presence of intrinsic absorption associated with the X-ray emission from the radio core is suggested by the lack of enhanced emission in the soft X-ray band at that same position. 

First, we extracted a spectrum within a small (radius of 1\farcs0), circular source region centred on the ATCA/{\it{HST}}/{\it{Gaia}} position. This was done to minimize the contamination from diffuse hot plasma around X-2, either from the nearby SNR or from the surrounding region. It provides as clean a spectrum as possible of the point-like ``X-ray core'' of X-2.  The background consisted of two circular regions (radius of 2\farcs5 each) a few arcsec away, not contaminated by other X-ray sources. We fitted the X-2 core spectrum in {\sc{xspec}} with the Cash statistics \citep{cash79}, as it contains only $\approx$78 net counts (net count rate of $(1.17 \pm 0.14) \times 10^{-3}$ ct s$^{-1}$ in 67.0 ks). At solar metallicity, we obtain a best-fitting thermal plasma temperature $kT = 0.95^{+0.31}_{-0.32}$ keV and a power-law photon index $\Gamma = 1.3^{+1.2}_{-1.0}$ (Table \ref{chandra}). We also tried adding an additional intrinsic absorption component in front of the \texttt{apec} component, but it does not improve the fit and its best-fitting values converges to zero. The total unabsorbed luminosity in the 0.3--10 keV band is $L_{\rm tot} = \left(8.8^{+12.8}_{-2.3}\right) \times 10^{37}$ erg s$^{-1}$, of which $\approx$8.2 $\times 10^{37}$ erg s$^{-1}$ are from the power-law component and $\approx$6 $\times 10^{36}$ erg s$^{-1}$ from thermal plasma. 

We then repeated the spectral analysis but with a larger source extraction radius: an ellipse with semi-major axes of 1\farcs25 and 1\farcs75 (position angle $= 0$, that is major axis oriented in the north-south direction), designed to include not only the unresolved core but also the extended soft X-ray emission. 
We used the same model applied to the X-2 core. We have $\approx$103 net counts (net count rate of $(1.53 \pm 0.16) \times 10^{-3}$ ct s$^{-1}$). The new best-fitting parameters are consistent with the previous ones, as the core dominates the emission; the best-fitting solar-metallicity thermal plasma temperature is $kT = 0.90^{+0.18}_{-0.22}$ keV and the power-law photon index is $\Gamma = 1.6^{+1.1}_{-0.9}$. The unabsorbed power-law luminosity is $\approx$8.6 $\times 10^{37}$ erg s$^{-1}$, and the unabsorbed thermal plasma luminosity is $L_{\rm{apec}} \approx 7 \times 10^{36}$ erg s$^{-1}$ ($N_{\rm{apec}} \approx 7.9 \times 10^{-7}$ in {\sc{xspec}} units). 

\begin{figure}[t!]
\hspace{-0.5cm}
\includegraphics[width=0.72\hsize,angle=270]{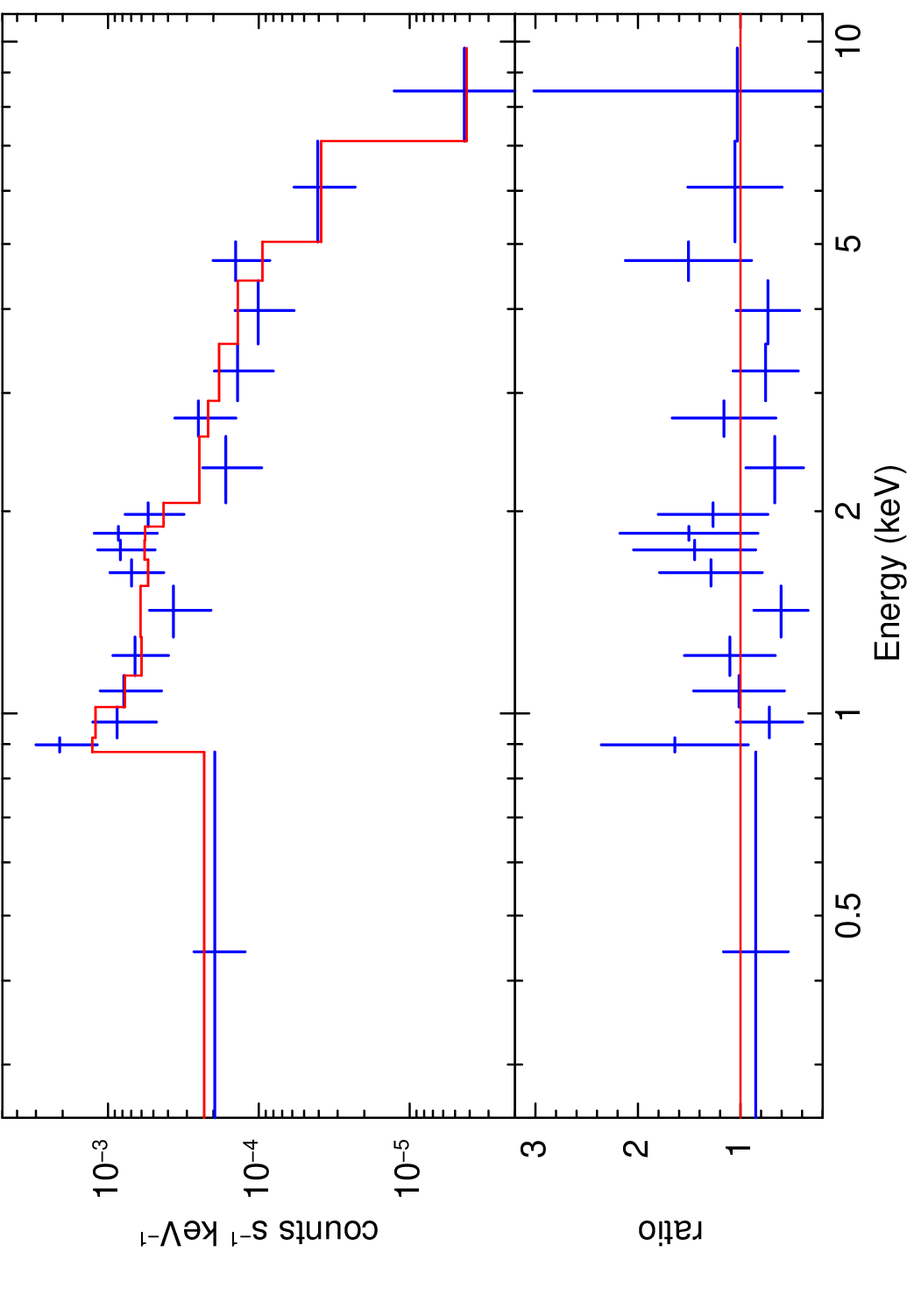}
   \vspace{0.9cm}
      \caption{{\it{Chandra}}/ACIS spectrum of X-2 (core plus slightly extended halo), fitted with \texttt{tbabs}$_{\mathrm{Gal}} \times$ (\texttt{apec} $+$ \texttt{tbabs}$_{\mathrm{int}} \times$ \texttt{pow}). Top panel: datapoints and best-fitting model; bottom panel: data/model ratios. The data have been grouped to a minimum signal-to-noise ratio of 2, for plotting purposes only. See Table \ref{chandra_0.3} for the full list of fit parameters.
      }
         \label{acis_spectrum}
   \end{figure}

Our third step was to take a more realistic (lower) metal abundance for the \texttt{apec} component. The metallicity of the two main starburst regions near X-2 is $12 + \log(O/H) \approx 8.03$--8.13 \citep{guseva11}, corresponding to $Z \approx 0.3Z_\odot$, somewhere between Large and Small Magellanic Cloud. This choice of lower metallicity does not improve our fit but provides a more realistic estimate of the physical parameters in the thermal plasma. 
The \texttt{apec} normalization is now 
$N_{\rm{apec}} = 2.3^{+1.2}_{-1.5} \times 10^{-6}$ 
and the unabsorbed thermal plasma luminosity is $L_{\rm{apec}} = 8.3^{+4.0}_{-5.4} \times 10^{36}$ erg s$^{-1}$ (Table \ref{chandra_0.3} and Figure \ref{acis_spectrum}).

Assuming a spherical geometry and a radius $R_{\rm{s}} \approx 1\arcsec$ for the diffuse X-ray emitting region, corresponding to a volume $V \approx 1.7 \times 10^{60}$ cm$^3$, and using the \texttt{apec} normalization derived from the low-metallicity model, we infer a density $n_e \approx 0.6 f^{-1/2}$ cm$^{-3}$, where $f$ is the filling factor of the hot gas. The corresponding pressure of this hot bubble of X-ray emitting plasma is $P/k \approx 1.5 \times 10^7$ cm$^{-3}$ K and the mass of the hot gas is $\approx$850 $f^{1/2}$ $M_\odot$. The thermal energy content of the hot bubble is $E_{\rm{th}} \approx 3 n_e kTV \approx 5 \times 10^{51} f^{1/2}$ erg. Let us assume that a jet or semi-relativistic, collimated outflow launched by the X-2 core is responsible for the 0.9-keV plasma. The shock velocity required to heat gas at such temperatures is $\approx$900 km s$^{-1}$. 
The age of the bubble is then $t \approx 3R_{\rm{s}}/5 \approx 4.9 \times 10^{11}$ s $\approx 15,700$ yr, in the Sedov approximation. Standard bubble theory \citep{weaver77} shows that the thermal energy content of the shocked gas between the reverse shock and the swept-up shell is $E_{\rm{th}} \approx (11/5) P_{\rm{k}} t$, where $P_{\rm{k}}$ is the kinetic power inflating the bubble and $t$ is the age of the bubble. 
Equating this value to the empirically derived thermal energy content of the hot plasma, we estimate $P_{\rm{k}} \approx 2 \times 10^{40} f^{1/2}$ erg s$^{-1}$. This kinetic power is comparable with the jet power inferred for other ULX bubbles such as NGC\,7793 S26 \citep{pakull10,soria10}, NGC\,5585 X-1 \citep{soria21}, M\,83 MQ1 \citep{soria14}.

Finally, in view of our finding of radio variability, we tested whether the hard X-ray emission from the radio core position varies significantly. In particular, we combined the first three {\it{Chandra}} observations (from 2002 to early 2004), corresponding to lower radio flux densities, and the later {\it{Chandra}} observations (from late 2004 to 2010), corresponding to higher radio flux densities. We analysed the two combined epochs separately. We find that the point-like, hard X-ray emission at the radio core position is present in both the early and the late spectra, without significant changes in the count rate. Thus, the observed factor-of-3 increase in radio luminosity is not reflected in similar variations of the X-ray core luminosity.

   \begin{figure*}[t!]
\hspace{-0.5cm}
\includegraphics[width=0.36\hsize,angle=270]{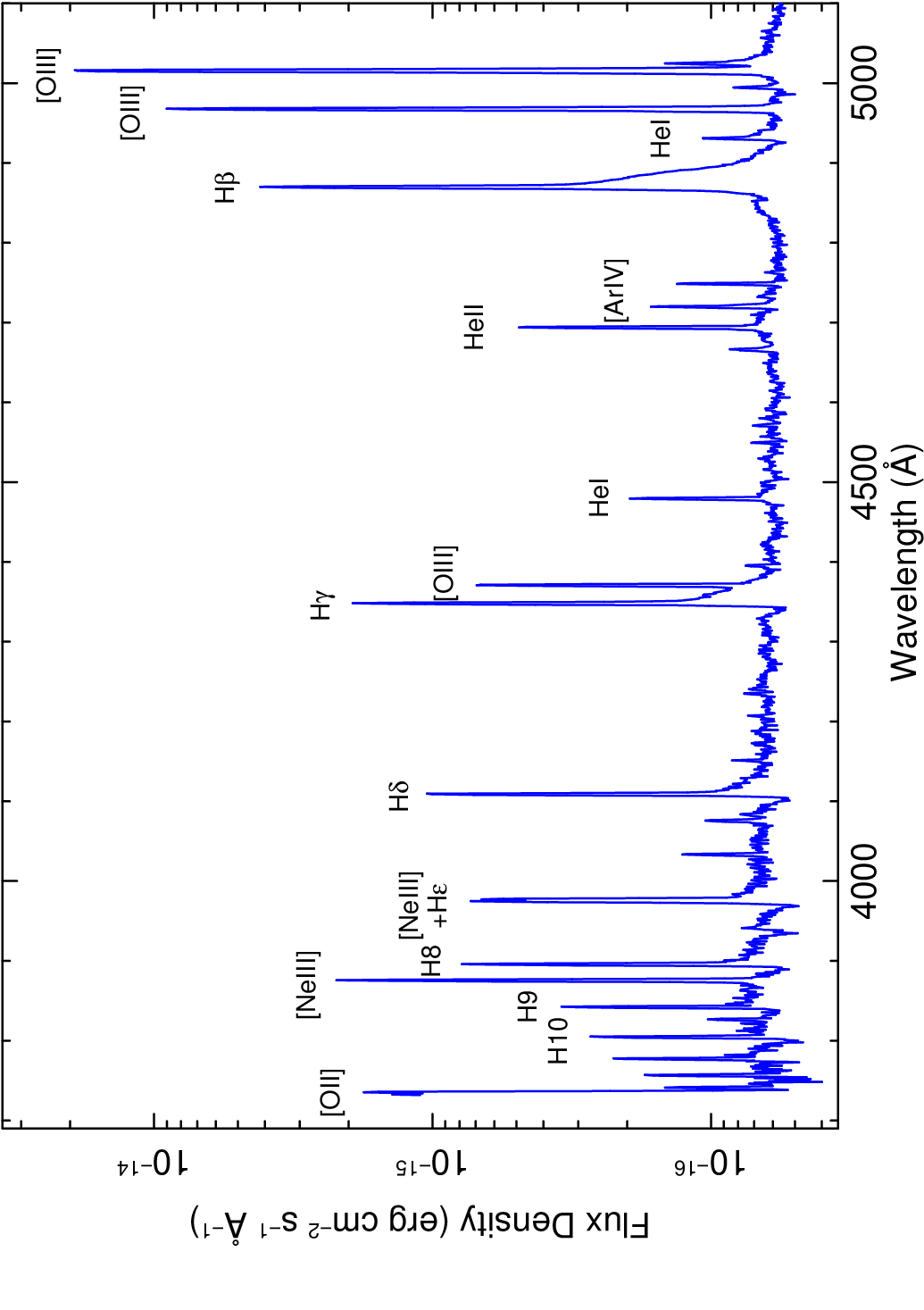}
\includegraphics[width=0.36\hsize,angle=270]{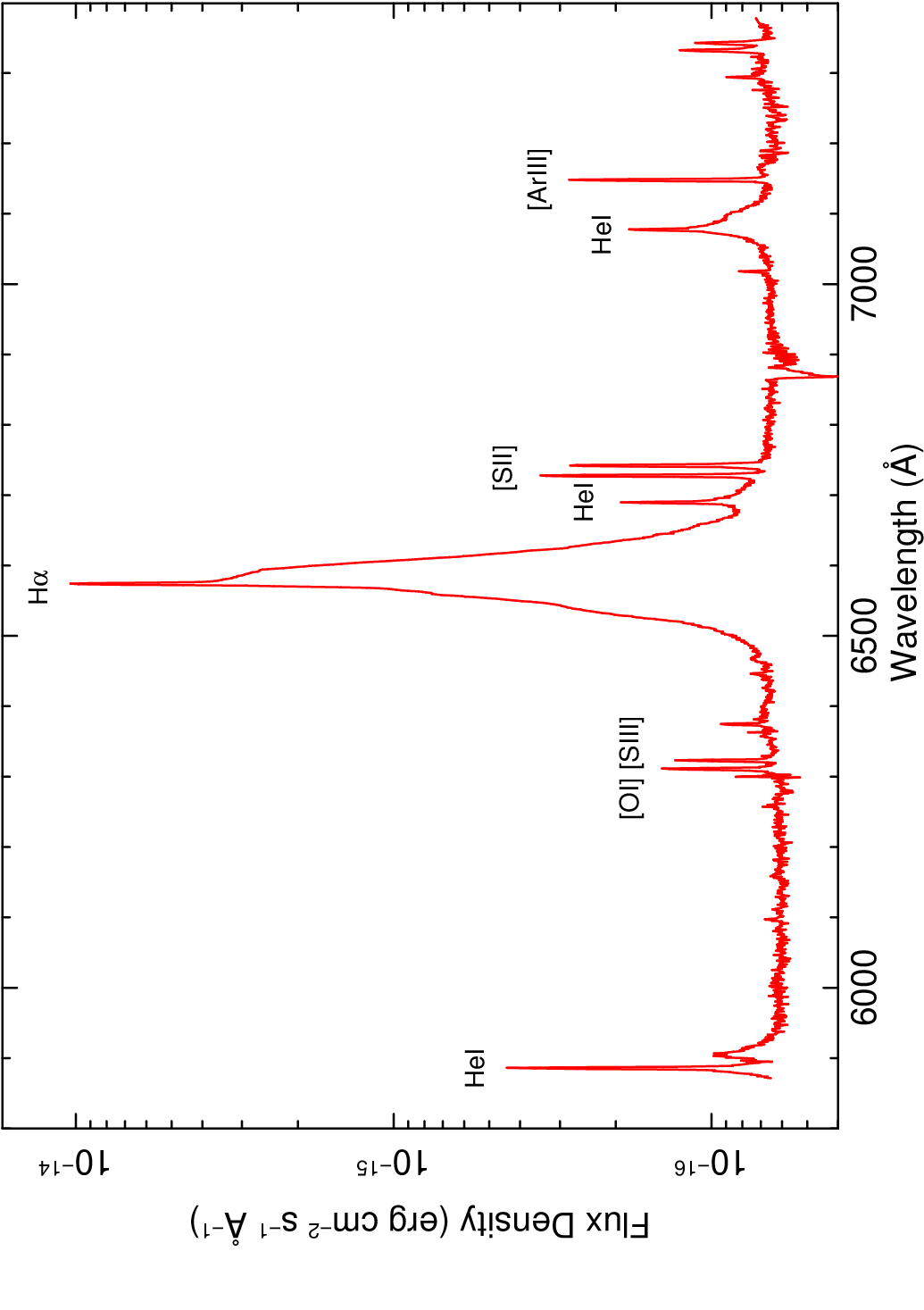}\\

\hspace{-0.5cm}
\includegraphics[width=0.36\hsize,angle=270]{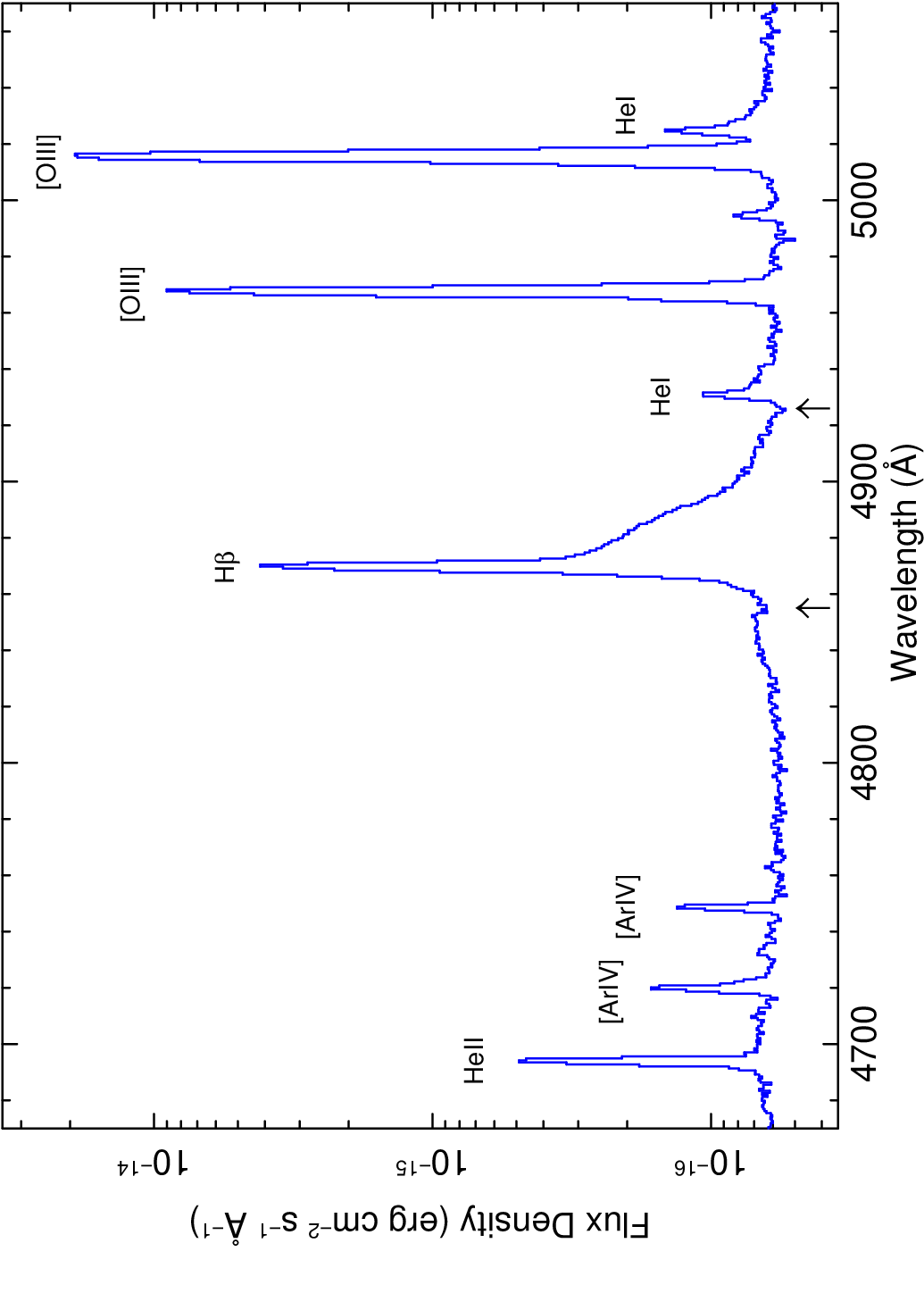}
\includegraphics[width=0.36\hsize,angle=270]{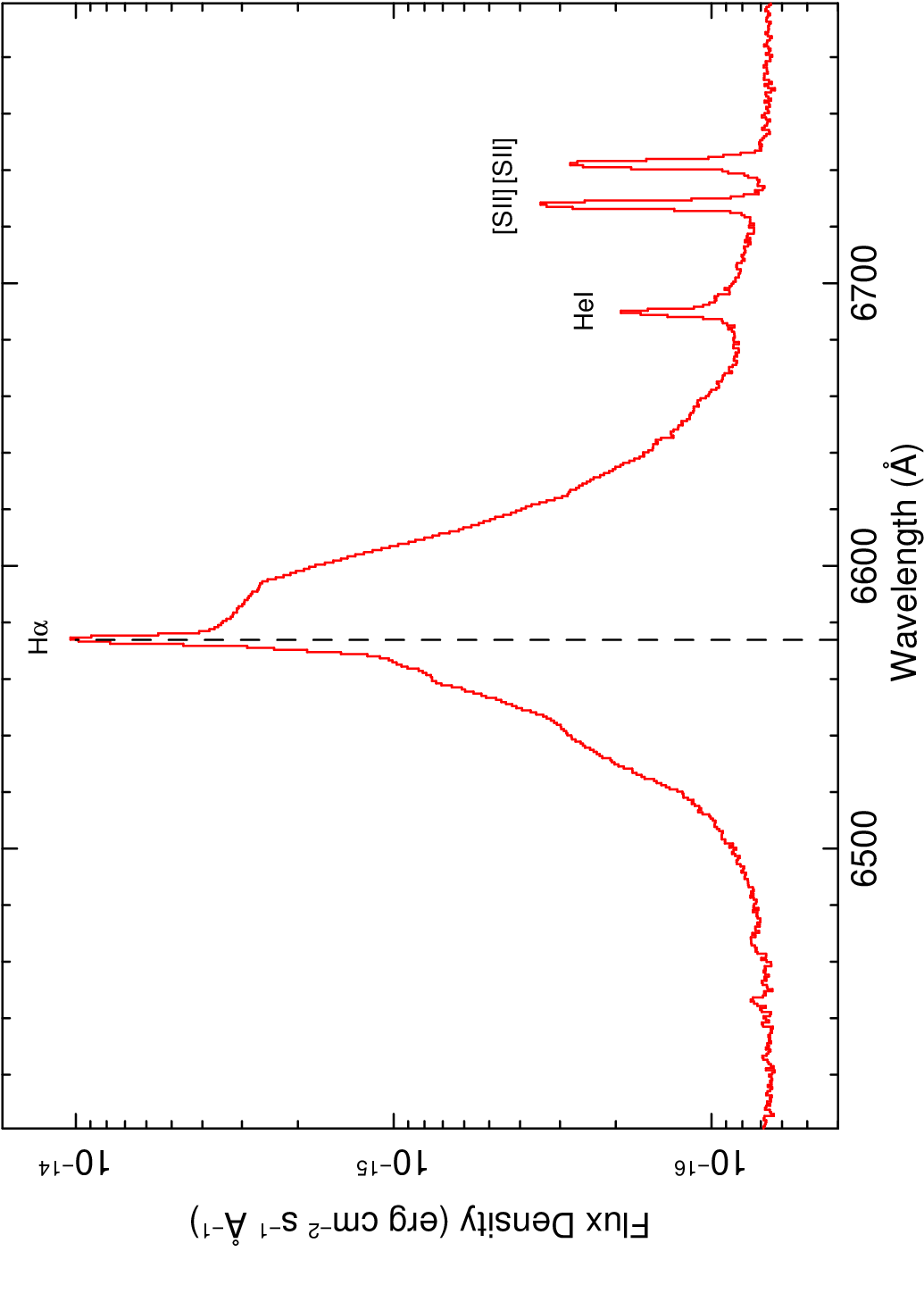}
   \vspace{0.9cm}
      \caption{Top panels: VLT FORS2 spectra from 2010 April 12, taken with the 1200B and 1200R grisms (spectral dispersion of 0.72 \AA\ and 0.74 \AA\ per pixel, respectively). The main lines are labelled. Bottom panel: zoomed-in view of the same spectra, around He\,{\footnotesize{II}} $\lambda$4686 -- [O\,{\footnotesize{III}}] $\lambda\lambda$4959,5007 for the blue, and H$\alpha$ -- [S\,{\footnotesize{II}}] $\lambda\lambda$6717,6731 for the red. In the blue spectrum, black arrows point to the most prominent P-Cygni features seen in the broad (outflow) lines. In the red spectrum, a dashed line marks the central position of the H$\alpha$ line, to highlight the asymmetric profile, in which a substantial fraction of the blue wing is cancelled by absorption (again, a P-Cygni effect). Note that the narrow H$\alpha$ on top of the broad component is saturated.
      }
         \label{fors2_2010}
   \end{figure*}

\subsection{Optical spectroscopy: FORS2}

The long-slit optical spectra of X-2 confirm that this source has unusual properties and is associated with a powerful, variable outflow. Here, we highlight a partial selection of the most interesting findings. We have chosen the 2010 FORS2 spectra as the main basis of our analysis, because it was taken with excellent seeing (Appendix A.3), because it covers both the red and the blue part of the spectrum (unlike the 2008 FORS1 spectra which only covered the blue band), and because its 1\farcs0 slit width guarantees that the X-2 emission is almost fully inside the slit (slit losses $\sim$10--20 per cent), near its lower edge. By contrast, with the 0\farcs7 slit used in 2011 and 2012, X-2 is centred at the lower edge of the slit or even slightly outside, and more than half of the X-2 flux is not falling on the slit. Exact estimates of the fraction of slit losses are not crucial for this work, because here we are mostly interested in the line profiles, line ratios, and order-of-magnitude estimates of the power of X-2. 

In addition to the flux from X-2, long-slit spectra inevitably contain some of the flux from the spatially extended He$^{++}$ emission region centred $\approx$0\farcs6 north of X-2 (Section 3.4). We do not have narrow-band images of the field taken during the 2008--2012 FORS1/FORS2 observations. However, given its size (radius $\approx$20 pc) it is plausible to assume that the emission spectrum of the ionized nebula cannot have changed substantially in 15 years; thus, we can take the 2024 MUSE spectrum as a good indicator of what the nebula contributed to the 2008--2012 FORS1/FORS2 spectra. The position angle of the slit in the FORS1/FORS2 observations makes it difficult to test the spatial extent of the emission lines and resolve the extended contribution from the point-like X-2 emission. Considering the slit width (1\farcs0) and the location of the X-2 core near the bottom of the slit ($\approx$0\farcs2 from its edge) rather than at the centre, we estimate that the 2010 FORS2 slit includes about 2/3 of the extended $\lambda$4686-emitting region seen in MUSE. In this section, we will examine the 2010 FORS2 spectrum; we will then examine the 2024 MUSE spectrum of the whole nebula (Section 3.7). From their comparison, we will show (Section 3.8) that the 2010 spectrum was qualitatively and quantitatively different from the 2024 nebular spectrum: it had much stronger lines, with broader components. We will argue that the difference between the two epochs is the direct contribution of X-2, which was extremely powerful in 2010 but had greatly weakened by 2024, as noted earlier (Section 3.3). We leave a full analysis, line modelling and comparisons of the numerous long-slit spectral datasets (including the NTT and SALT spectra) to further work. 

The most striking result from the long slit spectra is the presence of both a very broad component and a narrow component in the Balmer lines, He\,{\footnotesize{I}} lines, and He\,{\footnotesize{II}} $\lambda$4686. All forbidden lines only show a narrow emission component, a sign that the density in the outflow is too high to permit such lines. The 2010 FORS2 spectra (Figure \ref{vlt_2d} for the 2-d spectra, and Figure \ref{fors2_2010} for the 1-d spectrum of X-2) illustrate the strength and complexity of those two line components. In fact, a third component is also required for the fitting of the strongest lines: P-Cygni absorption of the blue wing, which is another hallmark of a dense outflow.

We fitted the Balmer line profiles from the 2010 FORS2 spectra (Figures \ref{fors2_2010}) with three Gaussians, using the {\sc{qfitsview}} software package. Two of the Gaussians were assumed to be in emission (broad and narrow component); the third one was in absorption (P-Cygni), removing a substantial chunk of the blue wing emission. The main properties of the three components for H$\alpha$, H$\beta$ and H$\gamma$ are summarized in Table \ref{line_parameters}. The fitted FWHMs $\approx$ 1600--1700 km s$^{-1}$ at the wind photosphere typically correspond to wind speeds at infinity $v_\infty \sim 2000$ km s$^{-1}$, by analogy with stellar winds in O and Wolf-Rayet stars. Even higher speeds are inferred from the half widths at zero intensity (HWZI), particularly for H$\alpha$, in which HWZI $\approx 4600$ km s$^{-1}$. However, such broad wings are likely enhanced by Thomson scattering from free electrons in the dense stellar wind, in addition to Doppler broadening from the outflow velocity \citep{castor70}.

\begin{figure}[t!]
\hspace{-0.5cm}
\includegraphics[width=0.73\hsize,angle=270]{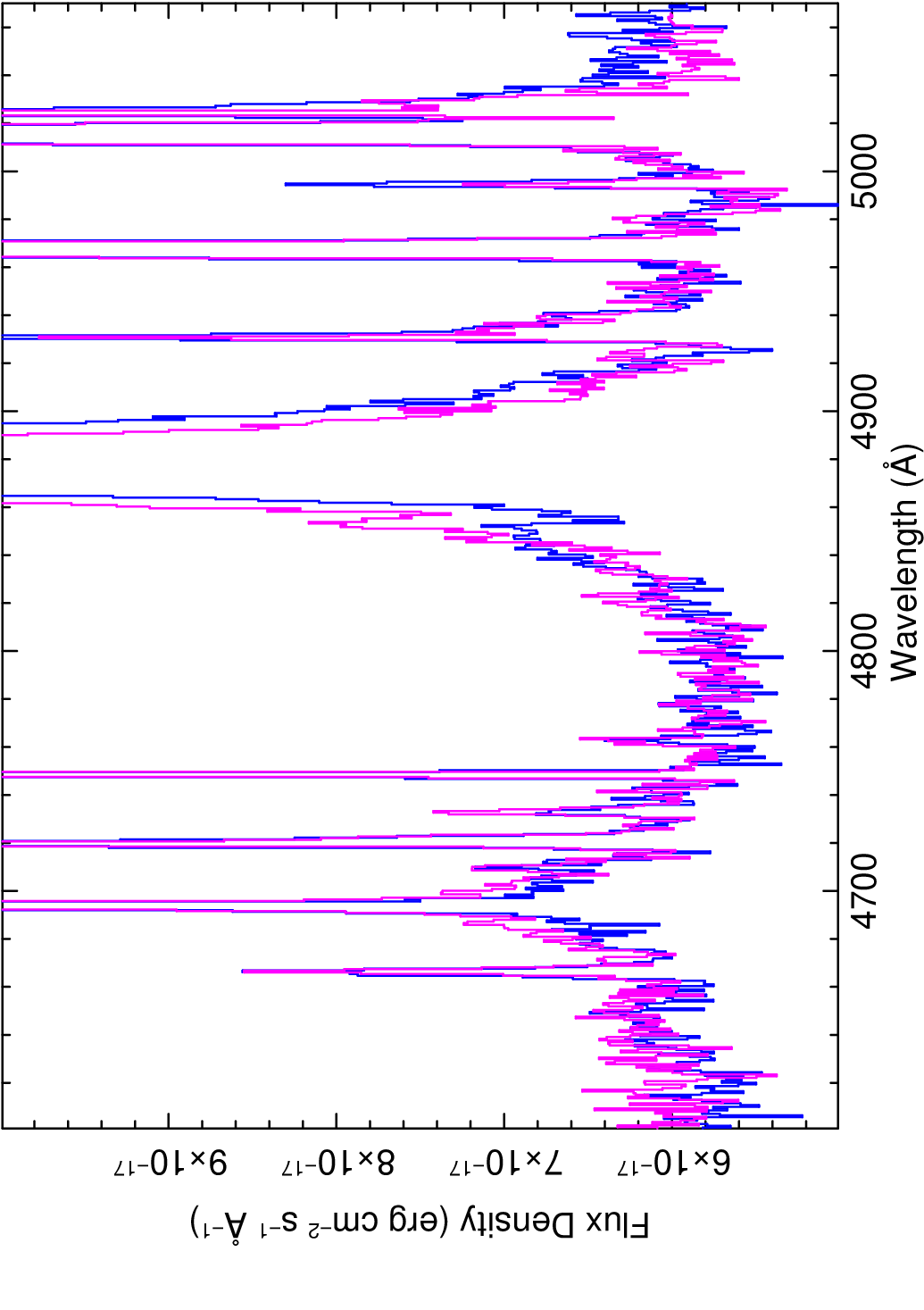}
   \vspace{0.9cm}
      \caption{Comparison between a blue section of the spectra from VLT/FORS2 in 2010 (blue) and 2012 (magenta), plotted in linear scale of flux density. The 2012 spectrum has been renormalized by a factor of 2.4 to compensate for the slit losses. The main lines with broad components (in addition to narrow component) in this section of the spectrum are H$\beta$, He\,{\footnotesize{I}} $\lambda$4921, He\,{\footnotesize{I}} $\lambda$5016 (all of them with P-Cygni absorption) and He\,{\footnotesize{II}} $\lambda$4686 (without P-Cygni absorption). The other lines, from [Fe\,{\footnotesize{III}}], [Ar\,{\footnotesize{IV}}] and [O\,{\footnotesize{III}}] (Table \ref{line_table_2010}) only show narrow components. The width and strength of the broad components was similar in 2010 and 2012, which suggests that the decline in the core activity started after that year.  
      }
         \label{fors2_2010_2012}
   \end{figure}

Broad components are also seen from He\,{\footnotesize{I}} and He\,{\footnotesize{II}}. In those cases, a two-component fit (broad and narrow) is generally sufficient for a good fit. A couple of representative examples are also reported in Table \ref{line_parameters}: they confirm a characteristic FWHM $\sim$2000 km s$^{-1}$. 
In particular, we highlight the broad component in He\,{\footnotesize{II}} $\lambda$4686, by replotting a zoomed-in section of the blue spectrum with linear flux density scale (Figure \ref{fors2_2010_2012}). In the same panel, we show the corresponding section of the 2012 FORS2 spectrum (renormalized by a factor of 2.4 to account for higher slit losses due to the smaller slit width). 
The broad components had approximately the same FWHM and EW in 2010 and 2012. This suggests that the decline of the core emission must have started after that date. Broad He\,{\footnotesize{II}} $\lambda$4686 is a hallmark of Wolf-Rayet stars; however, we will show (Section 3.7) that the broad $\lambda$4686 component had almost disappeared by 2024, which proves that its earlier origin was a variable accretion-powered outflow rather than a real Wolf-Rayet star. The similarity between optical line spectra of Wolf-Rayet stars and those of accretion disk winds is a typical property of X-ray binaries above the Eddington limit \citep{hunstead97,fabrika15}.

The steep Balmer decrement of the broad H\,{\footnotesize{I}} components (H$\alpha$/H$\beta$ $\approx 24$, H$\beta$/H$\gamma$ $\approx 4.6$) suggests that the Balmer lines are optically thick \citep{osterbrock06}. Instead, for the Balmer decrement of the optically thin components, we obtain H$\gamma$/H$\beta = 0.44$, H$\delta$/H$\beta = 0.24$ (Tables \ref{line_parameters} and \ref{line_table_2010}). This is consistent with Case-B recombination, with the addition of a mild reddening $E(B-V) \approx 0.1$ mag. The anomalously high ratio H$\alpha$/H$\beta \approx 2.58$ occurs because the narrow H$\alpha$ component is saturated in the 2010 spectra. For the same reason (saturation), the measured FWHM of the narrow H$\alpha$ component appears larger than the instrumental resolution.

The wealth of nebular lines (Table \ref{line_table_2010}) in the FORS2 spectra permits an estimate of the gas density and temperature based on several diagnostic line ratios \citep{osterbrock06}, with the PyNeb code \citep{luridiana15}. For the temperature in lower-ionization parts of the gas, we used the [O\,{\footnotesize{II}}] (3726$+$3729)/(7320$+$7330) ratio (observed value $\approx$20.5 in the 2010 spectrum), and the [S\,{\footnotesize{II}}] (4068$+$4076)/(6716$+$6731) ratio (observed value $\approx$0.12). For the higher-ionization gas component, the used the standard temperature diagnostic [O\,{\footnotesize{III}}] (4959$+$5007)/4363 (observed value $\approx$58, after taking into account that [O\,{\footnotesize{III}}] $\lambda$5007 is saturated but assuming that its flux is 3 times that of $\lambda$4959). For the densities, we used the [O\,{\footnotesize{II}}] 3729/3726 doublet (observed ratio of $\approx$1.30), the [S\,{\footnotesize{II}}] 6716/6731 doublet (observed ratio of $\approx$1.26), and the [Ar\,{\footnotesize{IV}}] 4711/4740 doublet \citep{keenan97} (observed ratio of $\approx$1.46). All three temperature diagnostics are consistent with a value of $T_e \approx (16,300 \pm 500)$ K. For the density, the [S\,{\footnotesize{II}}] ratio suggests $n_e \approx (180 \pm 50)$ cm$^{-3}$, and the [O\,{\footnotesize{II}}] ratio indicates $n_e \approx (220 \pm 50)$ cm$^{-3}$. The [Ar\,{\footnotesize{IV}}] ratio gives a mildly discrepant result, hitting the low-density limit $n_e \lesssim 100$ cm$^{-3}$. 

Finally, following the method of \cite{izotov19}, we checked the metal abundance of the photoionized gas, from the measured values of the $R_{23}$ and $O_{32}$ indices, where $R_{23} = $([O\,{\footnotesize{II}}] $\lambda\lambda$3726,3729 $+$ [O\,{\footnotesize{III}}] $\lambda$4959 $+$ [O\,{\footnotesize{III}}] $\lambda$5007)/H$\beta$, and  $O_{32} =$ [O\,{\footnotesize{III}}] $\lambda$5007/([O\,{\footnotesize{II}}] $\lambda\lambda$3726,3729). From Table \ref{line_table_2010}, we obtain $\log R_{23} \approx 0.97$, and $\log (R_{23} - 0.08 O_{32}) \approx 0.93$. This places us in the metallicity range where such indices are least constraining, in the range $7.7 \lesssim (12 + \log O/H) \lesssim 8.3$ \citep{izotov19}, consistent with the low (but not extreme) abundance estimated by \cite{guseva11} (Section 3.5). Other well-known metallicity indicators, such as the N2 and O3N2 indices defined by \cite{pettini04}, cannot be used in our case because [N\,{\footnotesize{II}}] $\lambda$6584 is completely swamped by the red wing of H$\alpha$.

\subsection{Optical spectroscopy: MUSE}

\begin{figure*}[t!]
\hspace{-0.5cm}
\includegraphics[width=0.36\hsize,angle=270]{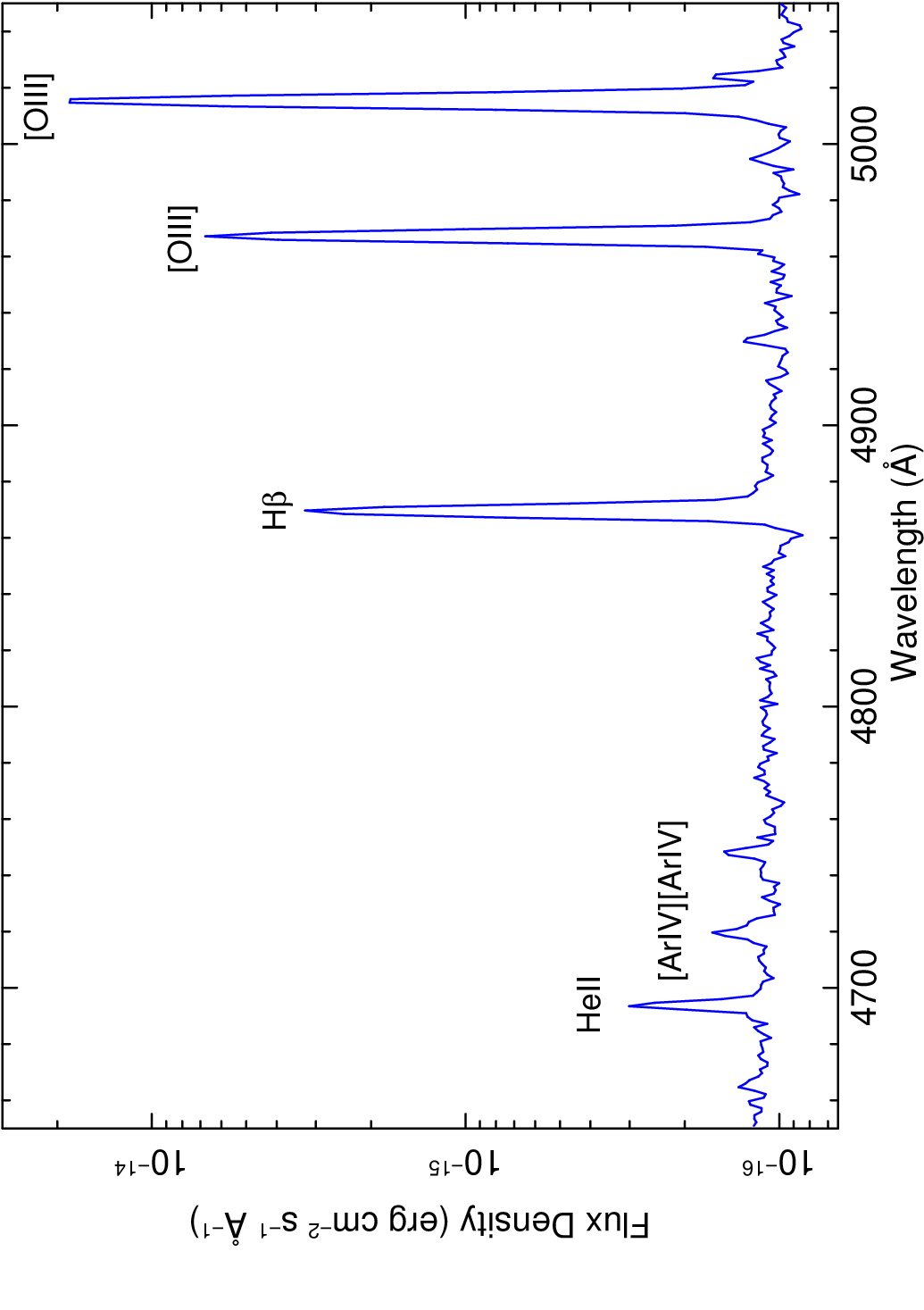}
\includegraphics[width=0.36\hsize,angle=270]{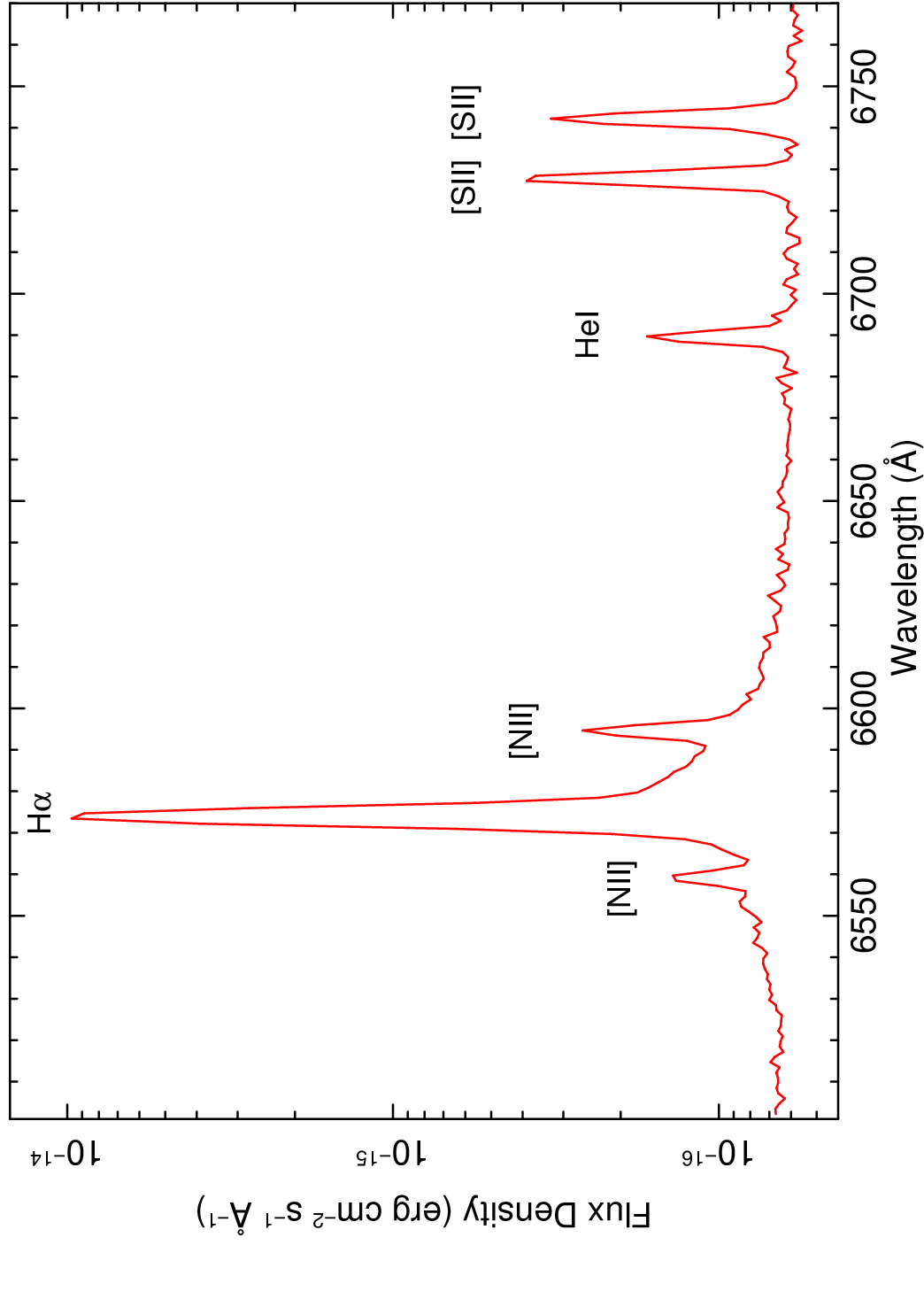}
   \vspace{0.9cm}
      \caption{Left panels: 2024 Jun 27 MUSE spectrum of the extended nebula near X-2 (including X-2 itself), zoomed in around He\,{\footnotesize{II}} $\lambda$4686 -- [O\,{\footnotesize{III}}] $\lambda\lambda$4959,5007. Right panel: the same MUSE spectrum, zoomed in around H$\alpha$ -- [S\,{\footnotesize{II}}] $\lambda\lambda$6717,6731. 
      The broad line components have weakened or disappeared, compared with 2010 (Figure \ref{fors2_2010}); a P Cygni feature remains visible at H$\beta$. The reason why the continuum flux is not fainter than the level shown in the 2010 spectrum (Figure \ref{fors2_2010}) is that the 2024 spectrum is taken from a larger region, including stellar contributions.
      }
         \label{muse_2024}
   \end{figure*}

Next, we analyse the 2024 MUSE spectra (May 1 and June 27) of the extended nebula, extracted from a circular region of radius 0\farcs8 centred on the peak of the He\,{\footnotesize{II}} $\lambda$4686 emission ($\approx$0\farcs6 north of X-2). We report (Table \ref{line_table_2024}) the observed line fluxes from June 27 because of its higher signal-to-noise ratio; the results from May 1 do not show significant differences. The spectra (Figure \ref{muse_2024}) are dominated by narrow lines, on top of a blue continuum (emitted by stars A and B and other massive stars around them). The broad components are strongly weakened compared with 2008--2012, as we shall quantify in detail in Section 3.8. The weak broad components still visible at H$\alpha$, H$\beta$ and He\,{\footnotesize{II}} $\lambda$4686 (Table \ref{line_table_2024}) are consistent with the emission from a metal-poor (weak wind) Wolf-Rayet: perhaps that same star A or star B that we identified as possible sources for the extended He$^{++}$ nebula. The same can be said for the P-Cygni feature $\approx$500 km s$^{-1}$ blueward of H$\beta$. To investigate the disappearance of such features, we also extracted and analysed spectra from small circular regions (0\farcs2, 0\farcs4, 0\farcs6) around the position of X-2 rather than the centre of the nebula; however, there is no significant increase in the broad components compared with the narrow ones.

The wavelength coverage is $\approx$4600--9300 \AA\ (Table \ref{line_table_2024}). This prevents us from using the three temperature diagnostics adopted for the FORS2 spectrum. However, we can use the [S\,{\footnotesize{III}}] (9069$+$9531)/6312 ratio, following \cite{diaz22} with the atomic data of \cite{fernandez18}. [S\,{\footnotesize{III}}] $\lambda$6312.1 and $\lambda$9068.6 are detected as nebular lines (Table \ref{line_table_2024}). [S\,{\footnotesize{III}}] $\lambda$9530.6 is outside our wavelength coverage; however, the theoretical intensity ratio of 9531/9069 $\approx$ 2.44 (both lines arise from the same upper level). Thus, we can determine an observed value of [S\,{\footnotesize{III}}] (9069$+$9531)/6312 $\approx$ 28.5, or 26.4 after dereddening. This corresponds to a characteristic temperature $T_e = (14,000 \pm 500)$ K, slightly lower than the temperature estimates from the 2010 FORS2 spectrum. 
For density estimates, we used the [S\,{\footnotesize{II}}] diagnostic ratio 6716/6731 $= 1.3 \pm 0.2$ (Table \ref{line_table_2024}), corresponding to $n_e \lesssim$ a few 100, consistent with the 2010--2012 estimates.
Finally, for estimates of the metal abundance, the MUSE spectra have the advantage that [N\,{\footnotesize{II}}] $\lambda 6584$ is resolved from the red wing of H$\alpha$ (much weaker than in 2010--2012, as discussed later). Therefore, we can use the $N_2$ index calibrated by \cite{pilyugin16} ($N_2 \equiv $[N\,{\footnotesize{II}}] $(6548+6584)/H\beta$). We estimate $7.5 \lesssim (12 + \log O/H) \lesssim 8.1$. Using instead  ([O\,{\footnotesize{III}}] $\lambda 5007/H\beta)/$([N\,{\footnotesize{II}}] $\lambda 6584/H\alpha)$ \citep{pettini04}, we obtain $7.5 \lesssim (12 + \log O/H) \lesssim 8.2$.

\subsection{Decline of X-2 from FORS1/FORS2 to MUSE}
We have already shown (Section 3.3, Figure \ref{muse_dimming}) that the optical continuum from the X-2 core dimmed by about 2 mag between 2010 and 2024. We will show here that a similar dimming affected the broad and narrow emission lines, between the 2010 FORS2 spectra and the MUSE spectrum (2024). 

The decrease in the broad-line components (proxy for the outflow power) is most dramatic.  H$\alpha$ retains a broad component (Figure \ref{muse_2024} and Table \ref{line_table_2024}), with a FWHM of $\approx$$(1100 \pm 200)$ km s$^{-1}$, but narrower than in 2010, when it was $\approx$1700 km s$^{-1}$. Its HWZI was $\approx$$(2300 \pm 150)$ km s$^{-1}$, half of the wing speed in 2010. The observed flux of the broad H$\alpha$ component in 2024 was $F({\mathrm{H}}\alpha,{\mathrm{broad}}) = (23 \pm 2) \times 10^{-16}$ erg cm$^{-2}$ s$^{-1}$, that is 50 times lower than in 2010. The broad component of H$\beta$ suffered a similar decline, with its FWHM reduced to $\approx$$(900 \pm 300)$ km s$^{-1}$ (it was $\approx$1600 km s$^{-1}$ in 2010) and its flux reduced to $(3 \pm 1) \times 10^{-16}$ erg cm$^{-2}$ s$^{-1}$, down from $\approx$$50 \times 10^{-16}$ erg cm$^{-2}$ s$^{-1}$ in 2010. Broad components of He\,{\footnotesize{I}} and He\,{\footnotesize{II}} are also down in FWHMs and fluxes, or not detected.
The signal-to-noise and spatial resolution of the MUSE data do not enable us to determine how much of the residual broad-line emission is still coming from the point-like X-2 counterpart and how much from the extended nebula ({\it{e.g.}}, from the candidate Wolf-Rayet stars A and B, 0\farcs5--0\farcs6 north of it). 
In any case, the fast outflow from X-2 has either disappeared or reduced dramatically. This may also be the reason for the dimming of the optical continuum, which we have attributed to an outflow photosphere.

Narrow line components also declined from 2010 to 2024, especially for the high-ionization lines. In 2024, the narrow component of He\,{\footnotesize{II}} $\lambda$4686 from the extended nebula (which we can assume constant) has an observed flux $f_{\rm{4686,narrow}} \approx 5 \times 10^{-16}$ erg cm$^{-2}$ s$^{-1}$ (Table \ref{line_table_2024}). Only $\approx$2/3 of this flux fell on the 2010 FORS2 slit. However, the narrow-component flux measured in 2010 was $f_{\rm{4686,narrow}} = (9.9 \pm 0.3) \times 10^{-16}$ erg cm$^{-2}$ s$^{-1}$ (Table \ref{line_table_2010}), not even including a $\sim$10--20 per cent slit loss. The difference between the two epochs is the unresolved narrow component of X-2, which is therefore estimated as $f_{\rm{4686,n,X-2}} \approx 7 \times 10^{-16}$ erg cm$^{-2}$ s$^{-1}$ in 2010, and is undetectable ($f_{\rm{4686,n,X-2}} \lesssim 1 \times 10^{-16}$ erg cm$^{-2}$ s$^{-1}$) in 2024.

The He\,{\footnotesize{II}} $\lambda$4686/H$\beta$ line ratio tells a consistent story. In 2024, the average ratio over the extended nebula was $\approx$ 0.05 (for the narrow components).
In the 2010 FORS2 spectrum, He\,{\footnotesize{II}} $\lambda$4686/H$\beta$ $\approx 0.10$ (Table \ref{line_table_2010}); however, if we take into account the dilution by the extended nebular component, the intrinsic He\,{\footnotesize{II}} $\lambda$4686/H$\beta$ ratio for X-2 was as high as $\approx$0.15. This ratio alone does not entirely rule out a hot Wolf-Rayet star as the ionizing source at X-2 \citep{pakull89,naze03a,naze03b}, but at least it is consistent with the presence of a harder ionization source (possibly a soft X-ray source) in 2010. As a comparison, the standard example of X-ray photoionized nebula around the Holmberg II X-1 ULX has He\,{\footnotesize{II}} $\lambda$4686/H$\beta$ $= 0.22 \pm 0.02$ \citep{kaaret04}.

To find a stronger proof of X-ray photoionization at X-2, we need to go back to the FORS1 spectrum taken in 2008, which has a more extended wavelength coverage in the near UV. 
The 2008 spectrum shows (Figure \ref{figneon}) [Ne\,{\footnotesize{V}}] $\lambda$3426 emission from both X-1 and X-2, with the latter being about 3 times stronger. To produce Ne$^{4+}$ in a plasma, photon energies $>97.1$ eV are required. Thus, [Ne\,{\footnotesize{V}}] $\lambda$3426 is an even stronger tracer of soft X-ray irradiation than He\,{\footnotesize{II}} $\lambda$4686 (ionization potential of 54.4 eV). The relation between ionizing photon flux and line emission is for [Ne\,{\footnotesize{V}}] $\lambda$3426 is not as simple as for He\,{\footnotesize{II}} $\lambda$4686; however, we can at least take the relative strength of the line in X-1 and X-2 as an indication of the potentially strong (but unseen by us) X-ray luminosity of X-2 at least at some epochs.

\begin{figure}[t!]
\hspace{-0.5cm}
\includegraphics[width=0.73\hsize,angle=270]{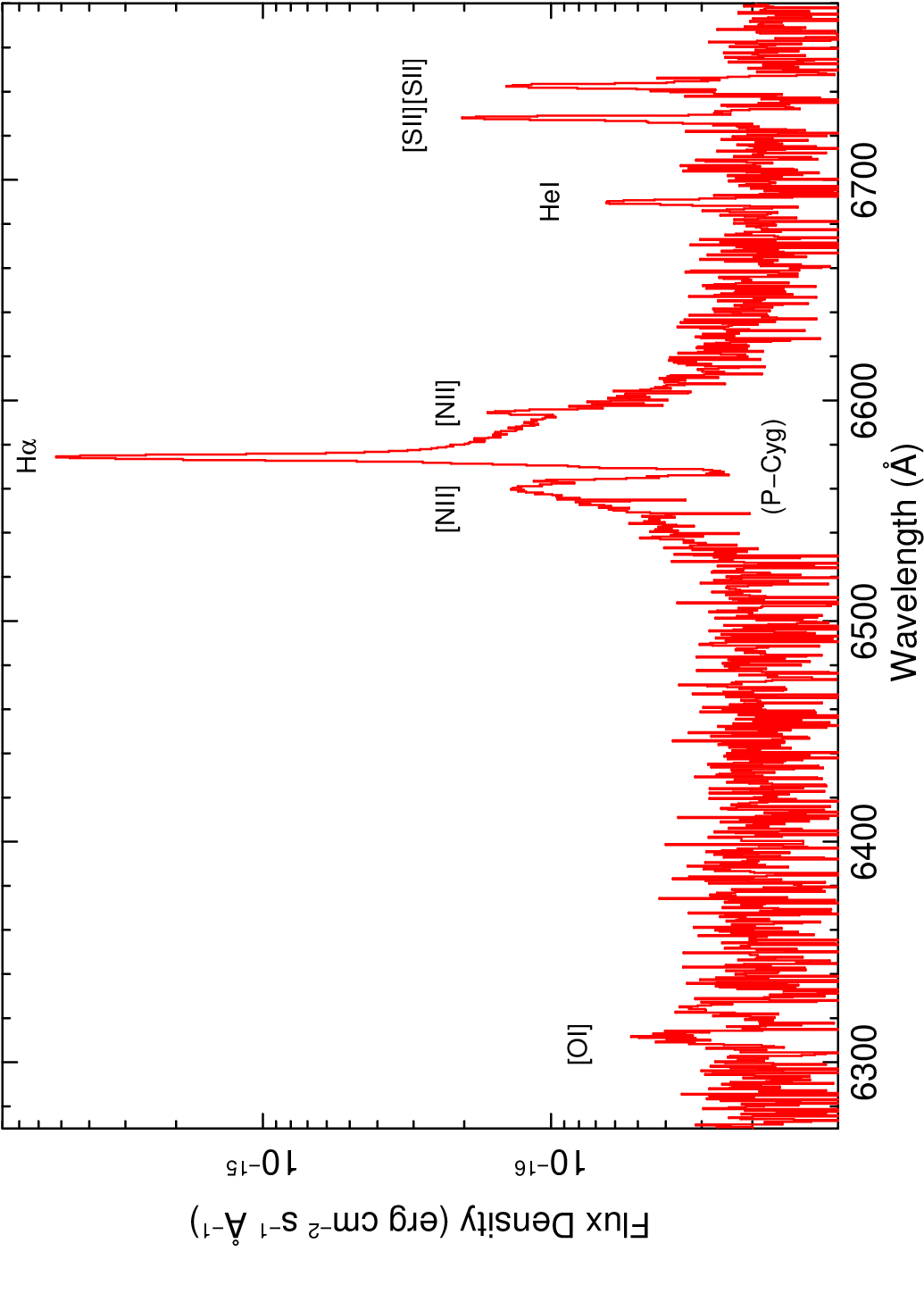}
   \vspace{0.9cm}
      \caption{Portion of the 1993 NTT/EMMI spectrum, around the H$\alpha$ region, showing that the outburst was already ongoing at that time. 
      }
         \label{emmi_1993}
   \end{figure}

\begin{figure}[t!]
\includegraphics[width=\hsize,angle=0]{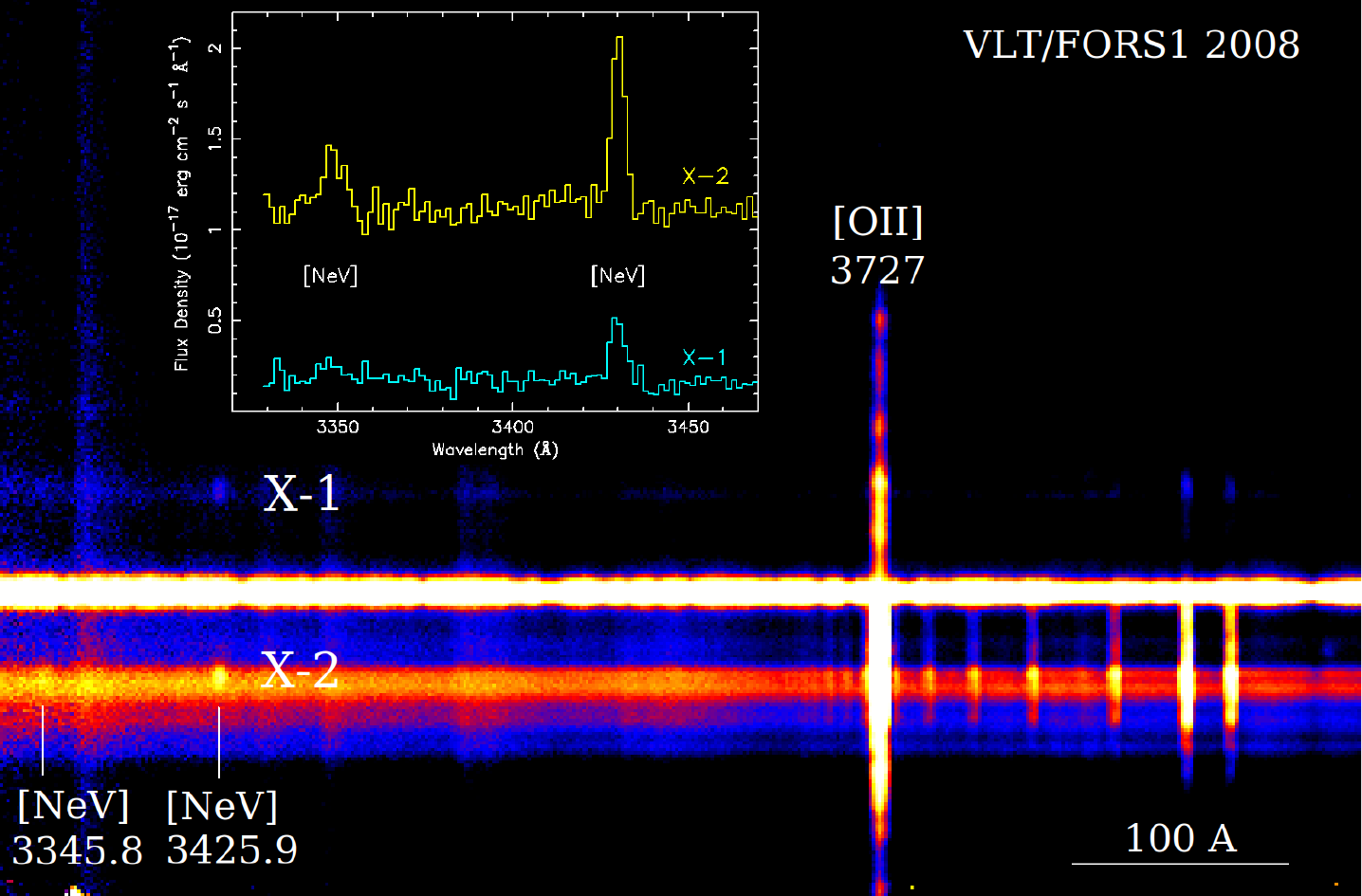}
      \caption{The 2008 FORS1 blue spectrum showing [Ne\,{\footnotesize{V}}] $\lambda$3426 emission from both NGC\,5408 X-1 and X-2, with the latter source being $\approx$3 times stronger.}
         \label{figneon}
   \end{figure}

Finally, we looked for empirical evidence regarding a possible start date of the outflow activity. The earliest spectrum of X-2 we could find is from 1993 (Section 2.2) and covers the H$\alpha$ wavelength band. It is clear (Figure \ref{emmi_1993}) that there is already a broad emission component, somewhat intermediate between the highest activity level of 2010--2012 (Figure \ref{fors2_2010}) and the lowest level of 2024 (Figure \ref{muse_2024}). The wavelength difference between the absorption dip and the narrow-line peak is $\approx$7 \AA\ $\approx$320 km s$^{-1}$. The continuum flux is close to the value measured in the {\it{HST}} observations. Thus, X-2 was already bright and launching an outflow as early as 1993.

\section{Discussion}
At first inspection, 2CXO J140318.4$-$412254 (NGC\,5408 X-2) may be dismissed as an ordinary high-mass X-ray binary in a starburst dwarf. However, with our combined radio, optical and X-ray study, we have shown that X-2 is a strange, extremely powerful source like no other currently known in the local universe. It has signatures of a powerful, flaring jet, with an optically-thin synchrotron spectrum and strong variability. It shows extremely broad lines with P-Cygni profiles, especially from the H\,{\footnotesize{I}} Balmer series and from He\,{\footnotesize{I}}, which suggest the presence of an optically thick outflow. It shows dramatic variability in the outflow lines (both in width and in intensity) from epoch to epoch. It shows a set of narrow photo-ionized lines from an unresolved compact nebula; they are also variable in flux. In addition, X-2 is located near a resolved (radius $\approx$20 pc) photoionized nebula centred about 0\farcs6 north-east of the core. This nebula is responsible for additional He\,{\footnotesize{II}} $\lambda$4686 emission. Finally, there is a marginally resolved region of soft X-ray emission (likely from thermal plasma at $kT \approx 0.9$ keV), approximately co-spatial with the He$^{++}$ nebula. 

Here, we put together a partial selection of what we consider the most important observational findings, trying to frame them into a coherent scenario.  Let us start from the optical images.  The yellow/red ``star'' is much too bright ($L_{\rm{bb}} \approx 4.4 \times 10^{39}$ erg s$^{-1}$ $\approx 1.1 \times 10^6 L_\odot$) and too large ($R_{\rm{bb}} \approx 1700 R_\odot$) to be consistent with ordinary red supergiants, and even more luminous than what is generally considered the upper limit for a hypergiant. If it were a cluster of young stars, rather than an individual one, we could explain its luminosity but not its red colour. Young star clusters in star-forming regions are blue, not red, because the contribution of OB stars and blue supergiants dominates over the comparatively rarer red supergiants. Most importantly, the strong dimming (2 mag) between the {\it{HST}} and VLT/MUSE images strongly suggests that it is a single object rather than a cluster of sources. 

Alongside the extreme optical brightness, X-2 has an outflow with speeds of $\approx$800--1000 km s$^{-1}$, if estimated from the wavelength difference between the centres of the broad lines and the corresponding P-Cygni features, or $\approx$1500--2000 km s$^{-1}$, if estimated from the FWHM of the broad lines. Even higher velocities are seen in the wings, but we do not have elements at this stage to distinguish between kinetic broadening and electron scattering.
A red supergiant or yellow hypergiant cannot produce such fast winds: their wind speeds are only $\approx$10--20 km s$^{-1}$ and $\approx$50--100 km s$^{-1}$, respectively. This is because ordinary stellar winds are roughly proportional to the escape velocity, which is lower in those classes of red stars, given their larger radii. Fast winds up to $\sim$2000 km s$^{-1}$ can be produced by Wolf-Rayet and O stars; however, the colours and optical spectra of X-2 are inconsistent with those stars. 
Optical spectra resembling X-2 are displayed by hot luminous blue variables (LBVs), in particular those found in dwarf galaxies. An example is the object number 3 in the low-metallicity galaxy DDO 68 \citep{pustilnik08,izotov09} which shows impressive P-Cygni profiles from the broad Balmer lines, but does not by far reach the extreme EW or FWHM displayed by the Balmer lines in X-2, at least at some epochs. Moreover, X-2 is redder (cooler) than typical LBVs \citep{smith15}.

Based on those findings, and on the presence of strong X-ray irradiation (quantified later in this Section), we disfavour a purely stellar model, and we argue that: i) the apparent size of the core in the optical continuum is the photosphere of an outflow; ii) the outflow is powered by accretion onto a compact object; iii) the accretion power is at least as large as the optical luminosity plus the kinetic energy of the outflow, making X-2 a likely super-critical accretor. The optical photons emerging from the photosphere at $T_{\rm{bb}} \approx 4600$ K may have been emitted in the X-ray band near the compact object and down-scattered in the outflow.

A jet is the third channel for the accretion power in X-2. Strong evidence for a jet comes from the unresolved core radio luminosity, exceptionally bright for a non-nuclear compact object. Rapid variability as well as a steep spectral index suggest that we are looking at a flaring jet, with the production of adiabatically expanding blobs. By analogy with the (lower) core radio luminosity of ULXs such as SS\,433, NGC\,5408 X-1 and Holmberg II X-1, the jet power could be as high as $\sim$10$^{40}$ erg s$^{-1}$. We cannot tell whether all the radio emission comes from the flaring jet, or, instead, there is a constant component from the interaction of the jet with dense circumstellar shells/envelope (on scales $\lesssim$20 pc) plus a variable component from the core.  Milliarcsec resolution VLBI observations are needed and are currently planned to determine the size and structure of the radio source.

Further indirect evidence for the presence of a strong kinetic power from X-2 is the detection of diffuse thermal plasma emission at $kT \approx 0.9$ keV, up to $\approx$1\arcsec north of the core. The marginally extended soft X-ray emission has a model-dependent luminosity $L_{\rm{apec}} \approx 8 \times 10^{36}\ {\rm erg~s^{-1}}$). This is considerably higher than the X-ray luminosities normally observed from colliding-wind binaries containing O and/or Wolf–Rayet stars. Even exceptionally X-ray-bright systems such as Cyg OB2\#8A, an O6Ifm $+$ O5.5III binary, reach only $L_{\rm{X}} \sim (1-2)$ $\times 10^{34}\ {\rm erg~s^{-1}}$ \citep{debecker06}, while the X-ray luminosities of WR+O colliding-wind systems are generally in the $10^{33}$--$10^{35}\ {\rm erg~s^{-1}}$ range \citep{zhekov12,rauw16}. The luminosity of the extended component is therefore difficult to reconcile with wind-wind shocks in a stellar binary, even allowing for the exceptionally powerful winds of WR stars. Instead, we suggest that the extended soft X-ray emission is associated with shocks produced by the outflow from X-2, either its jet or its fast accretion-driven wind, interacting with the surrounding nebular material.
X-ray-bright, thermal hot spots with similar luminosity and temperature are observed at the extremities of the jet in the NGC\,7793 S26 ULX bubble \citep{pakull10,soria10}. An unsolved problem to be addressed in further work (after higher resolution radio maps become available) is whether all the radio emission comes from the core or there is also a (sub-arcsec) resolved jet or hot spot in the direction of the region with thermal X-ray emission.

We showed that FORS1/FORS2 spectra capture at least three distinct sources of line emission: i) broad emission components with P-Cyg signatures, from a fast outflow, centred on X-2; ii) narrow, variable emission lines, also centred on X-2, probably from X-ray photoionization; iii) a diffuse, mostly narrow nebular component, slightly offset from X-2 (well visible in the MUSE images and spectra). We interpreted component iii) as the result of at least one candidate Wolf-Rayet star near X-2. Now we want to use the luminosity of the narrow, unresolved He\,{\footnotesize{II}} $\lambda$4686 line (component ii)) as a proxy for the true X-ray luminosity of X-2. We estimate that $L_{\rm{4686,n,X-2}} = (2.0 \pm 0.2) \times 10^{36}$ erg s$^{-1}$, after subtraction of the diluting contribution from the extended nebular emission.
We recall that $\approx$20 per cent of He$^{++}$ recombinations produce a $\lambda4686$ photon; therefore, each $\lambda4686$ photon requires $\approx$5 ionizing photons in the 54--300 eV range \citep{pakull86}. In the absence of any information on the spectral energy distribution of the ionizing photons  in that energy range, a simple assumption of an average energy of $\approx$130 eV per ionizing photon implies an ionizing luminosity $L_{\rm{54-300}}\approx$250 $L_{\rm{4686,n,X-2}} \approx$5 $\times 10^{38}$ erg s$^{-1}$ for the unresolved contribution from X-2 alone; this is a lower limit if X-2 also partly contributes to the extended He$^{++}$ nebula. 
For typical broadband ULX spectra (disk emission plus Comptonised component), this corresponds to 0.3--10 keV luminosities of several times 10$^{39}$ erg s$^{-1}$.

The source of this hard photoionizing flux at the X-2 position remains speculative. The bright optical counterpart, with $T_{\rm{bb}} \approx 4600$ K, cannot produce ionizing photons with energies $>$54 eV (let alone those with energies $>97$ eV, required to explain the [Ne\,{\footnotesize{V}}] lines). The point-like X-ray core has an apparent power-law luminosity of $\approx$8 $\times 10^{37}$ erg s$^{-1}$ (Section 3.4, Tables \ref{chandra}, \ref{chandra_0.3}), corresponding to $\approx$4.6 $\times 10^{36}$ erg s$^{-1}$ if extrapolated to the 54--300 eV band (after removing all intrinsic absorption from the power-law component); this luminosity is two orders of magnitude lower than required. The 0.9-keV thermal plasma component has a 0.3--10 keV luminosity slightly below $10^{37}$ erg s$^{-1}$, corrresponding to a 54--300 eV contribution $\lesssim$3 $\times 10^{36}$ erg s$^{-1}$.
A possible scenario is that we may be looking at the system along a high-inclination line of sight. In that case, the direct power-law component from the X-2 core may be completely obscured to us (a Compton-thick ULX). The power-law X-ray photons observed by us may be only a small fraction of the intrinsic X-ray emission, the fraction scattered or reflected by the outflow into our line of sight. A similar situation happens with the super-Eddington source SS\,433 in our Galaxy, where the observed hard X-ray luminosity ($L_{\rm{X}} \approx 10^{36}$ erg s$^{-1}$) is interpreted as a small fraction of photons reflected off the inner walls of an optically thick wind funnel \citep{middleton21}, while the face-on luminosity would be $>10^{39}$ erg s$^{-1}$. 
This scenario has the advantage of explaining at least qualitatively at the same time the origin of the optical continuum (photosphere of the optically thick outflow along our line of sight), the broad Balmer, He\,{\footnotesize{I}} and He\,{\footnotesize{II}} lines (another signature of the same outflow), and the high-ionization narrow lines, from a gas nebula that sees the direct illumination from EUV and X-ray photons escaping the system along the polar direction (low-density polar funnel).

\begin{table*} 
\caption{Summary of the main observational results for NGC\,5408 X-2 in various bands.} \label{tab:summary} 
\centering 
\setlength{\tabcolsep}{4pt} 
\begin{tabular}{@{} >{\raggedright\arraybackslash}m{2.7cm} >{\raggedright\arraybackslash}m{2.1cm} m{4.5cm} m{8.1cm} @{}} 
\hline\hline\\[-5pt]
Band & Component & Main measurements & Main observational properties \\ 
\hline \\[-5pt]
Radio & 
 & 
$S_{\rm 5\,GHz}\approx1.0$--$4.8$ mJy; $L_{\rm R}\approx1.4$--$7.6\times10^{35}$ erg s$^{-1}$; $\alpha\approx1.1$ 
& Unresolved, steep-spectrum source, with variability by a factor of $\sim5$ over the 2003--2009 observations. The rapid and non-monotonic variability implies a source size $\lesssim1$ pc. The radio source is coincident with the point-like optical and hard X-ray counterparts of X-2. \\
\ \\[-7pt]
\hline \\[-5pt]
Optical continuum & 
 & 
$M_{V,0}\approx-9.1$; $M_{I,0}\approx-10.6$; $T_{\rm bb}\approx4600$ K; $R_{\rm bb}\approx1700\,R_{\odot}$; $L_{\rm bb}\approx4.4\times10^{39}$ erg s$^{-1}$ 
& Unresolved red optical counterpart. The 2009--2010 optical/near-IR spectral energy distribution is consistent with a cool blackbody. The optical continuum declined by $\approx2$ mag between 2010 and 2024. \\ 
\ \\[-7pt]
\hline \\[-5pt]
\multirow{4.6}{*}{Optical lines} & 
Broad & 
FWHM $\approx1700$ km s$^{-1}$; ${\rm EW}({\rm H}\alpha)\approx1800$ \AA & Broad emission components from all the main permitted lines, including the H\,{\footnotesize{I}} Balmer series, He\,{\footnotesize{I}}, He\,{\footnotesize{II}}. P-Cygni profiles in the Balmer lines. Weak broad-line emission persisted in 2024 but had declined strongly relative to the earlier observations. \\ 
\ \\[-7pt]
& Narrow & 
$L_{4686}\approx2\times10^{36}$ erg s$^{-1}$; He\,{\footnotesize{II}}$\lambda$4686/H$\beta$ $\approx 0.15$ 
& Strong emission from He\,{\footnotesize{II}}$\lambda$4686 and [Ne\,{\footnotesize{V}}$]\lambda$3426 requires X-ray photo-ionization with $L_{\rm X}\sim10^{40}$ erg s$^{-1}$. Strong decline from 2012 to 2024.\\ 
\ \\[-7pt]
\hline \\[-5pt]
\multirow{4.1}{*}{X-rays} & 
Non-thermal & 
$L_{\rm pow}\approx8\times10^{37}$ erg s$^{-1}$;
$\Gamma = 1.6^{+1.1}_{-0.9}$  & 
Point-like hard X-ray source coincident with the radio/optical counterparts to within 0\farcs1. The spectrum is dominated by the power-law component. \\ 
\ \\[-7pt]
& Thermal & 
$L_{\rm apec}\approx8\times10^{36}$ erg s$^{-1}$;
$kT = 0.9^{+0.2}_{-0.2}$ keV  & 
Faint but resolved (diameter $\approx$50 pc) soft X-ray emission, spatially offset north of the core by $\approx$0\farcs7.  \\ 
\ \\[-7pt]
\hline 
\end{tabular} 
\end{table*}

We briefly outline three alternative scenarios where massive outflows can be launched in binary systems, associated with the release of accretion power from a compact object: super-Eddington disk winds, shell ejections from massive stars, and common envelope ejections.
The first scenario is based on an analogy with ultraluminous supersoft sources (ULSs). ULSs are the standard examples of super-Eddington accreting sources in which the observed emission is likely determined by the photosphere of an outflow, launched from near the spherization radius \citep{poutanen07,shen15,shen16,soria16,urquhart16,zhou19,yao19}. In the modelling of ULSs, characteristic blackbody temperatures $\sim$0.5--1 $\times 10^6$ K for the thermalising photosphere are obtained with scaled accretion rates ``at infinity'' $\Dot{m} \equiv \Dot{M}/\Dot{M}_{\rm{Edd}} \sim$ a few 100, for a 10-$M_\odot$ BH; most of this mass flow is ejected in the wind. An extrapolation of ULS outflow models to even cooler and larger photospheres requires scaled accretion rates $\Dot{m} \approx 10,000$, corresponding to physical rates $\Dot{M} \approx 10^{23}$ g s$^{-1}$ $\approx 2 \times 10^{-3} M_\odot$ yr$^{-1}$. This is an exceptionally high mass transfer rate for a stellar donor in a binary system, which may only be achieved for thermal-timescale mass transfer from the most massive stars. For example, assuming a thermal-timescale mass tranfer rate $-\Dot{M_2} \approx 3 \times 10^{-8} (M_2/M_\odot)^{2.8} M_\odot$ yr$^{-1}$ \citep{king99}, a donor star of mass $M_2 \gtrsim 50 M_\odot$ can provide the mass transfer required to generate a super-Eddington outflow with a large, cool photosphere. Such massive stars have radiative envelopes and may avoid the formation of a common envelope with the companion stellar-mass BH even at those mass transfer rates \citep{king99}. In summary, in this scenario the BH accretion disk powers the massive outflow, the radio jet, and the X-ray emission.

The second scenario applies to very strong stellar wind outflows or giant eruptions from LBV stars such as $\eta$ Carinae 
\citep{owocki16} or the supernova impostor SN\,2000ch in NGC\,3432 \citep{pastorello10}, or the even more enigmatic SN\,2010da/NGC\,300 ULX-1 \citep{chene25,beasor26}. In such cases, stellar winds can become optically thick, with an effective photospheric radius $R_{\rm{w}}$ much larger than the hydrostatic stellar radius. A fast wind may be launched from a relatively compact or progressively stripped star, and at the same time, the optical continuum may resemble a yellow/red hypergiant. For a spherically symmetric wind with speed $v_{\rm{w}}$, the photosphere is at $R_{\rm{w}} \approx (3/2)\,(\kappa \dot{M})/(4 \pi v_{\rm{w}})$, where $\kappa$ is the opacity \citep{owocki16}. Substituting plausible values for our scenario, $R_{\rm{w}} \approx 3670\, (\kappa/\kappa_{\rm{es}})\,\dot{M}_{{-2}}\,v_{\rm{w,8}}^{-1} \, R_\odot$, where $\kappa_{\rm{es}} \approx 0.34\, {\mathrm{cm}}^2\,{\mathrm{g}}^{-1}$ is the electron scattering opacity, $\dot{M}_{{-2}}$ is the mass outflow rate in units of $10^{-2} M_\odot$ yr$^{-1}$, and $v_{\rm{w,8}}$ is the wind speed in units of 1000 km s$^{-1}$. Assuming for simplicity $\kappa \approx \kappa_{\rm{es}}$, if $\dot{M}_{{-2}}/v_{\rm{w,8}} \approx 0.5$, we obtain a photospheric radius and
effective temperature $T_{\rm{eff}}$ similar to those inferred for the optical counterpart of X-2. On the other hand, \cite{owocki16} also showed that when $T_{\rm{eff}}$ falls below $\approx$6000 K (hydrogen recombination), the opacity drops ($\kappa \ll \kappa_{\rm{es}}$) and much higher values of $\dot{M}_{{-2}}/v_{\rm{w,8}} \sim 10$--100 are needed to bring $T_{\rm{eff}}$ down to $\approx$ 5000 K as observed from X-2. If the erupting massive star has a stellar-mass BH companion, the compact object will intercept a fraction of the outflow and accrete at super-critical rates. In summary, in this scenario the massive donor star launches the dense outflow, and the BH powers the radio jet and the X-ray flux. The observed duration of the great eruption of $\eta$ Carinae (1844--1858: \citealt{smith11}) may be a relevant analogy for the brightness evolution of X-2.

The third, most speculative possibility is that a common envelope between a convective-envelope red supergiant donor and a BH companion did form, and we are now witnessing the short-lived phase of common envelope evolution and ejection \citep{klencki21}. We know that such a phase must exist in nature, from the gravitational wave detection of binary BH mergers. Without a common envelope phase, the binary separation of two BHs formed from massive stellar progenitors is too large to permit a merger in a Hubble time. Instead, a common envelope dramatically reduces the binary separation; for example, in the simulations by \cite{belczynski16}, an 82-$M_\odot$ core helium-burning supergiant entering common envelope with a 35-$M_\odot$ BH companion will reduce the binary separation from $\approx$3800\,$R_\odot$ to $\approx$44\,$R_\odot$. In addition to orbital decay, the energy required for envelope ejection is thought to come from super-Eddington accretion onto the BH \citep{ricker08,macleod15}. Characteristic timescales for the orbital shrinking and envelope ejection can be as short as $\sim$10$^2$ yr \citep{ricker19}. It is also plausible that the accreting BH launches a jet from inside the envelope, which contributes to its ejection 
\citep{chamandy18,soker19,grichener21,hillel22}. In this scenario, we are witnessing (perhaps for the first time) this brief, highly energetic phase of massive binary evolution.

Finally, we are left with the mystery of the great optical dimming that took place over a few years in the mid-2010s. Further studies are needed to distinguish between various alternative scenarios. The first possibility is a decrease in the accretion rate onto the compact object. This could result for example from mass transfer instabilities from the donor star ({\it{e.g.}}, LBV-like activity, pulsations, changes in the wind structure from the donor, end of an $\eta$ Car-like great eruption), or from a long-period, highly eccentric orbit of the binary, with higher accretion rates near periastron. The second possibility is that a phase of increased accretion power and mass loss rate in the 2000s led to dust formation in the cooling outflow, which then obscured the system in the optical bands. Dust formation in cooling outflows is well documented in many classes of massive stars \citep{smith14}, from hot Wolf-Rayets \citep{lau22} to cool red supergiants \citep{smith09}. Dust obscuration has been invoked as an explanation for the optical dimming of $\eta$ Carinae \citep{smith10,smith11}, of LBVs and SN impostors \citep{kochanek11}, and, recently, of the red supergiant Betelgeuse \citep{dupree22}. Follow-up observations in the mid-IR will be the key to test for dust around X-2. Finally, in the framework of our most speculative scenario, another possibility for the dimming is that a common envelope feeding the high BH activity has been ejected over the last decade, and the system consists now of a BH orbiting a stripped star.

\section{Conclusions} 

We discovered and illustrated several peculiar properties of the highly energetic source NGC\,5408 X-2. It has the most luminous radio and optical counterparts of any X-ray binary or ULX, one of the strongest outflows, and a set of photoionized emission lines that require X-ray irradiation $\approx$10$^{40}$ erg s$^{-1}$ (much higher than what we see along our line of sight). Radio emission, optical continuum, line profile and intensity are all variable on a few years' timescale. We argued that this indicates a system going through a short-lived, non-equilibrium phase.  We speculated that the massive outflow responsible for the optical properties is either a disk wind launched during hyper-Eddington accretion, or a shell ejection from an LBV-type donor star, or (more speculatively), the long-sought phase of the common envelope between a BH and a red supergiant donor, in the process of being ejected. Much work remains to be done, beyond the scope of this paper, to model the variability of the system in two decades of multiband archival data, resolve or constrain the size of the radio emission, and determine the location of the X-ray photoionized gas (origin of high-ionization lines such as [Ne\,{\footnotesize{V}}] $\lambda$3426 and He\,{\footnotesize{II}} $\lambda$4686) with respect to the dense outflow (origin of the broad Balmer and He\,{\footnotesize{I}} lines). Further radio observations are planned, to determine whether there has been a post-2009 radio decline, following the optical decline.

\begin{acknowledgements}

We thank the anonymous referee for insightful and constructive comments that lead to substantial content and structure improvements.
We thank Jin Guo and Andres G\'urpide for insightful discussions about the properties of NGC\,5408 and the comparison between X-1 and X-2 
We also thank Roberta Amato, Pikky Atri, Miroslav Filipovic, Anastasia Kilina, Richard Plotkin, Kovi Rose, Chiara Salvaggio, Rongfeng Shen and Ryan Urquhart for comments and suggestions on ULXs and their radio/optical nebulae.
      This research has made use of data and/or software provided by the High Energy Astrophysics Science Archive Research Center (HEASARC), which is a service of the Astrophysics Science Division at NASA/GSFC.
This research is also based on observations made with the NASA/ESA Hubble Space Telescope obtained from the Space Telescope Science Institute, which is operated by the Association of Universities for Research in Astronomy, Inc., under NASA contract NAS 5-26555. 
      Some of the observations reported in this paper were obtained with the Southern African Large Telescope (SALT) under program(s) 2017-1-MLT-003 and 2017-1-DDT-007; we thank David Buckley for helping us to access the SALT archive. The optical spectroscopic work was based on observations made with ESO Telescopes at the La Silla Paranal Observatory. 
      This work has made use of data from the European Space Agency (ESA) mission
{\it Gaia} (\url{https://www.cosmos.esa.int/gaia}), processed by the {\it Gaia}
Data Processing and Analysis Consortium (DPAC,
\url{https://www.cosmos.esa.int/web/gaia/dpac/consortium}). Funding for the DPAC
has been provided by national institutions, in particular the institutions
participating in the {\it Gaia} Multilateral Agreement. 
Moreover, we used data obtained from the {\it Chandra} Data Archive and the {\it Chandra} Source Catalog, and software provided by the {\it Chandra} X-ray Center (CXC) in the {\sc ciao} application package. 
Optical spectroscopic data analysis made use of the python {\sc{mpdaf}} package \citep{Bacon_2016,Piqueras_2017}, and of QFitsView, developed by Thomas Ott at the Max-Planck-Institut für extraterrestrische Physik (MPE).
We also used the Vizier data archive and the Aladin sky atlas developed and maintained by CDS, Strasbourg Observatory, France.
RS was supported by the INAF grant 1.05.23.04.04.
RS also thanks the School of Physics at the University of Sydney, and the Institute of High Energy Physics (Beijing), for hospitality during part of this work (particular thanks to Feng Hua and Tao Lian).
     
\end{acknowledgements}

   \bibliographystyle{aa} 
   \bibliography{paper} 

\begin{appendix}
\nolinenumbers

\section{Detailed description of the observations}

\subsection{Radio: Australia Telescope Compact Array}
\label{app1}

We used several sets of archival Australia Telescope Compact Array (ATCA) observations in the 6-km array configuration (Table \ref{log_atca}). On 2000 March 31 and 2003 March 2 the data were recorded simultaneously at central frequencies of 4.8 GHz and 8.6 GHz, with a 128-MHz bandwidth at each frequency band. On 2003 December 6--7, NGC\,5408 was observed at central frequencies of 4.8 GHz and 2.4 GHz (128-MHz bandwidth at each frequency); for this observation, we used only the 4.8-GHz data as the lower frequency data did not provide adequate spatial resolution to resolve our target from the central source. On 2004 December 10, it was observed simultaneously at 4.8 GHz and 6.2 GHz (128-MHz bandwidth). Finally, on 2009 August 22, the data were taken with a bandwidth of 2 GHz around central frequencies of 5.5 GHz and 9 GHz following an upgrade to the Compact Array Broadband Backend (CABB) \citep{wilson11}. In most epochs, the primary flux density calibration was PKS B1934$-$638, and the phase calibrator was the nearby source PKS B1349$-$439; however, for the 2004 observations, flux and bandpass calibration used PKS B0823$-$500 and PKS B1320$-$446 as used as the phase calibrator. Results from the 2000 ATCA observations were previously reported by \cite{stevens02}, in the context of starburst regions in NGC\,5408; the 2003 and 2004 ATCA observations and 2003, 2004 and 2005 VLA observations were discussed by \cite{kaaret03,soria06,lang07} with a focus on the X-1 ULX. The 2009 ATCA dataset have so far been unpublished.

We processed the data following standard procedures within the Common Astronomy Software Application ({\sc{casa}}), version 5.1.2 \citep{casa22}. For imaging and to determine the radio centroid of X-2, we used the {\sc{casa}} task {\it{clean}}, and chose a Briggs robust parameter of $-2$ \citep{briggs95}, to maximise spatial resolution. The resulting synthesized beam sizes are listed in Table \ref{log_atca}.
To determine the centroid of X-2, we used the {\sc{casa}} task {\it{imfit}}, fitting a 2D Gaussian with a FWHM fixed to the parameters of the
synthesized beam. To measure flux densities, we imaged and used a Briggs robust parameter $r = 0$,  balancing sensitivity and resolution.

NGC\,5408 has also been observed by the VLA in different configurations (for full details, see \citealt{lang07}). Because of its Southern declination, the source is always very low on the horizon for the VLA with limited visibility. As such, the quality of the various datasets is not always reliable. We assessed the VLA datasets and chose to include results from two on them: the observations from 2003 December 20 (B configuration) and from 2005 January 29 (BnA configuration). We processed the 2003 VLA data with {\sc{casa}}, following the same steps used for the ATCA data analysis. Our VLA results agree with the maps presented by \cite{lang07}, although their study was mostly focused on X-1.

\begin{table}[t]
    \centering
    \caption{Summary of the ATCA 6-km and VLA observations used for this study, and corresponding flux densities of X-2. 
    }
    \begin{tabular}{lcccc}
    \hline\hline\\[-5pt]
        Date & $\nu_{\rm c}$ & Beam Size & PA & Flux Density  \\
         &   (GHz) & & & (mJy)  \\[2pt]
        \hline\\[-7pt]
        \multicolumn{5}{c}{ATCA}\\[2pt]
        \hline\\[-7pt]
        2000-03-31 & 4.8 & 2\farcs53 $\times$ 1\farcs50 & $-$6\fdg1  & $1.98 \pm 0.05$  \\
        2000-03-31 & 8.6 & 1\farcs36 $\times$ 0\farcs78 & $-$4\fdg8 & $0.99 \pm 0.06$  \\
        
        2003-03-02 & 4.8 & 2\farcs53 $\times$ 1\farcs63 & $-$1\fdg8  & $1.43 \pm 0.05$ \\
        2003-03-02 &  8.6 &  1\farcs35 $\times$ 0\farcs88 & $-$1\fdg13 & $0.69\pm 0.06$ \\
        
        2003-12-06 & 4.8  &   2\farcs59 $\times$ 1\farcs60 & 1\fdg23  & $1.0 \pm 0.1$ \\
        
        2003-12-07 & 4.8 &   2\farcs52 $\times$ 1\farcs58 & 0\fdg51   &  $1.05 \pm 0.05$ \\ 
        
        2004-12-10 & 4.8 &  2\farcs72 $\times$ 1\farcs62 & $-$2\fdg80  & $3.91 \pm 0.06$ \\
        2004-12-10 & 6.2 &  2\farcs29 $\times$ 1\farcs32 & $-$3\fdg00  & $3.05 \pm 0.06$ \\
        2009-08-22 & 5.5  &  1\farcs74 $\times$ 1\farcs18 & $-$6\fdg0 & $4.79 \pm 0.02$  \\
        2009-08-22 & 9.0  & $1\farcs11$ $\times$ 0\farcs79 & $-$7\fdg3 & $2.75 \pm 0.02$  \\
        \hline\\[-7pt]
        \multicolumn{5}{c}{VLA}\\[2pt]
        \hline\\[-7pt]
        2003-12-20 & 4.9  &   3\farcs45 $\times$ 0\farcs86  &  $-$1\fdg12  &  $1.13 \pm 0.03$  \\
        2004-02-09 & 4.9  &     5\farcs2 $\times$ 2\farcs97  &  9\fdg55   &  $<2.1$\tablefootmark{a}  \\
        2004-12-20 & 4.9  &    1\farcs09 $\times$ 0\farcs31   &  $-$0\fdg61  &  $2.85 \pm 0.03$  \\
        2005-01-29 & 4.9  & 1\farcs57 $\times$ 0\farcs95  &  9\fdg63 &  $3.10 \pm 0.04$   \\
        \hline
    \end{tabular}
    \tablefoot{ATCA observations from 2000--2004 were taken before the CABB upgrade; those from 2009 are post-CABB.
    \tablefoottext{a}{In this VLA observation, X-2 was detected but not fully resolved from the neighbouring sources, because of the lower spatial resolution.}
        }
    \label{log_atca}
\end{table}

\begin{table*}[t!]
\caption{Log of the {\it HST} observations covering NGC\,5408 X-2, and observed brightness of the source. }
\vspace{-0.4cm}
\label{tab:hst_obslog}
\begin{center}
\begin{tabular}{lcccccc}  
\hline \hline\\[-4pt]
Instrument & Filter & $\lambda_{\mathrm{c}}$  & Obs.~Date & Exp.~Time & Flux Density & $m$\\
 &  & (\AA) & & (s) &  ($10^{-17}$ erg cm$^{-2}$ s$^{-1}$ \AA$^{-1}$) & (Vegamag)\\[2pt]
\hline\\[-7pt]
WFPC2 & F606W & 5997  & 2000-07-04 & 600 & $5.34 \pm 0.27$ & 19.27 $\pm$ 0.05\\
WFPC2 & F814W & 8012  & 2000-07-04 & 600 & $6.50 \pm 0.33$  &  18.10 $\pm$ 0.05\\

WFPC2 & F336W & 3344  & 2009-04-04 & 4400 & 0.94 $\pm$ 0.05  & 21.37 $\pm$ 0.05 \\
WFPC2 & F439W & 4311  & 2009-04-04 & 1800 & 2.71 $\pm$ 0.14  &  20.97 $\pm$ 0.05\\
WFPC2 & F555W & 5439  & 2009-04-04 & 520 &  5.50 $\pm$ 0.27 &  19.55 $\pm$ 0.05\\
WFPC2 & F656N & 6564  & 2009-04-04 & 1800 &  128 $\pm$ 13 &  15.2 $\pm$ 0.1\\
WFPC2 & F814W & 8012  & 2009-04-04 & 520 &  7.25 $\pm$ 0.36 &  17.98 $\pm$ 0.05\\

WFC3-IR  & F105W & 10551 & 2010-05-02 & 298 & 7.74 $\pm$ 0.77  &  17.09 $\pm$ 0.10\\
WFC3-IR  & F160W & 15369 & 2010-05-02 & 498 & 3.20 $\pm$ 0.32 &  16.62 $\pm$ 0.10\\

WFC3-IR  & F105W & 10551 & 2010-05-15 & 298 & 7.74 $\pm$ 0.77  & 17.09  $\pm$ 0.10\\
WFC3-IR  & F160W & 15369 & 2010-05-15 & 498 & 3.29 $\pm$ 0.33  &  16.59 $\pm$ 0.10\\

WFC3-IR  & F105W & 10551 & 2010-09-12 & 298 &  7.37 $\pm$ 0.74 &  17.15 $\pm$ 0.10\\
WFC3-IR  & F160W & 15369 & 2010-09-12 & 498 & 3.01 $\pm$ 0.30  &  16.69 $\pm$ 0.10\\

\hline
\end{tabular}
\end{center}
\end{table*}

\subsection{Optical: New Technology Telescope (1993, 2006)}
\label{app2}

As part of a search for the optical counterpart of NGC\,5408 X-1, two long-slit optical spectra of X-2 were obtained on 1993 May 16 with the 3.6-m New Technology Telescope (NTT) using the EMMI spectrograph in medium-dispersion (DIMD) mode and grism \#6. The exposure time was 1800 s for each observation. The slit width was 1\farcs0, and it was oriented at a position angle of $+90^\circ$ ({\it{i.e.}}, east to west). The spatial dispersion along the slit was 0\farcs35 per pixel. The wavelength range was 6132--6845 \AA, with a linear dispersion of 0.407 \AA\ per pixel. The spectral resolution was $\approx$1550 at 6000\,\AA. Standard {\sc{midas}} procedures were used to correct for bias, flat-field. An Helium + Argon lamp allowed us to calibrate in wavelength the two spectra.

On 2006 May 1--3, NGC\,5408 was observed again with NTT EMMI, with several different slit positions to cover most of the star-forming region \citep{vaneymeren10}. The spectra taken on May 2 with grating \#7 ($3 \times 2700$-s exposures) serendipitously include X-2. That night, the 1\farcs0 slit was oriented at a position angle of $+54^\circ$ (north through east). The spatial dispersion along the slit was 0\farcs35 per pixel. The wavelength range was 5700--7250 \AA, with a linear dispersion of 0.82 \AA\ per pixel. The spectral resolution was $\approx$2600 at 6000\,\AA. The seeing varied between 1\farcs0 and 1\farcs4 during the May 2 series of exposures.
As for the 1993 observations, standard {\sc{midas}} procedures allowed to correct raw data for bias and flat-field, while an Helium + Argon lamp provided the wavelength calibration. Flux calibration was derived from the observation of LTT 6248 on May 3.

\subsection{Optical: Very Large Telescope (2008, 2010, 2011, 2012)}
\label{app3}

The ULX nebula around X-1 was the intended target of Very Large Telescope (VLT) FOcal Reducer and low dispersion Spectrograph 1 (FORS1) spectroscopic observations on 2008 April 10, with the 600B grism (3330--6350 \AA). The delivered seeing was 0\farcs70. Six blocks of 3 $\times$ 850-s exposures \citep{kaaret09} were taken and median-combined. The 1\farcs0 slit was oriented at a position angle of 105\fdg7 (Figure \ref{fig1}); thus, it covered also (on purpose) a bright reference star $\approx$8\arcsec\ to the west of X-1, and (serendipitously) X-2, $\approx$14\arcsec\ to the west of X-1. The CCD pixels were rebinned by 2 along both the wavelength and the spatial direction. After the rebinning, the spectral dispersion was 1.48 \AA\ per pixel; the spatial dispersion was 0\farcs252 per pixel. Wavelength calibration was obtained from a HeHgCd arc lamp. Observations of LTT 3218 and EG 274 standard stars provided the flux calibration.

X-1 and X-2 were observed again from the VLT with the FORS2 spectrograph on 2010 April 12 \citep{cseh11,cseh13}, with the 1200B (3660--5110 \AA) and 1200R (5750--7310 \AA) grisms. The exposure times were 3 $\times$ 850s for the blue spectra, and 3 $\times$ 850s for the red ones; the three sub-exposures for each grism were then combined together. The delivered seeing was 0\farcs51, the best seeing of all our available optical spectra.  As in the 2008 observations, the 1\farcs0 slit was oriented in such a way to cover X-1, the foreground reference star and X-2, at a position angle of 105\fdg2 (Figure \ref{fig1}). The CCD pixels were rebinned by 2 along both the wavelength and the spatial direction. The rebinned spectral dispersion was 0.72 \AA\ and 0.74 \AA\ per pixel for the blue and red spectrum, respectively; the rebinned spatial dispersion was 0\farcs252 per pixel. The spectral resolution $\lambda/\Delta \lambda$ was $\approx$1420 and $\approx$2140 at the blue and red central wavelengths, respectively. This corresponds to a resolution of $\approx$3.1 \AA\ $\approx$190 km s$^{-1}$ FWHM at H$\beta$ and $\approx$3.0 \AA\ $\approx$140 km s$^{-1}$ FWHM at H$\alpha$. The standard star LTT7379 was used for flux calibration. The HgCdHeNeA lamp provided the wavelength  calibration.

A third cycle of VLT FORS2 spectroscopic observations of X-1 and X-2 was carried out in 2011. Specifically, on April 3 ($3 \times 850$-s exposures), April 11 ($3 \times 850$ s), May 9 ($3 \times 850$ s), June 26 ($3 \times 850$ s), July 1 ($3 \times 850$ s) and July 2 ($1 \times 420$ s plus $3 \times 850$ s). All observations used the 1200B grism, covering the 3615--5065 \AA\ interval. The slit was 0\farcs7 wide, approximately matching the delivered seeing, which varied between 0\farcs6 and 0\farcs7 during the observations. The position angle of the slit was between 105\fdg2 and 105\fdg7 throughout the observations. After $2 \times 2$ binning, the spectral dispersion was 0.71 \AA\ per pixel, and the spatial dispersion 0\farcs252 per pixel. A HeHgCdNe lamp provided the wavelength calibration. The standard star LTT7379 provided the flux calibration.

The fourth and last cycle of FORS2 observations was carried out in 2012: on April 15, April 22, April 23, May 18, May 20. Exposure times were $3 \times 880$ s for each of those five observing blocks. The seeing was 0\farcs7, 0\farcs5, 0\farcs8, 0\farcs7 and 0\farcs7, respectively. For all observations, the 1200B grism was used, covering the 3615--5065 \AA\ interval. The slit was 0\farcs7 wide, at a position angle of 105\fdg2. Binning, spectral dispersion and spatial dispersion were the same as in 2011. As for other FORS2 observations, a HeHgCdNe lamp provided the wavelength calibration and the flux calibration was derived from observations of LTT 7987.

All FORS1 and FORS2 data were corrected for bias, flat-field and calibrated in wavelength and flux using standard {\sc{midas}} procedures.
In all observations, X-2 was not perfectly centred in the middle of the slit; this causes slit losses, different from epoch to epoch because of the different slit width, seeing and position angle. We estimate that slit losses in the 2010 FORS2 data were minimal, because in this epoch we had the largest slit (1\farcs0), the best seeing (0\farcs51), and the most favourable position angle. This estimate is supported by a comparison of the continuum flux from the 2010 spectrum with the 2009 Hubble Space Telescope {\it{HST}} fluxes (Appendix~\ref{app6} and Section 3.2). At other epochs, {\it{e.g.}} in 2012, at least half of the light of X-2 fell off the slit. For this reason, we cannot do a time variability analysis based on the VLT data; however, our study of the profile and equivalent width (EW) of the spectral lines is unaffected.

\subsection{Optical: South African Large Telescope (2017)}
\label{app4}

From the Southern African Large Telescope’s (SALT) archive, we extracted flat- and bias-corrected data, reduced with the SALT science pipeline\footnote{http://pysalt.salt.ac.za} \citep{Crawford_2010}. X-2 was observed twice in 2017 with the Robert Stobie Spectrograph (RSS). Runs took place on May 19 (proposal ID 2017-1-MLT-003) and on August 18 (proposal ID 2017-1-DDT-007). A total of 720\,s and 420\,s long low resolution spectra were obtained in May 19 and August 18 respectively. High resolution spectra were acquired during the last night with an exposure time of 1080\,s. In all cases, a slit width of 1\farcs25 was used with the same orientation on the sky (position angle $= 161^{\circ}$). The PG0900 low resolution grating covers a 3200--9000 \AA\ range with a resolving power of $\approx$1200 at 6000 \AA\ while the PG2300 high resolution grating spans the 3800--7000 \AA\ range with a resolving power of $\approx$4000 at 6000 \AA. Xe and Cu+Ar arc lamps provided the wavelength calibration, carried out with standard {\sc{midas}} procedures.

   \begin{figure*}[t!]
   \centering
   \includegraphics[width=\hsize]{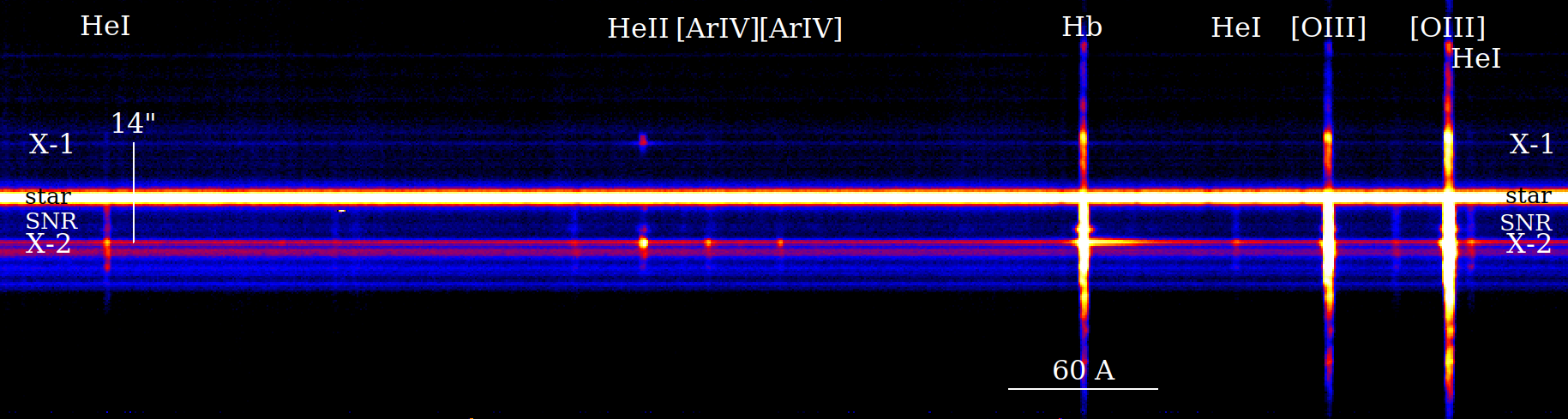}\\
   \includegraphics[width=\hsize]{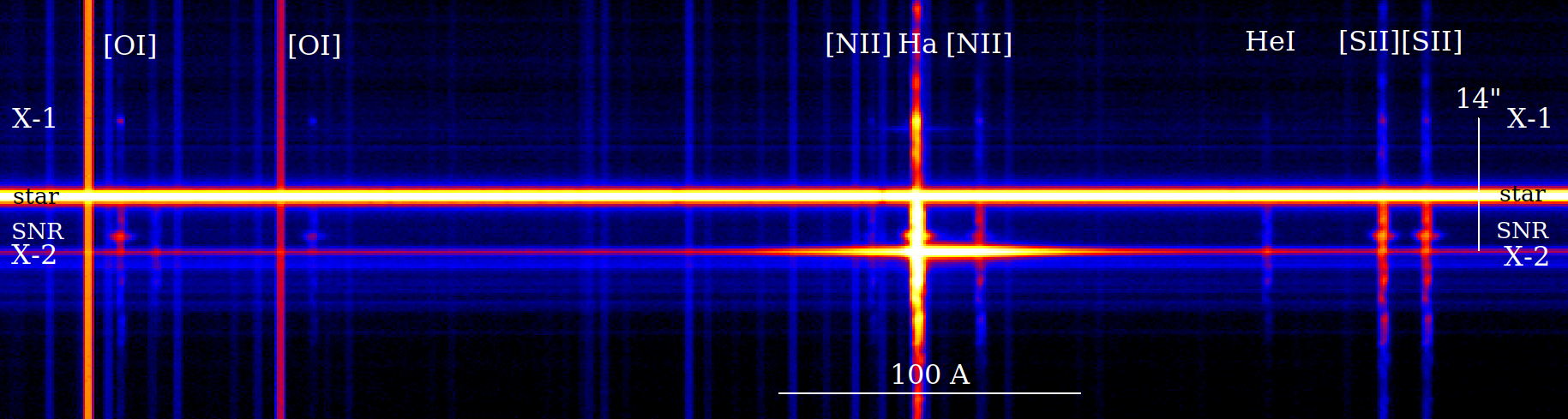}
      \caption{Top panel: portion of the VLT/FORS2 2D spectrum across NGC\,5408, taken on 2010 April 12 with the 1200B grism, slit width of 1\farcs0, exposure time of 850\,s. A 2 $\times$ 2 binning was applied to both the spatial and wavelength direction. The X axis is the wavelength direction, with a spectral dispersion of 0.72 \AA\ per pixel. The spectral resolution $\lambda/\Delta \lambda$ is $\approx$1420, which corresponds to a resolution of $\approx$3.1 \AA\ $\approx$190 km s$^{-1}$ FWHM at H$\beta$. The Y axis represents the spatial direction (position angle $= 105^{\circ}$; slit position shown in Figure 1), with a dispersion of 0\farcs25 per pixel. Both X-1 and X-2 were on the slit; X-2 stands out for its broad H$\beta$ emission (as opposed to the narrow, nebular H$\beta$ emission of X-1), and for much stronger emission from the high-ionization lines He\,{\footnotesize{II}} $\lambda$4686 and [Ar\,{\footnotesize{IV}}] $\lambda\lambda$4711,4740.
      Bottom panel: as in the top panel, for the red section of the spectrum, taken with the 1200R grism. The spectral dispersion is 0.74 \AA\ per pixel. The spectral resolution is $\approx$2140, which corresponds to a resolution of $\approx$3.0 \AA\ $\approx$140 km s$^{-1}$ FWHM at H$\alpha$. Forbidden lines from [O\,{\footnotesize{I}}], [N\,{\footnotesize{II}}], [S\,{\footnotesize{II}}] are narrow (nebular emission) but H$\alpha$ is exceptionally broad (full width at zero intensity of $\approx$6000 km s$^{-1}$), which suggests an origin from a fast, dense outflow.
      }
         \label{vlt_2d}
   \end{figure*}

\subsection{Optical: VLT Multi Unit Spectroscopic Explorer (2024)}
\label{app5}

The Multi Unit Spectroscopic Explorer (MUSE, \citealt{Bacon_2010}) observed the center of the NGC\,5408 galaxy as part of ESO program 0113.C-2093(A). ESO-VLT-U4 observations took place on 2024 May 1, June 22 and June 27. MUSE was in Wide Field Mode, {\it{i.e}}, with a pixel size of 0\farcs2 and a field of view of $\approx$1\arcmin$\times$1\arcmin, covering both NGC\,5408 X-1 and X-2. Spectra were obtained in extended mode, {\it{i.e}}, without the filter blocking possible second order contamination in the red, but offering a full wavelength interval of 4600 to 9350 \AA, which includes the He\,{\footnotesize{II}} $\lambda$4686 high excitation line. The resolving power ranges from $\approx$1700 in the blue up to $\approx$3500 at longer wavelengths. MUSE pipeline-reduced data \citep{Weilbacher_2020} were retrieved from the ESO archive database.

The best seeing (0\farcs65) was obtained during the short (560s) 2024 May 1 run. The June 27 run has a lower sky background and a longer exposure time (1680s), but the seeing was worse (0\farcs97). Instead, the June 22 observation was discarded because the sky background level was too high (bright time, one day after full moon). The two datasets retained for analysis were obtained at a relatively high airmass (1.7 and 1.9). We checked the MUSE reduction pipeline built-in correction for differential refraction \citep{Weilbacher_2020} by comparing stellar positions in blue and red images, which appear consistent within 0\farcs05. Further analysis of comparison stars in the May 1 and June 27 images suggests that the latter dataset was not obtained in photometric conditions; observed fluxes are systematically lower by $\approx$28 per cent in the latter images, which we attribute to thin clouds affecting the observation. 

Continuum and emission lines flux maps as well as spectra were computed with the {\sc{mpdaf}} python package \citep{Bacon_2016,Piqueras_2017}. We built emission line maps by removing a mean continuum map from the map made in the selected wavelength range. The continuum maps are averages of small intervals of $\approx$ 50\,\AA\ at shorter and longer wavelengths, excluding other emission or absorption features as seen in the mean field spectrum.

\subsection{Optical: Hubble Space Telescope}
\label{app6}

The Hubble Space Telescope ({\it HST}) observed NGC\,5408 several times in 2000 and 2009 with the Wide Field Planetary Camera 2 (WFPC2), and in 2010 with the Wide Field Camera 3 (WFC3). In most cases, X-1 was the target and X-2 was sometimes outside the field of view. This is unfortunately the case for all WFC3 Ultraviolet and Visible (UVIS) channel observations from 2010, which were taken with a 512 $\times$ 512 pixel subarray around X-1; instead, all the WFC3 Infrared (IR) images were in full frame and did include X-2. In the WFPC2 observations, X-2 was located in the WFC3 chip in 2000 and in the WF4 chip in 2009. We list in Table \ref{tab:hst_obslog} all the {\it HST} observations that cover X-2.

We downloaded the data from the Mikulski Archive for Space Telescopes (MAST). We checked and, when necessary, improved the astrometric alignment of every image, based on the {\it{Gaia}} catalogue. This is straightforward because there are about 50 catalogued {\it{Gaia}} sources within 1\arcmin\ of X-2, that appear point-like and not saturated in the WFPC2 and WFC3 images. In particular, we found that the WFC3-IR images as downloaded from MAST are already well aligned onto the {\it{Gaia}} frame, while the 2000 and 2009 WFPC2 images needed to be systematically shifted by $\approx$0\farcs5. After the re-alignment, the {\it HST} positions coincide with the {\it Gaia} positions within an uncertainty (1-$\sigma$ scatter) of $\approx$0\farcs1. X-2 itself is a {\it{Gaia}} source, at $\alpha_{\rm J2000} = 14^{\rm h} 03^{\rm m} 18\fs424$, $\delta_{\rm J2000} = -41\degr 22\arcmin 55\farcs23$.

We measured the brightness of X-2, doing aperture photometry with SAOImage's {\sc{ds9}} package, version 8.6. For the WFPC2 images, we used a source extraction radius of 0\farcs2, corresponding to an encircled energy fraction of 0.80--0.81, for F336W and F439W (where X-2 is fainter and more contaminated by nearby sources) and 0\farcs3 for the other filters, corresponding to an encircled energy fraction of 0.88--0.90  \citep{holtzman95a}; the local background was estimated from the annulus between 0\farcs4 and 0\farcs7. For the WFC3-IR images, we used a source extraction radius of 0\farcs39, corresponding to an encircled energy fraction of 0.85 for F105W and 0.83 for F160W\footnote{{\url{https://www.stsci.edu/hst/instrumentation/wfc3/data-analysis/photometric-calibration/ir-encircled-energy}}}, while the background was taken from the annulus between 0\farcs8 and 1\farcs3. 

As an independent check of the aperture photometry, we performed point spread function (PSF) fitting photometry on the drizzled {\it{HST}} images. For each filter, a region centred on the source was fitted with a two-dimensional circular Moffat PSF plus a constant local background. The PSF normalization and width, together with a small adjustment of the source centroid, were optimized by nonlinear least-squares fitting; for the lower signal-to-noise sources, the Moffat shape parameter was fixed to the value determined from the higher signal-to-noise images. The net count rates derived from PSF photometry are perfectly consistent with the aperture photometry values, for all bands.

To convert from count rates to fluxes and magnitudes, for the WFPC2 images we used the PHOTFLAM values in the fits headers (essentially identical to the values in \citealt{holtzman95b} and \citealt{casertano97}). While the mosaiced, resampled, drizzled images ($\ast\_$drz.fits) from the MAST archive provide the best image quality and astrometry, we also used the flat-fielded images ($\ast\_$flt.fits) of the individual chips (WF3 and WF4) to correct for the charge transfer efficiency losses, which are a function of chip location, number of source and background counts, and epoch of observation \citep{dolphin09}. For the WFC3-IR images, we applied the PHOTFLAM values in the fits headers, without further corrections.

   \begin{figure*}[t!]
   \centering
   \includegraphics[width=0.49\hsize]{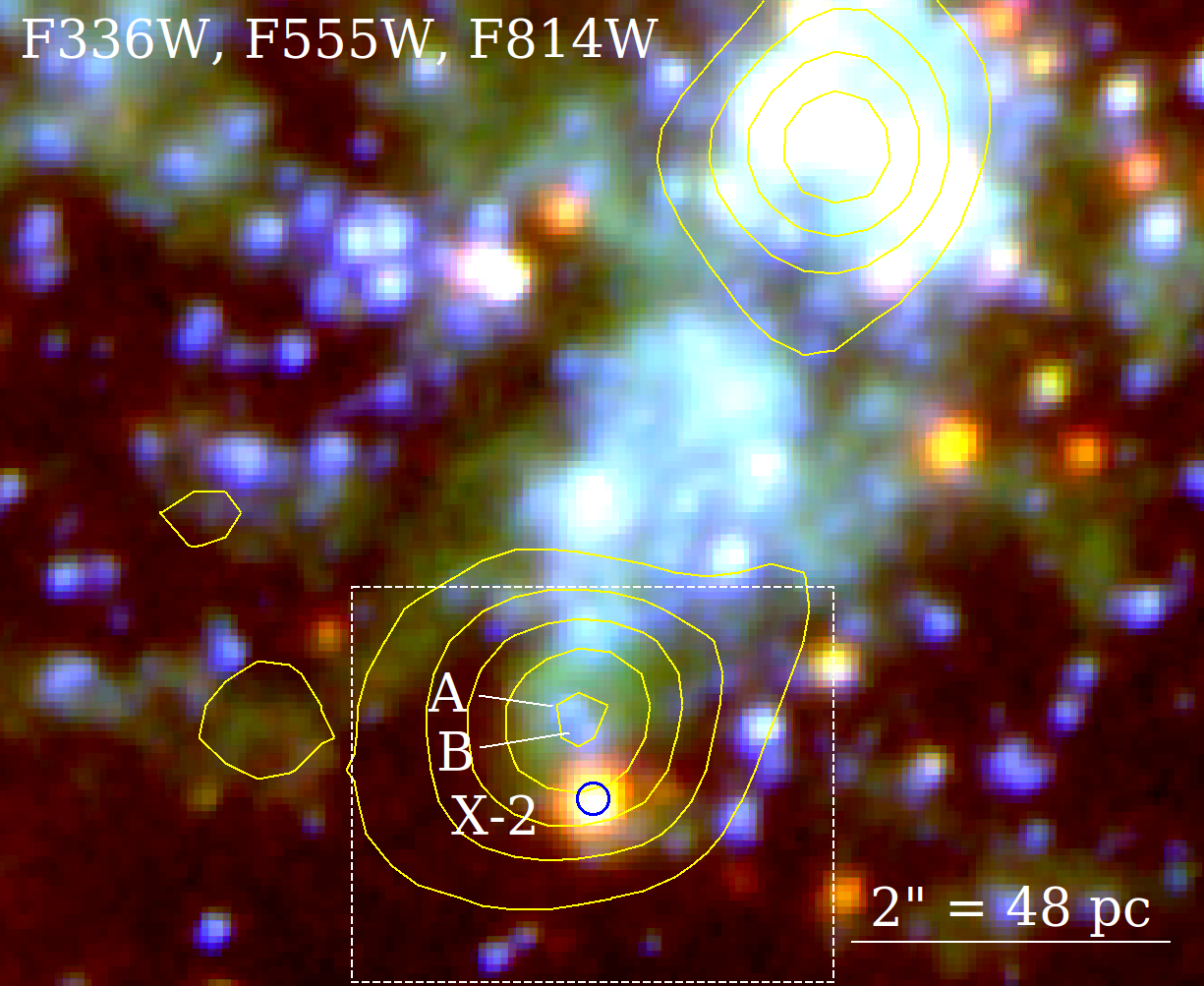}
   \includegraphics[width=0.49\hsize]{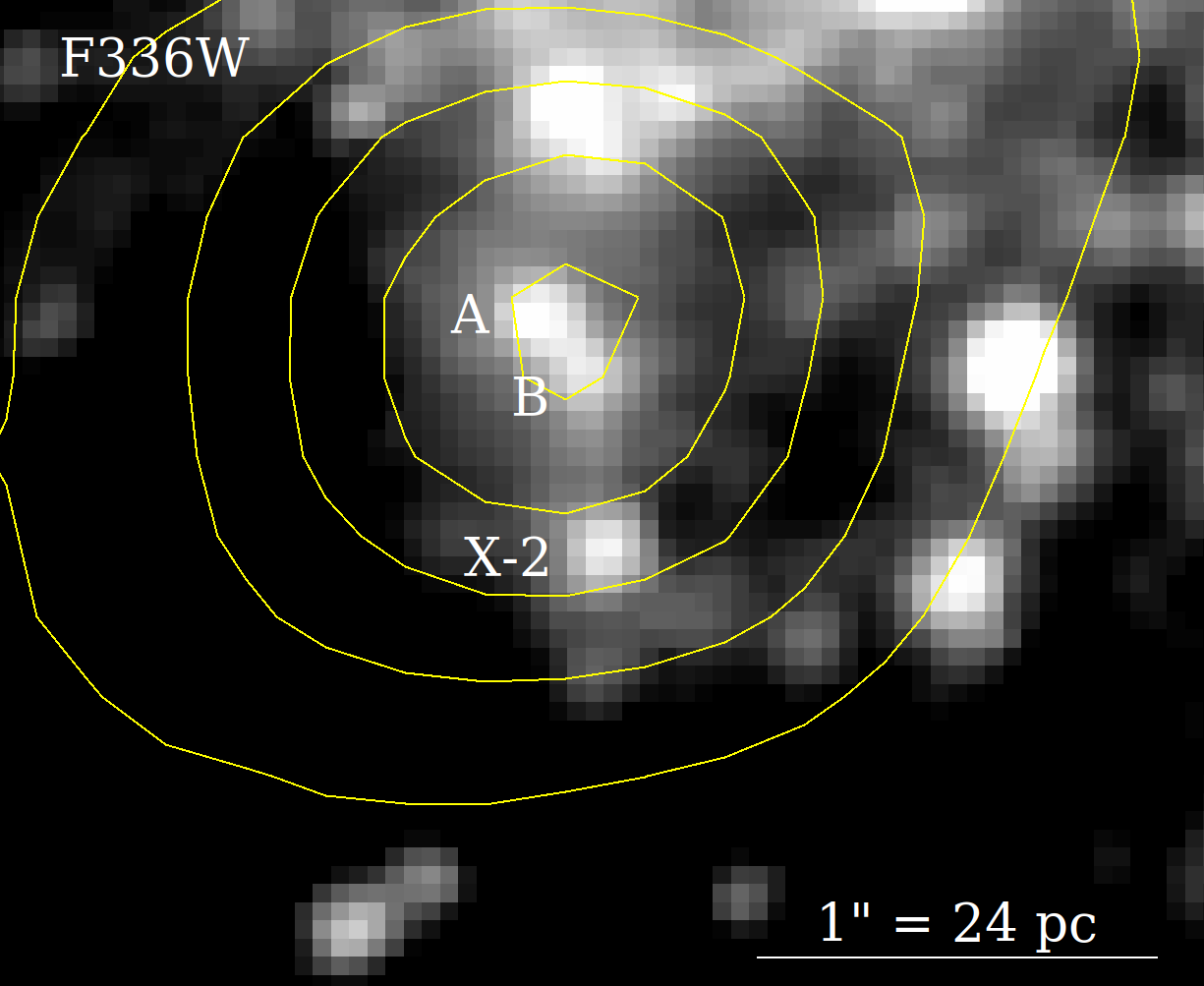}
      \caption{Left panel: comparison between optical continuum and He\,{\footnotesize{II}} $\lambda$4686 emission. The true-colour optical image is from {\it{HST}}/WFPC2 (red = F814W, green = F555W, blue = F336W). The yellow contours (linear scale with arbitrary zeropoint) are from the VLT/MUSE He\,{\footnotesize{II}} $\lambda$4686 emission map (combined dataset from 2024 May 1 and June 27). The blue circle (coincident with the optical counterpart of X-2) marks the ATCA radio position. The blue stars A and B are plausible candidates for the ionizing flux responsible for the  He\,{\footnotesize{II}} $\lambda$4686 emission. The F336W image of the dashed white box is shown zoomed-in on the right. 
      Right panel: {\it{HST}}/WFPC2 zoomed-in view of the surroundings of X-2, in the F336W filter. Yellow contours are the He\,{\footnotesize{II}} $\lambda$4686 emission map. Stars A and B are the same as in the left panel.
      }
         \label{hst_col_4686}
   \end{figure*}

\begin{table}[t!]
\caption{Summary of the {\it Chandra} observations.}
\vspace{-0.4cm}
\begin{center}
\begin{tabular}{lccc}  
\hline \hline\\[-4pt]    
ObsID & Instrument & Obs.~Date &  Exp.~Time\\ 
&&&(ks)  \\[2pt]
\hline  \\[-7pt]
2885 & ACIS-S & 2002-05-07 & 4.7 \\ 
4555 & ACIS-S & 2003-12-20 & 4.7\\ 
4556 & ACIS-S & 2004-02-09 & 4.5\\
4557 & ACIS-S & 2004-12-20 & 4.7 \\ 
4558 & ACIS-S & 2005-01-29 & 4.7\\ 
11032 & ACIS-S & 2010-05-02 & 11.0\\
11033 & ACIS-S & 2010-05-15 & 11.1\\
11034 & ACIS-S & 2010-05-28 & 10.9\\
13018 & ACIS-S & 2010-09-12 & 10.9\\
\hline
\end{tabular} 
\label{log_chandra}
\end{center}
\end{table}

\subsection{X-rays: Chandra}
\label{app7}

NGC\,5408 was observed by {\it Chandra}/ACIS-S on eight epochs between 2002 and 2010, for a total of 67 ks (Table \ref{log_chandra}). Roll angles varied from epoch to epoch, but X-2 was always located within $\approx$30\arcsec of the aimpoint on the S3 chip.

After downloading the data from the public archives, we reprocessed and analysed them with the Chandra Interactive Analysis of Observations ({\sc ciao}) software version 4.15 \citep{fruscione06}, Calibration Database 4.12.0. We created new level-2 event files with the {\sc ciao} task {\it chandra\_repro}, and re-aligned and stacked images with {\it reproject\_obs}. We then used {\it specextract} to create spectra and associated response and ancillary response files (combined for all datasets); we set the {\it specextract} parameters ''weight = no'' and ``correctpsf = yes'', suitable to point-like sources. We selected local background regions at least 4 times the size of the source regions.

A key aspect of this work is the association between the optical and the X-ray emission of X-2. For this, we need accurate astrometry and also the best possible spatial resolution. With the default astrometry of the ACIS-S event files from the archives, the 90\% uncertainty radius of each observation is $\approx$0\farcs9\footnote{\url{https://cxc.harvard.edu/cal/ASPECT/celmon/}}. By taking the average of eight observations, we expect that the centroid of a point-like source in the stacked image has an uncertainty of $\approx$0\farcs3. To further reduce this uncertainty, we looked for associations between {\it Chandra} and {\it HST} sources. 
We found three {\it Chandra} point-like sources with an unambiguous point-like counterpart in WFC3-IR (not including the optical star at the position of X-2), at the outskirts of NGC\,5408. Their optical/X-ray alignment is good within $\approx$0\farcs2, for all three sources.

To increase the spatial resolution, we applied the sub-pixel binning {\it edser} algorithm in {\sc ciao} to the spatial region around X-2. This is possible because X-2 is sufficiently close to the aimpoint, so that the PSF of point sources is smaller than the pixel size (radius of 50\% encircled-energy $\approx$0\farcs4). 
For spectral analysis, we defined source and background regions with {\sc{ds9}}, and built spectra and associated response and ancillary response files with the standard {\sc ciao} task {\it{specextract}}. We rebinned the data to 1 count per bin, with the {\sc ftools} \citep{blackburn95} task {\it grppha}. We modelled the spectra with xspec \citep{arnaud96} version 12.14.1, fitting them with the Cash statistics \citep{cash79} because of the low number of counts.

   \begin{figure*}[t!]
   \centering
   \includegraphics[width=0.49\hsize]{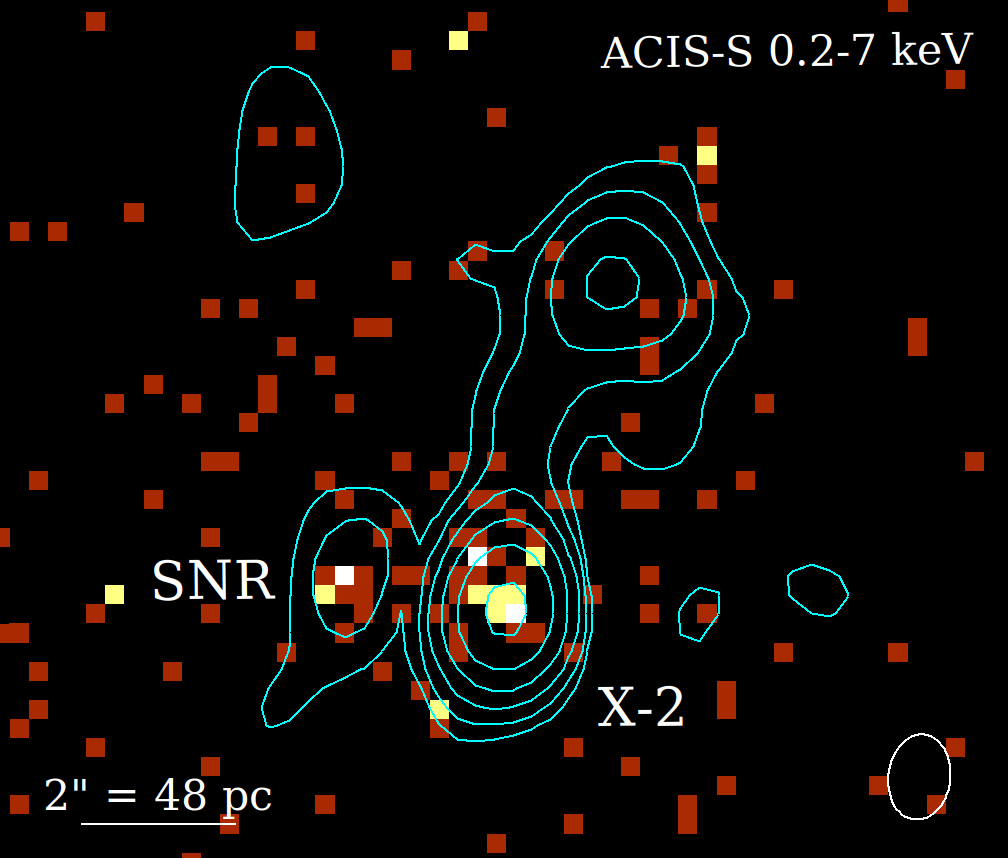}
   \includegraphics[width=0.49\hsize]{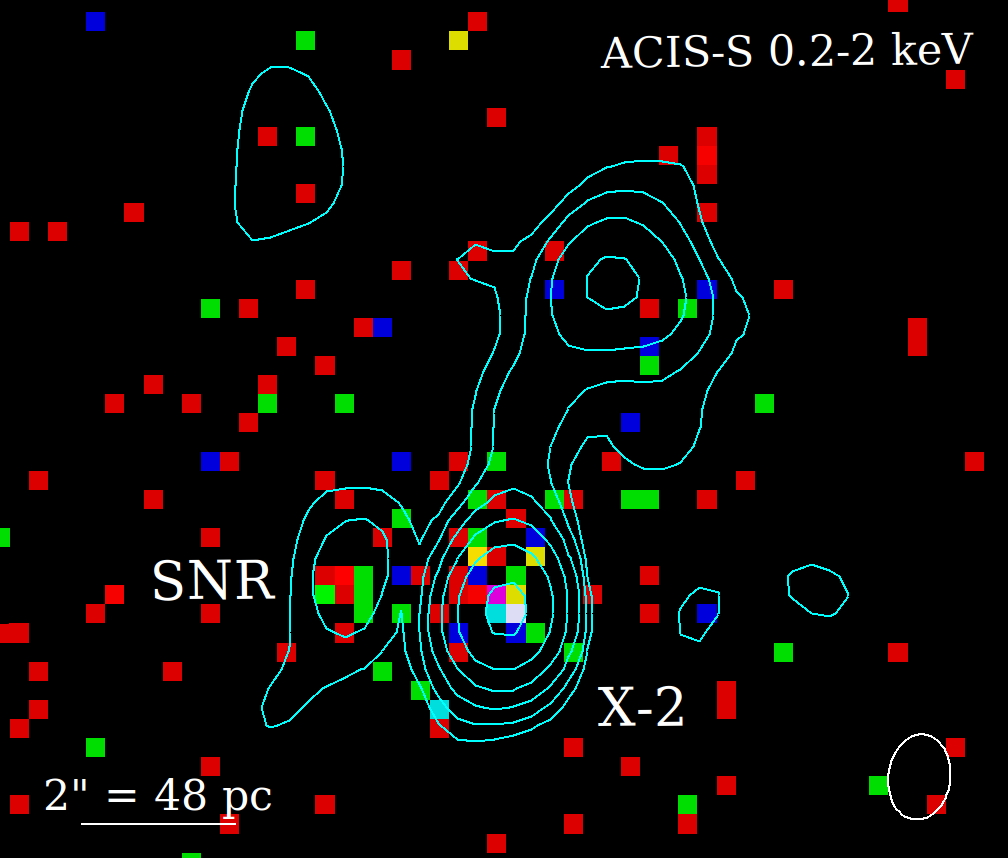}\\
      \caption{Comparison between X-ray and radio emission.  Left: combined {\it{Chandra}}/ACIS-S image in the 0.2--7.0 keV band; the pixel size is $\approx$0\farcs25 (binning factor of 0.5). Right: same as in the top left, but colour-coded: red corresponds to 0.2--1.2 keV, green to 1.2--2.4 keV, blue to 2.4--7.0 keV. 
      }
         \label{chandra1}
   \end{figure*}

\begin{table}[t]
    \centering
    \caption{Radio spectral index in the four ATCA observations that were taken simultaneously in two bands, centred at frequencies $\nu_1$ and $\nu_2$.
    }
    \begin{tabular}{lcc}
    \hline\hline\\[-5pt]
        Date & $\nu_1/\nu_2$ & $\alpha$    \\[2pt]
        \hline\\[-7pt]
        2000-03-31 & 4.8/8.6   & $1.19 \pm 0.11$  \\
        2003-03-02 & 4.8/8.6  & $1.25 \pm 0.16$ \\
        2004-12-10 & 4.8/6.2 &  $0.97 \pm 0.10$ \\
        2009-08-22 & 5.5/9.0   & $1.13 \pm 0.02$   \\
        \hline
    \end{tabular}
    \label{alpha_atca}
\end{table}

\begin{table}
\caption{Best-fitting blackbody parameters for the optical counterpart of X-2.
}
\vspace{-0.4cm}
\begin{center}
\begin{tabular}{lr}  
\hline \hline\\[-5pt]    
Parameter & Value \\ [2pt]
\hline  \\[-7pt]
$E(B-V)$ & 0.36$^{+0.03}_{-0.03}$ mag \\[4pt] 
$T_{\mathrm{bb}}$ & 4585$^{+750}_{-550}$ K\\[4pt]  
$L_{\mathrm{bb}}$ & $4.39^{+0.17}_{-0.17} \times 10^{39}$ erg s$^{-1}$\\[4pt] 
 & $\approx$1.15$^{+0.05}_{-0.05} \times 10^{6} L_\odot$\\[4pt] 
$R_{\mathrm{bb}}$ & $1185^{+25}_{-25} \times 10^{11}$ cm \\[4pt]  
 & $\approx$1693$^{+33}_{-33}$ $R_\odot$\\[4pt] 
\hline
\end{tabular} 
\tablefoot{Fit based on the broad-band {\it HST} observations of 2009--2010. Errors are 90\% confidence limits. The best-fitting total reddening includes a line-of-sight Galactic component $E(B-V) \approx 0.06$ mag.}
\label{hst_sed}
\end{center}
\end{table}

\begin{table}[t!]
\caption{Best-fitting parameters for the combined {\it Chandra} spectrum of X-2 (point-like core). 
}
\label{chandra}
\centering
\begin{tabular}{lc}
\hline\hline\\[-5pt]
Parameter & Value \\[2pt]
\hline\\[-7pt]
$N_{\rm H,Gal}$ & $[0.53 \times 10^{21}]$ cm$^{-2}$ \\
$kT_{\rm apec}$ & $0.95^{+0.31}_{-0.32}$ keV \\
$Z_{\rm apec}/Z_\odot$ & [1] \\
$N_{\rm apec}$\tablefootmark{a} & $\left(7.0^{+5.7}_{-5.7}\right) \times 10^{-7}$ \\
$N_{\rm H,int}$ & $\left(9.4^{+19.0}_{-\ast}\right) \times 10^{21}$ cm$^{-2}$ \\
$\Gamma$ & $1.25^{+1.23}_{-1.02}$ \\
$N_{\rm pow}$\tablefootmark{b} & $\left(2.6^{+12.8}_{-2.0}\right) \times 10^{-6}$ \\[3pt]
\hline\\[-7pt]
C-stat/d.o.f. & 52.9 / 68 \\
\hline\\[-7pt]
$F_{\rm tot}$ & $\left(2.3^{+1.4}_{-0.7}\right) \times 10^{-14}$ erg cm$^{-2}$ s$^{-1}$ \\
$F_{\rm apec}$ & $\left(0.2^{+0.1}_{-0.1}\right) \times 10^{-14}$ erg cm$^{-2}$ s$^{-1}$ \\
$F_{\rm pow}$ & $\left(2.1^{+1.2}_{-0.6}\right) \times 10^{-14}$ erg cm$^{-2}$ s$^{-1}$ \\
$L_{\rm tot}$ & $\left(8.8^{+12.8}_{-2.3}\right) \times 10^{37}$ erg s$^{-1}$ \\
$L_{\rm apec}$ & $\left(0.6^{+0.3}_{-0.3}\right) \times 10^{37}$ erg s$^{-1}$ \\
$L_{\rm pow}$ & $\left(8.2^{+13.1}_{-2.6}\right) \times 10^{37}$ erg s$^{-1}$ \\[3pt]
\hline
\end{tabular}
\tablefoot{Spectrum extracted within a source region of 1\farcs0 radius, fitted with \texttt{tbabs}$_{\mathrm{Gal}} \times$ (\texttt{apec} $+$ \texttt{tbabs}$_{\mathrm{int}} \times$ \texttt{pow}). Absorbed fluxes and unabsorbed luminosities are in the 0.3--10 keV band. Metallicity is fixed at the solar value. Errors are given at the 90\% confidence level for one interesting parameter.
\tablefoottext{a}{Normalization of the \texttt{apec} component, defined in {\sc{xspec}} as $\frac{10^{-14}}{4\pi [D_{\rm A}(1+z)]^2} \int n_{\rm e} n_{\rm H} \, dV$, where $D_A$ is the angular diameter distance to the source (cm), $dV$ is the volume element (cm$^3$), and $n_e$ and $n_H$ are the electron and H densities (cm$^{-3}$).}
\tablefoottext{b}{Units of photons keV$^{-1}$ cm$^{-2}$ s$^{-1}$ at 1 keV.}
}

\end{table}

\begin{table}[t!]
\caption{Best-fitting parameters for the combined {\it Chandra} spectrum of X-2 (point-like core plus slightly extended region).
}
\label{chandra_0.3}
\centering
\begin{tabular}{lc}
\hline\hline\\[-5pt]
Parameter & Value \\[2pt]
\hline\\[-7pt]
$N_{\rm H,Gal}$ & $[0.53 \times 10^{21}]$ cm$^{-2}$ \\
$kT_{\rm apec}$ & $0.91^{+0.18}_{-0.23}$ keV \\
$Z_{\rm apec}/Z_\odot$ & [0.3] \\
$N_{\rm apec}$\tablefootmark{a} & $\left(2.3^{+1.2}_{-1.5}\right) \times 10^{-6}$ \\
$N_{\rm H,int}$ & $\left(9.1^{+14.7}_{-8.6}\right) \times 10^{21}$ cm$^{-2}$ \\
$\Gamma$ & $1.57^{+1.08}_{-0.86}$ \\
$N_{\rm pow}$\tablefootmark{b} & $\left(3.8^{+13.4}_{-2.6}\right) \times 10^{-6}$ \\[3pt]
\hline\\[-7pt]
C-stat/d.o.f. & 67.9 / 84 \\
\hline\\[-7pt]
$F_{\rm tot}$ & $\left(2.2^{+0.9}_{-0.6}\right) \times 10^{-14}$ erg cm$^{-2}$ s$^{-1}$ \\
$F_{\rm apec}$ & $\left(0.23^{+0.12}_{-0.10}\right) \times 10^{-14}$ erg cm$^{-2}$ s$^{-1}$ \\
$F_{\rm pow}$ & $\left(1.9^{+0.8}_{-0.5}\right) \times 10^{-14}$ erg cm$^{-2}$ s$^{-1}$ \\
$L_{\rm tot}$ & $\left(9.3^{+14.9}_{-2.3}\right) \times 10^{37}$ erg s$^{-1}$ \\
$L_{\rm apec}$ & $\left(0.8^{+0.4}_{-0.5}\right) \times 10^{37}$ erg s$^{-1}$ \\
$L_{\rm pow}$ & $\left(8.4^{+15.5}_{-2.6}\right) \times 10^{37}$ erg s$^{-1}$ \\[3pt]
\hline
\end{tabular}
\tablefoot{Spectrum extracted from an ellipse with semi-major axes of 1\farcs25 and 1\farcs75 (position angle $= 0$), fitted with \texttt{tbabs}$_{\mathrm{Gal}} \times$ (\texttt{apec} $+$ \texttt{tbabs}$_{\mathrm{int}} \times$ \texttt{pow}). Absorbed fluxes and unabsorbed luminosities are in the 0.3--10 keV band. Metallicity is fixed at 0.3 $Z_\odot$. Errors are given at the 90\% confidence level for one interesting parameter. Normalizations are defined as in Table \ref{chandra}.}
\end{table}

\begin{table*}[t!]
    \centering
    \caption{Main parameters of a selection of lines with broad components observed in the VLT/FORS2 spectra from 2010. 
    }
    \begin{tabular}{lccccccc}
    \hline\hline\\[-5pt]
        Line & Centre & Flux & EW & FWHM & FWHM & HWZI & HWZI  \\
            &   (\AA)   & (10$^{-16}$ CGS)
            &  (\AA)   &  (\AA)  & (km s$^{-1}$)  
            &  (\AA)  & (km s$^{-1}$) \\
        \hline\\[-7pt]
        H$\alpha$ (broad) &  $6582 \pm 1$  & $1238 \pm 20$  & $1770 \pm 30$  &  $37.5 \pm 0.4$  &  $1710 \pm 20$  &  $100 \pm 10$ & $4560 \pm 500$  \\
        H$\alpha$ (narrow) & $6574.0 \pm 0.1$  & $>$$(253 \pm 4)\tablefootmark{a}$  & $>$$(360 \pm 10)\tablefootmark{a}$  &  $<$$(2.98 \pm 0.03)\tablefootmark{a}$  &  $<$$(136 \pm 1)\tablefootmark{a}$  & -- & -- \\
        H$\alpha$ (P Cyg) &  $6565.3 \pm 0.5$  & $-140 \pm 10$  & $200 \pm 15$  &  $13.3 \pm 0.9$  &   $600 \pm 40$  & -- & --  \\
        \hline\\[-7pt]
        H$\beta$ (broad) &  $4874 \pm 3$  & $51 \pm 10$  & $85 \pm 21$  &  $25.5 \pm 4.2$  &  $1570 \pm 190$  &  $50 \pm 5$ & $3100 \pm 300$  \\
        H$\beta$ (narrow) & $4869.9 \pm 0.1$  & $97.8 \pm 1.4$  & $163 \pm 3$  &  $(2.36 \pm 0.03)\tablefootmark{b}$  &  $(145 \pm 2)\tablefootmark{b}$  & -- & -- \\
        H$\beta$ (P Cyg) &  $4860.0 \pm 1.0$  & $-9 \pm 3$  & $-15 \pm 5$  &  $9.5 \pm 4.0$  &   $590 \pm 250$  & -- & --  \\
        \hline\\[-7pt]
        H$\gamma$ (broad) &  $4355 \pm 4$  & $11 \pm 5$  & $18 \pm 8$  &  $24 \pm 10$  &  $1650 \pm 690$  &  $35 \pm 5$ & $2400 \pm 350$  \\
        H$\gamma$ (narrow) & $4348.1 \pm 0.1$  & $43.9 \pm 0.9$  & $69 \pm 2$  &  $(2.21 \pm 0.03)\tablefootmark{b}$  &  $(150 \pm 2)\tablefootmark{b}$  & -- & -- \\
        H$\gamma$ (P Cyg) &  $4340.1 \pm 1.6$  & $-2 \pm 1$  & $-3 \pm 2$  &  $6.5 \pm 3.0$  &   $450 \pm 200$  & -- & --  \\
        \hline\\[-7pt]
        He\,{\footnotesize{II}} $\lambda$4685.7 (b) &  $4696 \pm 4$  & $3.6 \pm 1.7$  & $6.0 \pm 2.8$  &  $30 \pm 10$  &  $1900 \pm 640$  &  $45 \pm 5$ & $2800 \pm 300$  \\
        He\,{\footnotesize{II}} $\lambda$4685.7 (n) & $4694.0 \pm 0.1$  & $9.9 \pm 0.3$  & $16.5 \pm 0.5$  &  $(2.09 \pm 0.05)\tablefootmark{b}$  &  $(134 \pm 3)\tablefootmark{b}$  & -- & -- \\
        \hline\\[-7pt]
        He\,{\footnotesize{I}} $\lambda$7065.2 (b) &  $7082.6 \pm 0.6$  & $14.5 \pm 1.1$  & $20.7 \pm 1.6$  &  $36 \pm 2$  &  $1500 \pm 100$  &  $36 \pm 5$ & $1500 \pm 200$  \\
        He\,{\footnotesize{I}} $\lambda$7065.2 (n) & $7077.4 \pm 0.1$  & $2.5 \pm 0.1$  & $3.6 \pm 0.2$  &  $3.01 \pm 0.12$  &  $128 \pm 5$  & -- & -- \\
        \hline
    \end{tabular}
    \tablefoot{We fitted the H\,{\footnotesize{I}} Balmer lines with two Gaussians (broad and narrow emission component) plus a third Gaussian in absorption for the P-Cygni absorption. The He\,{\footnotesize{I}} and He\,{\footnotesize{II}} lines were fitted with one broad and one narrow Gaussian component. The observed flux was not  corrected for slit losses; such losses were relatively small in the 2010 dataset, probably in the range of $\sim$10--20 per cent.
    \tablefoottext{a}{The narrow component is saturated in the FORS2 detector.}
        \tablefoottext{b}{Consistent with the instrumental resolution.}
        }
    \label{line_parameters}
\end{table*}

\begin{table}[t!]
    \centering
    \caption{Fluxes and FWHMs for the narrow-line components observed in the 2010 FORS2 spectrum.
    }
    \begin{tabular}{lcc}
    \hline\hline\\[-5pt]
     Line ID & Flux & FWHM    \\

             & &  (\AA)\\
        \hline\\[-7pt]
                
        [O\,{\footnotesize{II}}] $\lambda$3726.0 &  $32.5 \pm 4.3$  & $2.01 \pm 0.24$ \\

        [O\,{\footnotesize{II}}] $\lambda$3728.8 &  $42.2 \pm 4.3$  & $2.07 \pm 0.19$ \\
                
        H\,{\footnotesize{I}} $\lambda 3734.4$ (H13)&  $1.85 \pm 0.45$  & $2.01 \pm 0.39$  \\
        H\,{\footnotesize{I}} $\lambda 3750.2$ (H12)&  $2.64 \pm 0.27$  & $2.03 \pm 0.16$  \\
        H\,{\footnotesize{I}} $\lambda 3770.6$ (H11)&  $3.12 \pm 0.36$  & $1.75 \pm 0.16$  \\
        H\,{\footnotesize{I}} $\lambda 3797.9$ (H10)&  $4.99 \pm 0.34$  & $2.14 \pm 0.11$  \\
        H\,{\footnotesize{I}} $\lambda 3835.4$ (H9) &  $6.0 \pm 1.3$  & $1.99 \pm 0.36$ \\
        
        [Ne\,{\footnotesize{III}}] $\lambda 3868.8$ &  $43.9 \pm 0.85$  & $1.87 \pm 0.03$  \\
        
        H\,{\footnotesize{I}} $\lambda 3889.1$ (H8)&  $16.4 \pm 1.0$  & $2.09 \pm 0.11$  \\
        
        [Ne\,{\footnotesize{III}}] $\lambda 3967.5$\tablefootmark{a} &  $15.6 \pm 0.9$  & $2.0 \pm 0.2$  \\
        H\,{\footnotesize{I}} $\lambda 3970.1$ (H$\epsilon$)\tablefootmark{a} &  $10.6 \pm 0.5$  & $2.0 \pm 0.2$ \\
        He\,{\footnotesize{I}} $\lambda 4026.2$ &  $1.42 \pm 0.11$  & $2.07 \pm 0.11$ \\
        
        [S\,{\footnotesize{II}}] $\lambda 4068.6$ &  $1.20 \pm 0.12$  & $2.64 \pm 0.24$ \\
        
        [S\,{\footnotesize{II}}] $\lambda 4076.3$\tablefootmark{b} &  $0.50 \pm 0.12$  & $3.7 \pm 0.7$ \\
        
        H\,{\footnotesize{I}} $\lambda 4101.7$ (H$\delta$)&  $23.5 \pm 1.0$  & $2.09 \pm 0.06$  \\
        He\,{\footnotesize{I}} $\lambda 4143.8$ &  $0.31 \pm 0.06$  & $1.6 \pm 0.3$ \\
        H\,{\footnotesize{I}} $\lambda 4340.5$ (H$\gamma$)&  $44.4 \pm 0.9$  & $2.21 \pm 0.03$  \\
      
        [O\,{\footnotesize{III}}] $\lambda 4363.2$ &  $14.6 \pm 0.9$  & $1.99 \pm 0.10$ \\  
        He\,{\footnotesize{I}} $\lambda 4387.9$ &  $0.35 \pm 0.07$  & $2.1 \pm 0.3$ \\
        He\,{\footnotesize{I}} $\lambda 4471.5$ &  $3.15 \pm 0.13$  & $2.18 \pm 0.06$ \\

        [Fe\,{\footnotesize{III}}] $\lambda 4658.1$ & $0.76 \pm 0.07$  & $3.06 \pm 0.21$ \\

        He\,{\footnotesize{II}} $\lambda 4685.7$ &  $10.1 \pm 0.3$  & $2.09 \pm 0.05$ \\
        
        [Ar\,{\footnotesize{IV}}] $\lambda 4711.4$\tablefootmark{c} &  $2.42 \pm 0.16$  & $2.20 \pm 0.11$ \\
        
        [Ar\,{\footnotesize{IV}}] $\lambda 4740.2$ &  $1.66 \pm 0.08$  & $1.94 \pm 0.07$ \\
        
        H\,{\footnotesize{I}} $\lambda 4861.3$ (H$\beta$)&  [100]\tablefootmark{d}  & $2.36 \pm 0.03$  \\
        He\,{\footnotesize{I}} $\lambda 4921.9$ &  $0.91 \pm 0.10$  & $2.13 \pm 0.16$ \\

                [O\,{\footnotesize{III}}] $\lambda 4958.9$ &  $213 \pm 5$  & $2.18 \pm 0.04$ \\

        [Fe\,{\footnotesize{III}}] $\lambda 4986.7$ & $0.67 \pm 0.05$  & $2.49 \pm 0.14$ \\
                
                [O\,{\footnotesize{III}}] $\lambda 5006.8$\tablefootmark{e} &  $\sim$640   & $<$2.47\\
                
        He\,{\footnotesize{I}} $\lambda 5015.7$ &  $1.56 \pm 0.15$  & $2.28 \pm 0.16$ \\

        He\,{\footnotesize{II}} $\lambda 5411.5$ &  $0.35 \pm 0.11$  & $2.7 \pm 0.7$ \\

        He\,{\footnotesize{I}} $\lambda 5875.6$ &  $8.1 \pm 0.4$  & $2.22 \pm 0.09$ \\

        [O\,{\footnotesize{I}}] $\lambda 6300.3$ &  $2.11 \pm 0.16$  & $2.50 \pm 0.15$ \\
        
        [S\,{\footnotesize{III}}] $\lambda 6312.1$ &  $1.49 \pm 0.15$  & $2.21 \pm 0.18$ \\
        
        [O\,{\footnotesize{I}}] $\lambda 6363.8$ &  $0.76 \pm 0.11$  & $2.57 \pm 0.33$ \\
                
        [Ar\,{\footnotesize{V}}] $\lambda 6435.1$ &  $0.26 \pm 0.06$  & $2.8 \pm 0.5$ \\
        H\,{\footnotesize{I}} $\lambda 6562.8$ (H$\alpha$)\tablefootmark{f}&  $\sim$300  & $<3$  \\
        He\,{\footnotesize{I}} $\lambda 6678.2$ &  $2.26 \pm 0.10$  & $2.26\pm 0.07$ \\

        [S\,{\footnotesize{II}}] $\lambda 6716.4$ &  $7.96 \pm 0.45$  & $2.52 \pm 0.11$ \\
        
        [S\,{\footnotesize{II}}] $\lambda 6730.8$ &  $6.29 \pm 0.35$  & $2.68 \pm 0.11$ \\
                        
        [Ar\,{\footnotesize{V}}] $\lambda 7005.4$ &  $0.33 \pm 0.04$  & $1.84 \pm 0.15$ \\
        He\,{\footnotesize{I}} $\lambda 7065.2$ &  $2.60 \pm 0.14$  & $3.01\pm 0.12$ \\  
        
        [Ar\,{\footnotesize{III}}] $\lambda 7135.8$ &  $5.57 \pm 0.30$  & $2.39 \pm 0.10$ \\
        
        He\,{\footnotesize{I}} $\lambda 7281.3$ &  $0.48 \pm 0.07$  & $2.27\pm 0.25$ \\  
        
        [O\,{\footnotesize{II}}] $\lambda\lambda$7318.9,7320.0 &  $2.01 \pm 0.10$  & $3.43 \pm 0.14$ \\ 
        
        [O\,{\footnotesize{II}}] $\lambda\lambda$7329.7,7330.7 &  $1.63 \pm 0.11$  & $3.77 \pm 0.19$ \\
        \hline
    \end{tabular}
    \tablefoot{Fluxes are given as a fraction of $F({\mathrm{H}}\beta,{\mathrm{narrow}}) \equiv 100$. FWHMs are consistent with the instrumental resolution, except in a few cases where lines are blended, as indicated below. 
        \tablefoottext{a}{Partly blended.}
        \tablefoottext{b}{Likely contamination from O\,{\footnotesize{II}} $\lambda 4075.9$.}
        \tablefoottext{c}{Possible contamination from He\,{\footnotesize{I}} $\lambda 4713.1$.}
        \tablefoottext{d}{$F(H\beta,{\mathrm{narrow}}) = (97.8 \pm 1.4) \times 10^{-16}$ erg cm$^{-2}$ s$^{-1}$, not including corrections for slit losses, which we estimate at $\sim$10--20 per cent.}
                \tablefoottext{e}{The actual measurement is $560 \pm 14$, but the line is mildly saturated.}
        \tablefoottext{f}{Saturated.}
}
    \label{line_table_2010}
\end{table}

\begin{table}[t!]
    \centering
    \caption{Fluxes and FWHMs of the lines observed in the 2024 MUSE spectrum of the extended He$^{++}$ nebula. (0\farcs8 radius centred between star A and star B). Fluxes are given as a fraction of $F({\mathrm{H}}\beta,{\mathrm{narrow}}) \equiv 100$.  
    }
    \begin{tabular}{lcc}
    \hline\hline\\[-5pt]
     Line ID & Flux & FWHM    \\

             & &  (\AA)\\
        \hline\\[-7pt]
                [Fe\,{\footnotesize{III}}] $\lambda 4658.1$ & $0.9 \pm 0.3$  & $3.8 \pm 0.7$ \\
        He\,{\footnotesize{II}} (n) $\lambda 4685.7$ &  $4.9 \pm 0.3$  & $2.7 \pm 0.1$ \\
        He\,{\footnotesize{II}} (b) $\lambda 4685.7$ &  $1.2 \pm 0.6$  & $8.9 \pm 2.7$ \\
        
        [Ar\,{\footnotesize{IV}}] $\lambda 4711.4$\tablefootmark{a} &  $2.2 \pm 0.4$  & $4.7 \pm 0.7$ \\
        
        [Ar\,{\footnotesize{IV}}] $\lambda 4740.2$ &  $1.2 \pm 0.5$  & $2.8 \pm 0.9$ \\
        
        H\,{\footnotesize{I}} $\lambda 4861.3$ (H$\beta$, n)&  [100]\tablefootmark{b}  & $3.05 \pm 0.01$  \\
        H\,{\footnotesize{I}} $\lambda 4861.3$ (H$\beta$, b)&  $3 \pm 1$  & $15 \pm 5$  \\
        H\,{\footnotesize{I}} $\lambda 4861.3$ (H$\beta$, PCyg)&  $-2.3 \pm 1.0$  & $11.8 \pm 4.4$  \\

        [O\,{\footnotesize{III}}] $\lambda 4958.9$ &  $196 \pm 3$  & $2.86 \pm 0.02$ \\
        
        [Fe\,{\footnotesize{III}}] $\lambda 4986.7$ & $0.86 \pm 0.25$  & $3.4 \pm 0.7$ \\
                
        [O\,{\footnotesize{III}}] $\lambda 5006.8$ &  $585 \pm 3$   & $2.73 \pm 0.01$\\

        He\,{\footnotesize{I}} $\lambda 5015.7$ &  $1.2 \pm 0.4$  & $3.0 \pm 0.4$ \\
        
        He\,{\footnotesize{II}} $\lambda 5411.5$ &  $0.5 \pm 0.1$  & $3.3 \pm 0.8$ \\

        He\,{\footnotesize{I}} $\lambda 5875.6$ &  $9.9 \pm 1.2$  & $2.80 \pm 0.03$ \\
        
        [O\,{\footnotesize{I}}] $\lambda 6300.3$ &  $2.47 \pm 0.12$  & $2.74 \pm 0.10$ \\
        
        [S\,{\footnotesize{III}}] $\lambda 6312.1$ &  $1.85 \pm 0.12$  & $2.75 \pm 0.14$ \\
        
        [O\,{\footnotesize{I}}] $\lambda 6363.8$ &  $0.99 \pm 0.12$  & $3.08 \pm 0.24$ \\

        [N\,{\footnotesize{II}}] $\lambda 6548.1$ &  $1.60 \pm 0.25$  & $2.65 \pm 0.29$ \\

        H\,{\footnotesize{I}} $\lambda 6562.8$ (H$\alpha$, n) &  $306 \pm 2$  & $2.87 \pm 0.02$  \\
        H\,{\footnotesize{I}} $\lambda 6562.8$ (H$\alpha$, b) &  $22 \pm 2$  & $25 \pm 0.5$  \\
        H\,{\footnotesize{I}} $\lambda 6562.8$ (H$\alpha$, PCyg) &  $-4.3 \pm 1.9$  & $8 \pm 2$  \\

        [N\,{\footnotesize{II}}] $\lambda 6583.5$ &  $4.69 \pm 0.12$  & $2.78 \pm 0.04$ \\
        
        He\,{\footnotesize{I}} $\lambda 6678.2$ &  $2.96 \pm 0.12$  & $2.68\pm 0.03$ \\

        [S\,{\footnotesize{II}}] $\lambda 6716.4$ &  $10.0 \pm 1.4$  & $2.68 \pm 0.03$ \\
        
        [S\,{\footnotesize{II}}] $\lambda 6730.8$ &  $7.9 \pm 1.4$  & $2.87 \pm 0.04$ \\
                        
        He\,{\footnotesize{I}} $\lambda 7065.2$ &  $2.59 \pm 0.08$  & $4.16\pm 0.34$ \\  
        
        [Ar\,{\footnotesize{III}}] $\lambda 7135.8$ &  $7.28 \pm 0.10$  & $2.39 \pm 0.10$ \\
             
        He\,{\footnotesize{I}} $\lambda 7281.3$ &  $0.49 \pm 0.10$  & $2.27\pm 0.25$ \\  
    
        [O\,{\footnotesize{II}}] $\lambda\lambda$7318.9,7320.0 &  $2.35 \pm 0.13$  & $3.4 \pm 0.2$ \\ 
        
        [O\,{\footnotesize{II}}] $\lambda\lambda$7329.7,7330.7 &  $1.98 \pm 0.13$  & $3.2 \pm 0.2$ \\
        
        [Ar\,{\footnotesize{III}}] $\lambda$7751.1 &  $1.85 \pm 0.10$  & $2.64 \pm 0.12$ \\

        H\,{\footnotesize{I}} $\lambda 8467.3$ (Pa17) &  $0.33 \pm 0.06$  & $2.56 \pm 0.36$  \\

        H\,{\footnotesize{I}} $\lambda 8502.5$ (Pa16) &  $0.54 \pm 0.06$  & $2.76 \pm 0.26$  \\
        
        H\,{\footnotesize{I}} $\lambda 8545.4$ (Pa15) &  $0.64 \pm 0.10$  & $2.99 \pm 0.38$  \\

        H\,{\footnotesize{I}} $\lambda 8598.4$ (Pa14) &  $0.67 \pm 0.10$  & $2.85 \pm 0.31$  \\ 

        H\,{\footnotesize{I}} $\lambda 8665.0$ (Pa13) &  $0.80 \pm 0.07$  & $2.65 \pm 0.19$  \\

        H\,{\footnotesize{I}} $\lambda 8750.5$ (Pa12)\tablefootmark{c} &  $1.85 \pm 0.19$  & $4.3 \pm 0.4$  \\

        H\,{\footnotesize{I}} $\lambda 8862.8$ (Pa11) &  $1.43 \pm 0.22$  & $2.75 \pm 0.33$  \\

        H\,{\footnotesize{I}} $\lambda 9014.9$ (Pa10) &  $1.91 \pm 0.14$  & $3.03 \pm 0.17$  \\

        [S\,{\footnotesize{III}}] $\lambda$9068.6 &  $15.3 \pm 0.4$  & $2.82 \pm 0.05$ \\

        H\,{\footnotesize{I}} $\lambda 9229.0$ (Pa9) &  $2.69 \pm 0.05$  & $3.11 \pm 0.04$  \\

        \hline
    \end{tabular}

\tablefoot{Spectrum extracted from a circular region of radius 0\farcs8, centred between star A and star B. Fluxes are given as a fraction of $F({\mathrm{H}}\beta,{\mathrm{narrow}}) \equiv 100$.  
\tablefoottext{a}{Possible contamination from He\,{\footnotesize{I}} $\lambda 4713.1$.}
        \tablefoottext{b}{$F({\mathrm{H}}\beta,{\mathrm{narrow}}) = (104 \pm 0.5) \times 10^{-16}$ erg cm$^{-2}$ s$^{-1}$.}
        \tablefoottext{c}{Possible contamination from N\,{\footnotesize{I}} $\lambda 8747.4$.}
        }
    \label{line_table_2024}
\end{table}

\end{appendix}

\end{document}